\documentclass[fleqn,usenatbib]{mnras}

\usepackage{newtxtext,newtxmath}

\usepackage[T1]{fontenc}
\usepackage{ae,aecompl}

\usepackage{graphicx}	
\usepackage{amsmath}	
\usepackage{amssymb}	

\title[CRTS changing-state quasars]{Understanding extreme quasar optical variability with CRTS: II. Changing-state quasars}

\author[M. J. Graham et al.]{Matthew J. Graham,$^1$\thanks{E-mail:mjg@caltech.edu (MJG)} 
Nicholas~P.~Ross,$^2$
Daniel Stern,$^3$ 
Andrew~J.~Drake,$^1$ 
\newauthor
Barry~McKernan,$^{4,5,6}$
K.~E.~Saavik~Ford,$^{4,5,6}$
S.~G.~Djorgovski,$^1$ 
\newauthor
Ashish~A.~Mahabal,$^1$  
Eilat~Glikman,$^7$ 
Steve Larson,$^8$
\& Eric Christensen$^8$
\\
$^{1}$California Institute of Technology, 1200 E. California Blvd, Pasadena, CA 91125, USA \\
$^{2}$ Institute for Astronomy, University of Edinburgh, Royal Observatory, Blackford Hill, Edinburgh EH9 3HJ, UK \\
$^{3}$Jet Propulsion Laboratory, California Institute of Technology, 4800 Oak Grove Drive, Pasadena, CA 91109, USA \\
$^{4}$ Department of Science, CUNY Borough of Manhattan Community College, 199 Chambers Street, New York, NY 10007, USA\\
$^{5}$ Deparment of Astrophysics, American Museum of Natural History, Central Park West, New York, NY 10028, USA\\
$^{6}$ Physics Program, CUNY Graduate Center, 365 5th Avenue, New York, NY 10016, USA \\
$^{7}$Department of Physics, Middlebury College, Middlebury, VT 05753, USA\\
$^{8}$University of Arizona, Department of Planetary Sciences, Lunar and Planetary Lab, Tucson, AZ 85721, USA\\
}

\date{Accepted XXX. Received YYY; in original form ZZZ}

\pubyear{2019}

\hypersetup{draft}
\begin{document}
\label{firstpage}
\pagerange{\pageref{firstpage}--\pageref{lastpage}}
\maketitle

\begin{abstract}
We present the results of a systematic search for quasars in the Catalina Real-time Transient Survey exhibiting both strong photometric and spectroscopic variability over a decadal baseline. We identify 73 sources with specific patterns of optical and mid-IR photometric behavior and a defined spectroscopic change. 
These ``Changing-State'' quasars (CSQs) form a higher luminosity sample to complement existing sets of ``Changing-Look'' AGN and quasars in the literature. 
The CSQs (by selection) exhibit larger photometric variability than the CLQs. The spectroscopic variability is marginally stronger in the CSQs than CLQs as defined by the change in H$\beta$/[\ion{O}{III}] ratio. 
We find 36 sources with declining H$\beta$ flux, 37 sources with increasing H$\beta$ flux and discover seven sources with $z > 0.8$, further extending the redshift arm.
Our CSQ sample compares to the literature CLQ objects in similar distributions of H$\beta$ flux ratios and differential Eddington ratios between high (bright) and low (dim) states. Taken as a whole, we find that this population of extreme varying quasars is associated with changes in the Eddington ratio and the timescales imply cooling/heating fronts propagating through the disk.

\end{abstract}




\section{Introduction}

Quasar variability is generally regarded as a stochastic process. The summation of activity associated with accretion disk instabilities, ionizing continua, jets, stellar activity close to the core, and dust clouds are all potential contributors. Sparse studies, either in terms of sample size or temporal sampling, have produced a simple statistical model, the damped random walk \citep{kelly09}, which aims to describe this variability \citep[but see][for counter-discussions]{zu13,graham14,kasliwal15, kozlowski17,smith18, moreno18}. The growing availability of large collections of rich multiepoch data is, however, enabling a much more phenomenological approach. Systematic studies of the quasar population (or substantial fractions thereof) are now possible with different characterizations of variability in terms of discriminative features or statistical models. These aim to capture specific patterns of behavior associated with particular underlying physical processes. In this way, we have identified sources exhibiting periodic activity \citep{graham15a,graham15b}, major flaring \citep{graham17, drake18}, and extreme broad line variability \citep{stern17, stern18, ross18}.

Recent investigations of spectroscopic variability, primarily from dual epoch SDSS spectroscopy, have reported a number of objects with emerging or disappearing broad emission lines (BELs, prototypically H$\beta$) in their optical spectra and often large, order of magnitude changes in the optical photometry \citep{lamassa15, ruan16, runnoe16, macleod16, gezari17, runco16, yang17, assef18, stern18, wang18, ross18, macleod19}. Such {\em changing-look quasars} (CLQs) are consistent with a change of spectral type (broad-lined to narrow-lined or vice versa) and may, in principle, be associated with a large change of obscuration, accretion rate, or accretion disk luminosity. Microlensing is also a potential cause of the CLQ phenomena. The term ``changing-look quasar" is borrowed from X-ray astronomy where large changes in X-ray luminosity have been shown to be typically associated with varying absorption, e.g., \cite{matt03, rivers15a, rivers15b}. Similar significant spectral variability has also been known for many years in a number of local low-luminosity AGN \citep[e.g.,][]{khachikian71,tohline76,penston84,cohen86,bischoff99,aretxaga99,eracleous01,shappee14,denney14,li15,parker16}. We note that significant photometric and spectroscopic variability have also been detected in quasars with absorption lines systems, e.g., the extreme BAL QSO reported by \cite{stern17}, but such objects are not normally considered as CLQs.

Although this type of behavior has so far seemed rare, it may well be that data sets are only now sufficient in size and temporal coverage to effectively detect such activity. \cite{runco16} studied 102 local ($0.02 \le z \le 0.1$) Seyfert galaxies with $M_{BH} > 10^7 M_\odot$ and found that $\sim$66\% of objects showed variability (change in values) in width or flux in H$\beta$ over a 3--9 year timeframe and H$\beta$ completely disappeared in three sources. From a study of SDSS DR7 and DES Y3A1 data for 8,640 quasars, Rumbaugh et al. (2017, hereafter, R17)
 found that $\sim 10\%$ exhibited extreme variability ($|\Delta g| > 1$ mag) sometime within a 15-year baseline (the actual distribution of the restframe baseline over which the maximum $g$-band variability was observed peaks around 1000 days but is largely insensitive to timescales beyond $\sim$1500 days). Correcting for selection incompleteness, they speculate that 30--50\% of all quasars may actually show such behavior on these timescales. 

The link between extreme spectroscopic and photometric variability is not clear, however. Despite R17's suggestion that extreme variable quasars (EVQs, $|\Delta g| > 1$) are good candidates for CLQs, a large amplitude photometric variation alone is not enough to identify them. On the one hand, we have previously reported extreme variability that is not associated with significant spectroscopic changes \citep{graham17}. On the other hand, while \cite{macleod16} report on a sample of 10 CLQs with $|\Delta g| > 1$, \cite{yang17} present a sample of 21 CLQs, 15 of which have $|\Delta g| < 0.5$ (their full sample spans $0.03 \le |\Delta g| \le 1.89$). Therefore, observations show that the spectroscopic CLQ and photometric EVQ phenomena are not directly correlated. What does seem to be indicated is that there are specific {\it patterns} of variable behavior that are likely associated with CLQs: for example, Lawrence et al. (2016, hereafter, L16) identify a sample of 15 quasars, including one CLQ, showing slow steady changes over several years, which they attribute to a mixture of changes in accretion state and microlensing. It is also unclear whether CLQs stand out from {\it single} epoch spectroscopy. L16 report that slow blue hypervariables have weaker \ion{Mg}{II} and [\ion{O}{III}] emission lines. However, R17 find that EVQs have lower Eddington ratios and larger \ion{Mg}{II} and [\ion{O}{III}] equivalent widths (EWs) than control quasars matched in redshift and luminosity. Although L16 and R17 form a superset containing some CLQs, they do suggest that quasars with lower accretion rates are more susceptible to changes in accretion rate and exhibiting more extreme behavior. 

Strong correlations are reported as well between CLQ behavior and mid-infrared (MIR) variability \citep{sheng17,yang17,assef18} with optical and MIR colors also changing with flux variation: a bluer-when-brighter chromaticism in the optical and redder-when-brighter in the mid-infrared. Given the pc-scale size of the MIR-emitting region, this clearly indicates that the strong variability is not due to an obscuring screen. Instead the chromatic trends are likely due to less host galaxy contributions and a stronger inner accretion disk contribution when quasars are more luminous. 

Further evidence against obscuration being the primary cause of CLQs comes from optical polarimetric studies. If the disappearance of the broad emission lines originates from the obscuration of the quasar core by dusty clouds moving in the torus, high linear optical polarization would also be expected. Measurements of the polarization of CLQs \citep{hutsemekers17, hutsemekers19, marin17} are less than 1\%, which suggests that the phenomenon is not due to obscuration but physical changes in the accretion disk and/or accretion rate. Such low polarization degrees indicate as well that these quasars are seen under inclinations close to the system axis. Finally, imaging of the host galaxies of four faded CLQs \citep{charlton19} suggests that these are predominantly disrupted or merging galaxies that resemble AGN hosts, rather than inactive galaxies.


In this work, we present a search for quasars showing photometric and spectroscopic variability consistent with a change of state of activity. We will refer to these as {\em changing-state quasars (CSQs)} rather than changing-look quasars. Though the latter has been the conventional term in the literature to date, it is ill-defined with no clear phenomenology, be it photometric or spectroscopic, associated with it beyond ``significant" variability. In the optical community, literature often ascribes variability to either changes in obscuration (as in the X-ray community) or changes in accretion rate, and it has been a challenge to identify physical mechanisms that would lead to accretion rate changes. The significant MIR variability associated with the optical variability in these sources largely rules out the obscuration scenario as already noted. More recently, several papers have noted that the variability events occur on thermal timescales \citep{stern18, ross18, noda18, parker19}, implying that the luminosity changes are likely associated with rapid changes in the temperature of the accretion disk. In contrast, changes in the accretion rate would be expected on the viscous timescale, which is more than an order of magnitude longer (i.e., decades/centuries rather than years). CSQ is therefore the more appropriate term but we will continue to refer to those objects already identified in the literature as CLQs (see Table~\ref{tab:knownclqs} and Fig.~\ref{fig:knownclq} for those covered by the data used in this work).

This paper is structured as follows: in Section 2, we summarize the data sets used and in Section 3, we present the selection technique and criteria  for identifying CSQs. We discuss our results in Section 4. Section 5 considers implications for the physical mechanisms behind the variability. We assume a standard WMAP 9-year cosmology ($\Omega_\Lambda = 0.728$, $\Omega_M = 0.272$, $H_0 = 70.4$ km s$^{-1}$ 
Mpc$^{-1}$; \cite{jarosik}) and magnitudes are approximately on the Vega system.

\section{Data sets} 

A systematic search for sources showing the particular behavioral patterns associated with CSQs requires a data set with a long temporal baseline and also a high sampling rate. Some candidates may be identified from relatively few epochs of data spread over a roughly decadal baseline, but such data sets will typically have insufficient resolution or sensitivity to detect specific forms. The Catalina Real-time Transient Survey \citep[CRTS;][]{drake09} represents the best data currently available with which to systematically define sets of quasars with particular temporal characteristics.

\subsection{CRTS}
\label{crts}

The CRTS archive\footnote{http://catalinadata.org} contains the Catalina Sky Survey data streams from three telescopes --  the 0.7 m Catalina Sky Survey (CSS) Schmidt and 1.5 m Mount Lemmon Survey (MLS) telescopes in Arizona and  the 0.5 m Siding Springs Survey (SSS) Schmidt in Australia. These surveys, operated by the Lunar and Planetary Laboratory at the University of Arizona, were designed to search for near-Earth objects, but have proven extremely valuable for astrophysics topics ranging from Galactic transients \citep{drake14} to distant quasars \citep{graham14,graham15b,graham17}. CRTS covers up to $\sim$2500 deg$^2$ per night, with 4 exposures per visit, separated by 10 min. The survey observes over 21 nights per lunation. The data are broadly calibrated to Johnson $V$ (see \citealt{drake13} for details) and the current CRTS data set contains time series for approximately 400 million sources to $V \sim 20$ above Dec $> -30$ from 2003 to 2016 May (observed with CSS and MLS) and 100 million sources to $V \sim 19$ in the southern sky from 2005 to 2013 (from SSS).

The error model used for CRTS DR2 is incorrect: errors at the brighter magnitudes are overestimated and those at fainter magnitudes ($V > 18$) are underestimated \citep{palaversa13, drake14}. In this analysis, we employ the improved error model derived in \cite{graham17}; the actual CRTS error model will be fixed in a future release. We also note that none of the sources we consider have fewer than 10 observations in their light curve. We apply the same preprocessing steps described in \cite{graham15b} to all light curves, which remove outlier photometry points and combine all exposures for a given night to give a single weighted value for that night. We also remove sources associated with nearby bright stars or identifiable as blends from a combined multimodality in their magnitude and observation position, i.e., the spatial distribution of all points in a light curve is best described by $n > 1$ Gaussians. 

\subsection{WISE}

In this paper we use MIR $W1$ (3.4$\mu$m) and $W2$ (4.6 $\mu$m) {\em WISE} data from the beginning of the mission in 2010 January through 2017 December, corresponding to the fourth year of {\em NEOWISE} operations\footnote{http://irsa.caltech.edu/wise}. Note that there is a gap between 2011 February and 2013 September when the satellite was in hibernation. For most sky positions, there are $\sim$12 observations of a source over a $\sim$1 day period with a six-month gap between repeat visits. We combine all exposures for a given 24 hour period with a signal-to-noise ratio greater than five to produce a single value using the same method as for CRTS data. 

\subsection{Spectroscopically confirmed quasars}

The Million Quasars (MQ) catalogue\footnote{http://quasars.org/milliquas.htm} v5.2 contains all spectroscopically confirmed type 1 QSOs (577,146), AGN (30,062) and BL Lacs (1,615) in the literature up to 2017 August 5. Previous versions have formed the basis for the results of \cite{graham15b} and \cite{graham17}. MQ (v5.2) also contains 1,297,111 photometric quasar candidates from SDSS or {\em WISE}. We crossmatched MQ against the CRTS data set with a 3\arcsec \, matching radius and find that 1,411,364 sources are covered by the full CRTS. Of these, 268,202 do not have enough observations $(n < 10)$, leaving a data set of 1,143,162 quasars and quasar candidates. We also remove 3,724 known blazars based on the class designation in MQ and the BZCAT v5.0 catalog of blazars \citep{massaro15}. 

Table~\ref{tab:clqselect} gives a summary of this superset of MQ sources to which we now apply our selection criteria to identify CSQs.


\section{Selecting Changing State Quasars}
\label{sec:swv}

\subsection{Optical photometry selection}
Photometrically, we expect to see a gradual change in magnitude associated with monotonically varying broad emission line strengths and/or continuum changes. This would show as a strong localized trend or enhanced variability strength over some timescale in the time series. 

\begin{table}
\centering
\caption{A summary of the selection criteria employed to identify CSQs. Note that sources associated with nearby bright stars or identifiable as blends are removed before crossmatching against MQ.}
\label{tab:clqselect}
\begin{tabular}{lr}
\hline
Selection & Total \# \\
\hline
Number of MQ sources with CRTS light curve & 1,411,364 \\
with more than 10 observations & 1,143,162 \\
and not a known blazar &  1,139,438 \\
and outside 95\% contour in BB/SWV$_{1}$ space & 65,816 \\
and $\Delta |W1|$ or $ \Delta |W2| > 0.2$ & 47,451  \\ 
and $z < 0.95$ & 14,412 \\ 
and has SDSS spectrum & 7,576 \\ 
and has second epoch spectrum after $\ge 500$ days & 466 \\
and H$\beta$ / [\ion{O}{III}] ratio changes by > 30\% & 73 \\
\hline
\end{tabular}
\end{table}

\begin{figure}
\includegraphics[width = 3.5in]{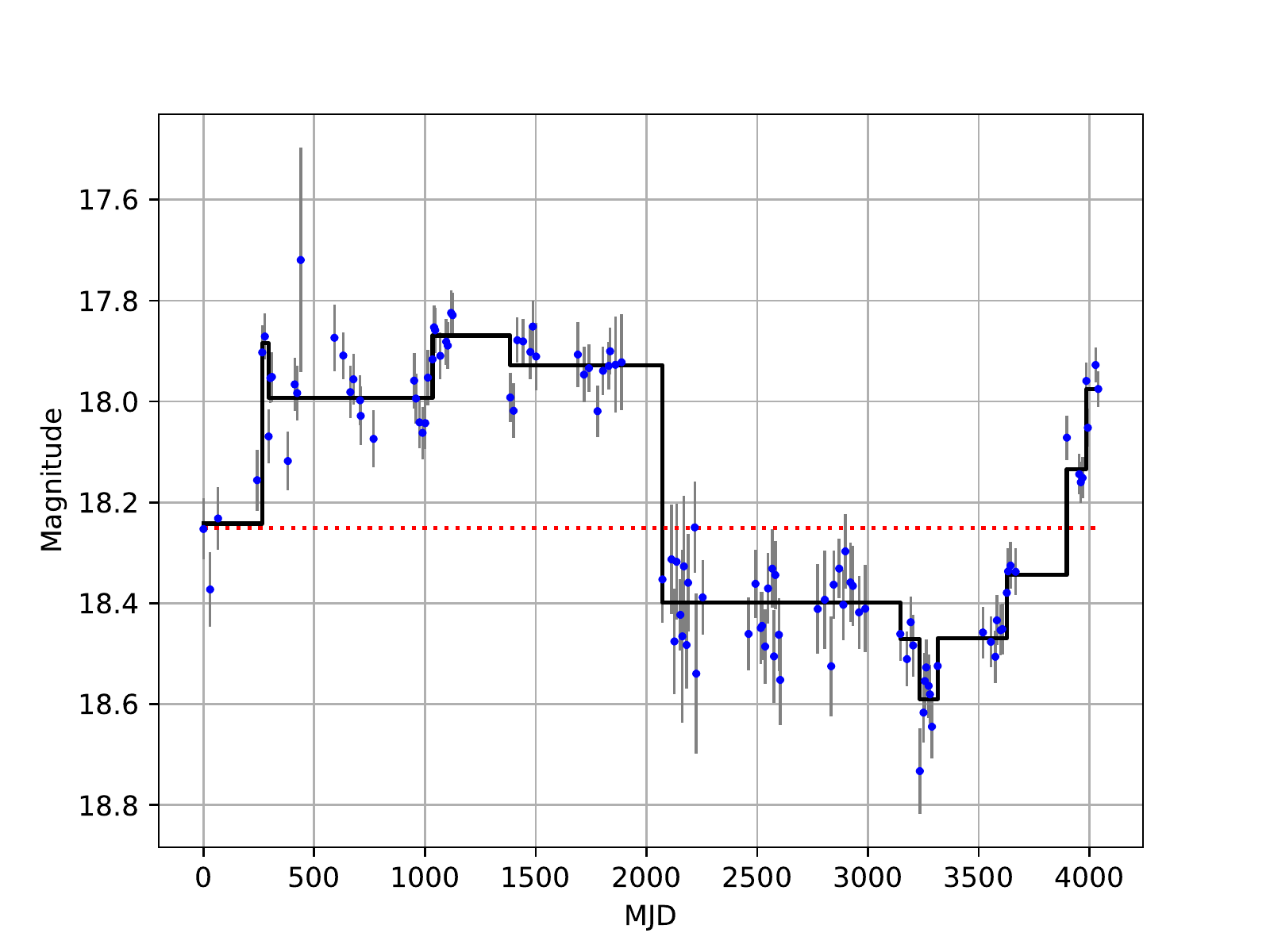}
\caption{A Bayesian block (BB) representation of a quasar time series. The red dotted line indicates the median magnitude. The difference between the first and last segments is $\Delta \mathrm{BB} = 0.3$ mag.}
\label{fig:bb}
\end{figure}

To identify local variability in a time series, we use a Bayesian blocks (BB) representation \citep{scargle13} which provides an optimal segmentation of the data in terms of a set of discontinuous piecewise constant components (see Fig.~\ref{fig:bb}). This approximation makes it easier to detect significant changes of behavior in the presence of irregular sampling, noise, and gaps. In particular, it is more sensitive to coherent magnitude variations over time characterized by the difference between the first and last piecewise segments of the BB than fitting a linear trend model to the data. We assume it unlikely that a quasar will undergo a transition and return to its initial state within the timeframe of our light curves. We also distinguish between this pattern of behavior and flaring as we identified in \cite{graham17}.

To detect stronger variability on particular timescales, we use the Slepian wavelet variance \citep[SWV;][]{graham14} which provides a measure of the relative contributions of variability at specific timescales to the total variability in a time series. We have determined the median observed SWV for quasars in magnitude bins of width $\Delta m = 0.25$. For a given source $j$, we then calculate the quantity:

\[ SWV_{1,j} = \sum_i \left[ \log_2 SWV_{i,j}(\tau_{i,j}) - \log_2 \overline{SWV} (\tau_{i,j}) \right] \]
 
\noindent
i.e., the sum of the differences in dyadic log-log space between the source SWV and the appropriate median SWV interpolated at the timescales of the source SWV. Note that dyadic logs are used since the wavelet bands are defined in terms of base-2 widths.

\begin{figure}
\includegraphics[width = 3.5in]{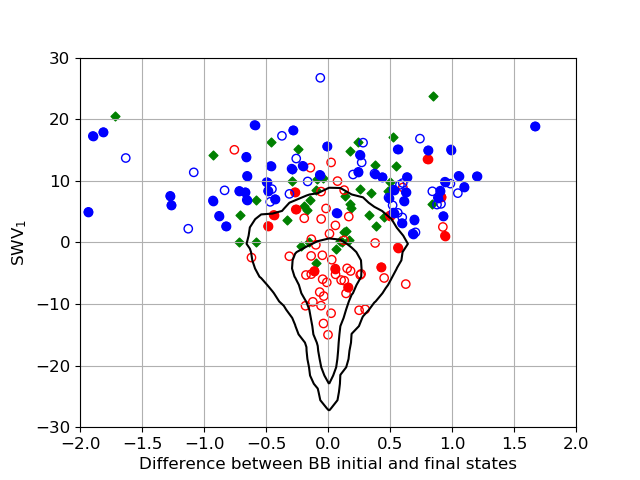}
\caption{The distribution of the magnitude difference between initial and final Bayesian block components and the similarity (aggregate distance) to the median Slepian wavelet variance for the quasars in our data set. Known CLQs are denoted by red circles, and blue circles indicate CSQs identified in this work. Open symbols indicate low luminosity sources with $M_V < -23$. The contours indicate the 68th and 95th percentile levels respectively  for a population of 128,000 spectroscopically confirmed quasars with $V < 19$. The green diamonds show quasars which have exhibited significant flaring activity \citep{graham17}.}
\label{fig:bbdiff}
\end{figure}

Fig.~\ref{fig:bbdiff} shows the distribution of the BB difference and $SWV_1$ for the quasars in our data set and for known CLQs. We note that using this characterization places the majority (82\%) of the known CLQs within the 95th percentile contour. These objects have predominantly been identified from their spectroscopic variability, e.g., from dual epoch SDSS spectra where the source is classified as a galaxy in one epoch and a quasar in the other. They will thus be in a quiescent galaxy state for at least a fraction of the time period covered by their CRTS light curve (see Fig.~\ref{fig:knownclq}) and show less photometric variability over this period than the median quasar at the same magnitude. Since our focus is on extreme variability, we only consider sources outside the 95th percentile contour and with $SWV_1 > 0$ as candidate objects. The latter criterion selects objects with above median variability.

\subsection{Mid-infrared Selection}
Strong ($> 0.4$ mag) MIR variability has been shown to be a characterizing property of CLQs \citep{sheng17, assef18, stern18} and so we also consider this as a discriminating feature. From the maximum $\Delta W1$ ($\Delta W2$) distribution of MIR variability for quasars over the seven years of {\em WISE} observations, $\Delta m = 0.4$ corresponds to the 82$^{\mathrm{nd}}$ (88$^{\mathrm{th}}$) percentile and 3\% (3\%) of sources vary by more than a factor of two in flux, or $\Delta m > 0.75$, over the period. \cite{yang17} argue that a reasonable selection criterion is $|\Delta(W1 - W2)| > 0.1$ when $|\Delta(W1)| > 0.2$ but they define $\Delta(W1)$ as the magnitude difference between the brightest epoch in a {\em WISE} time series and either the first or last epoch, depending on whether the CLQ is turning on or off. This means that unless the MIR flux is monotonically varying, 
$\Delta(W1)$ will be some fraction of the total $W1$ variability. It also relies on prior knowledge of whether the CLQ is turning on or off to determine the appropriate magnitude difference to measure. We employ the absolute magnitude difference in a {\em WISE} time series, i.e., $W_{\rm max} - W_{\rm min}$ - as a selection criterion and require either $|\Delta W1| > 0.2$ or $|\Delta W2|$ > 0.2 (see Table~\ref{tab:clqselect}).


\begin{figure}
\includegraphics[width = 3.5in]{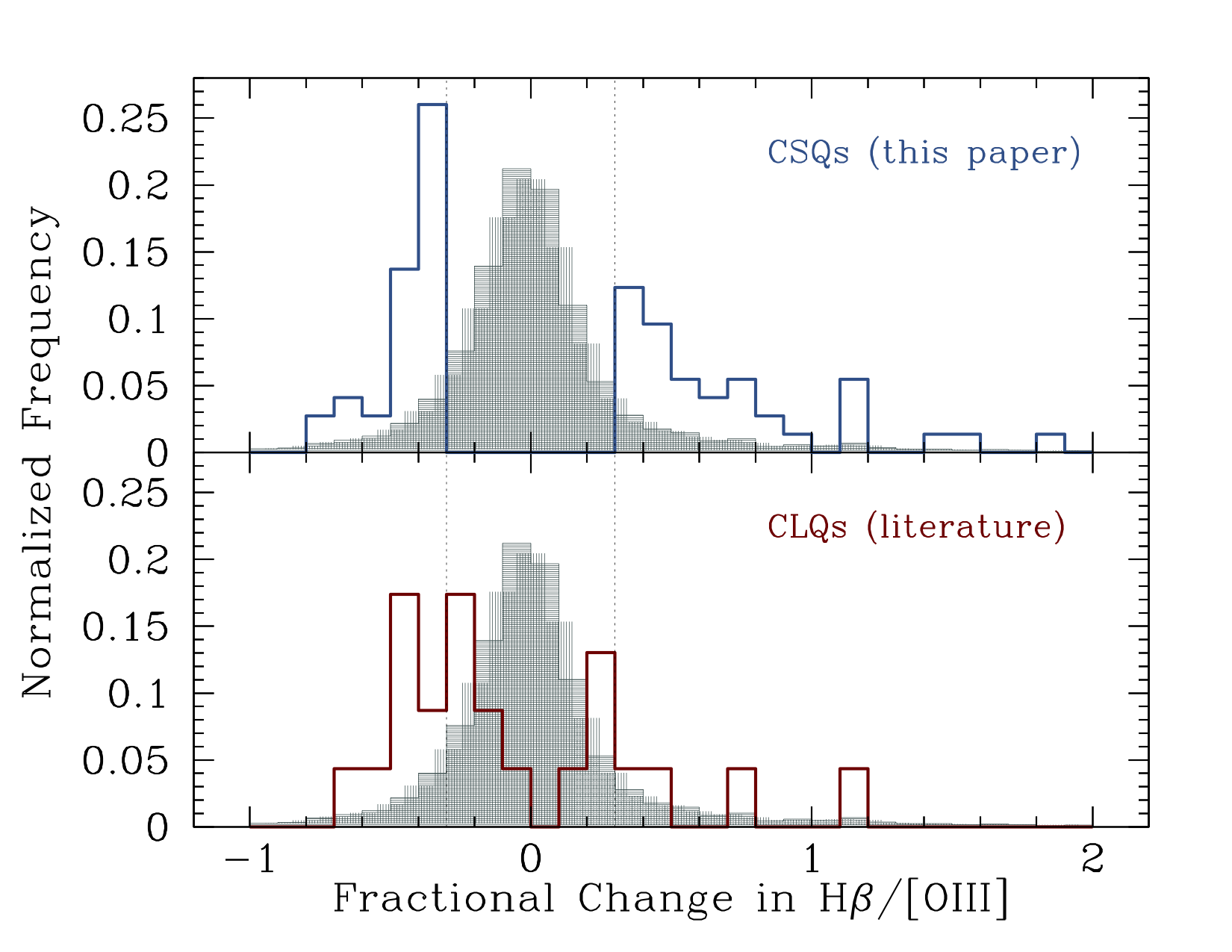} 
\caption{The distribution of the fractional change in the H$\beta$/[\ion{O}{III}] ratio for SDSS sources (gray) with multiepoch spectra and H$\beta$ coverage. The red (bottom panel) indicates the distribution for known CLQs from the literature with publicly available spectra and the blue (upper panel) indicates the CSQs selected here. The dotted lines show the spectroscopic variability selection criteria.}
\label{fig:hbeta}
\end{figure}

\begin{figure*}
\begin{minipage}{0.45\textwidth}
    \includegraphics[width = 3.4in]{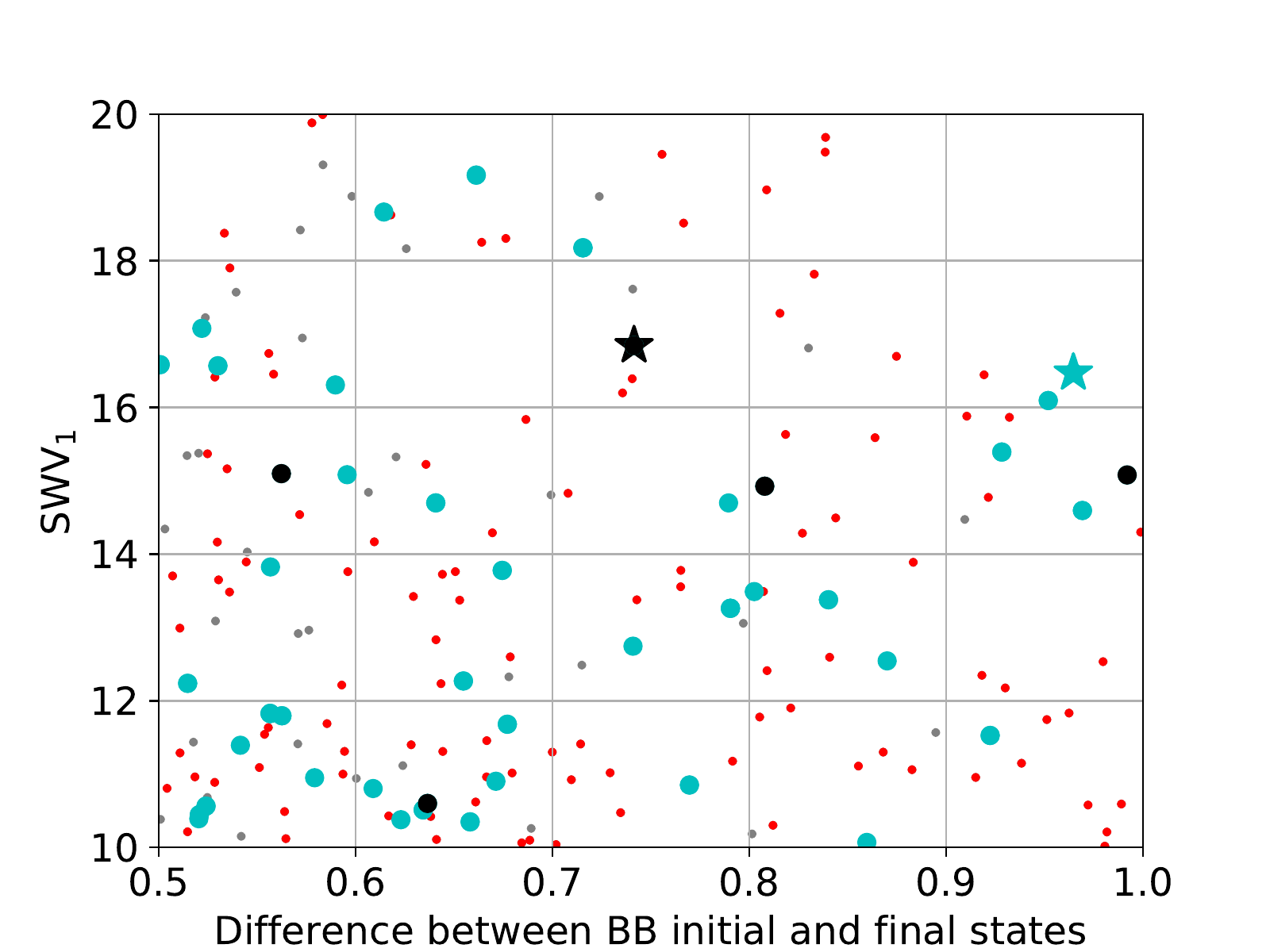}
\end{minipage}
\begin{minipage}{0.45\textwidth}
      \includegraphics[width = 3.5in]{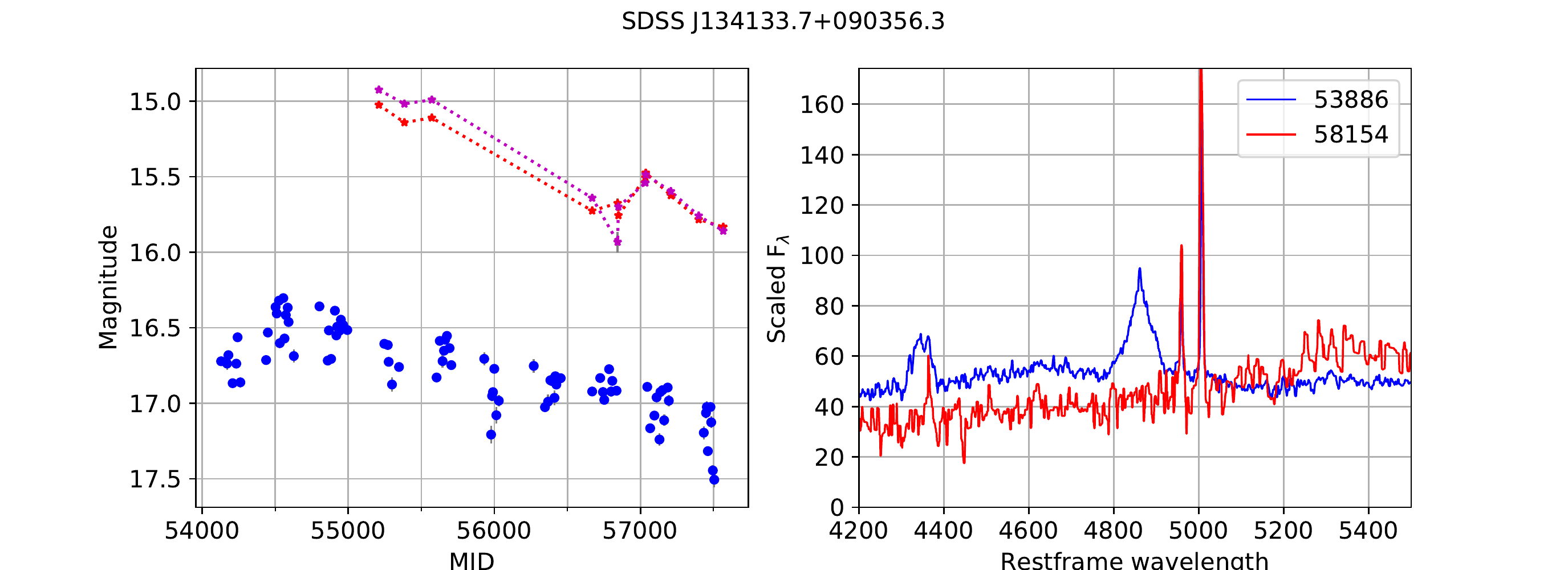} 
      \includegraphics[width = 3.5in]{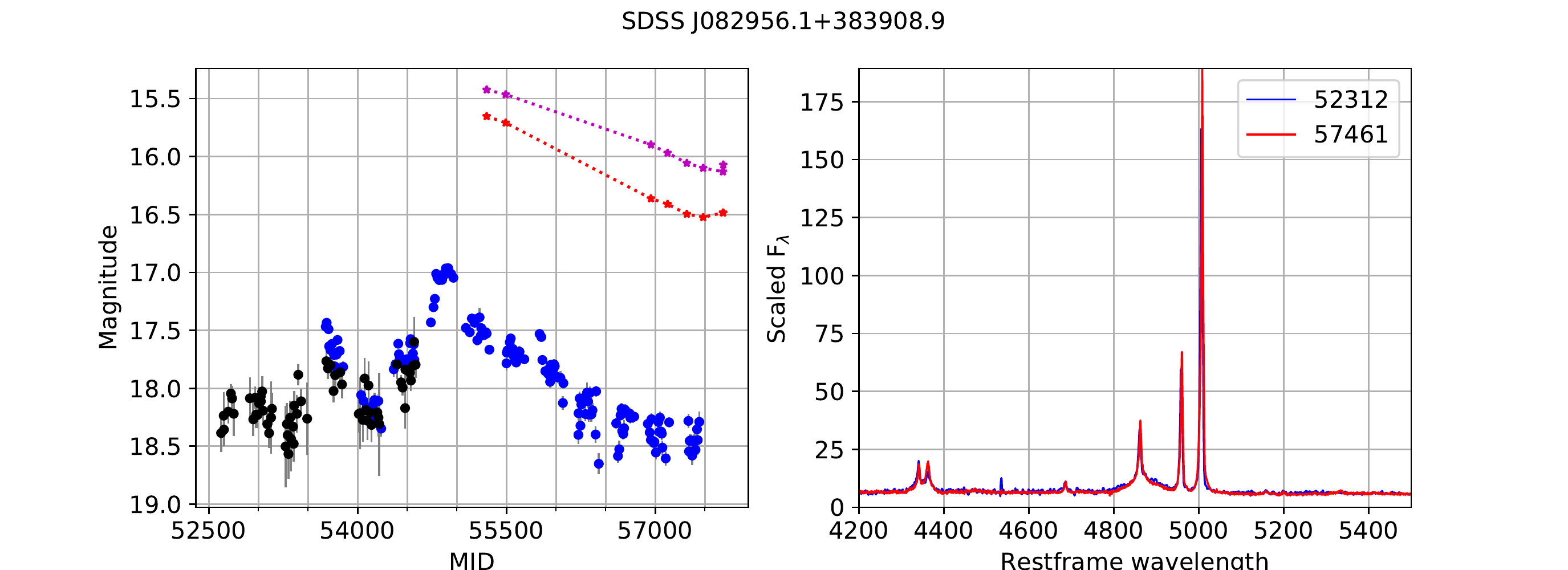} 
\end{minipage}
\caption{Left: A zoom-in of a region of Fig.~2 showing sources progressively matching the CSQ selection criteria so that each is a subset of the former: red indicates $|\Delta W_{1,2}| > 0.2$, cyan the presence of a second epoch spectrum, and black significant spectral change in the H$\beta$/[\ion{O}{III}] ratio. Right: the light curves and spectral region around H$\beta$ for two candidate CSQs (stars in the left panel). The top shows the data for a CSQ (the black star in the left plot) whilst the bottom shows the same for a source showing no spectral variability (the cyan star in the left plot). It is possible that a spectrum taken at MJD $\sim$54500 might have shown some variation. The light curves show data from CRTS (blue), LINEAR (black), and WISE (red).}
\label{fig:example}
\end{figure*}

\begin{table*}
\centering
\caption{CSQs selected in CRTS and associated features including the median CRTS magnitude ($V_m$), the optical amplitude (Amp) , absolute change in $W1$, Bayesian block change ($\Delta$BB), Slepian wavelet variance measure ($SWV_1$), and the change in flux ratio of H$\beta$ to [\ion{O}{III}]. SMBH virial mass estimates are calculated as described in Sec.~\ref{sec:luminosity}.
}
\label{tab:clqs}
\begin{tabular}{lccccccccc}
\hline
Name &  $V_m$ & $z$ & $\log(M_{BH})$ & $\log (L_V)$ & Amp & $\Delta$BB & $SWV_1$ & $|\Delta W1|$ & $\Delta ($H$\beta$ / [\ion{O}{III}$])$ \\
 & (mag) & & ($M_{\sun}$) & (erg s$^{-1}$) & (mag) & (mag) & & (mag) & \\
\hline
SDSS J002353.5$-$025159.1  & 16.94 & 0.246 & 7.5 & 44.01 & 0.44 & 0.81 & 15 & 0.44 & -0.31 \\
SDSS J011919.3$-$093721.6  & 19.04 & 0.383 & 7.4 & 43.60 & 0.79 & 0.84 & 8.3 & 0.50 & -0.39 \\
SDSS J022014.6$-$072859.3  & 17.02 & 0.213 & 7.8 & 43.83 & 0.31 & -0.30 & 12 & 0.29 & 0.75 \\
SDSS J024533.6$+$000745.2  & 18.64 & 0.655 & 8.2 & 44.27 & 0.62 & -0.49 & 9.8 & 0.27 & 0.91 \\
SDSS J025003.0$+$010930.7  & 17.44 & 0.194 & 7.2 & 43.61 & 0.46 & 0.27 & 13 & 1.00 & 0.41 \\
SDSS J025410.1$+$034912.5  & 19.03 & 0.774 & 8.1 & 44.34 & 0.75 & -0.88 & 4.3 & 0.53 & 0.37 \\
SDSS J025505.7$+$002523.5  & 18.47 & 0.353 & 7.6 & 43.79 & 0.77 & 1.20 & 11 & 0.63 & -0.40 \\
SDSS J025619.0$+$004501.0  & 19.17 & 0.723 & 7.9 & 44.19 & 0.87 & 0.38 & 11 & 0.64 & -0.62 \\
SDSS J074542.3$+$421404.5  & 17.94 & 0.268 & 7.5 & 43.71 & 0.56 & -0.83 & 8.4 & 0.44 & 0.65 \\
SDSS J075440.3$+$324105.1  & 17.97 & 0.411 & 7.7 & 44.13 & 0.72 & 0.26 & 14 & 0.30 & 0.40 \\
SDSS J075728.3$+$245510.1  & 18.53 & 0.187 & 7.6 & 43.13 & 0.73 & -1.08 &  11 & 0.34 & 1.19 \\
SDSS J080138.7$+$423355.2  & 19.93 & 0.771 & 7.4 & 43.92 & 0.98 & 0.60 & 3.1 & 0.72 & -0.35 \\
SDSS J080500.3$+$340225.6  & 18.32 & 0.401 & 7.3 & 43.93 & 0.70 & -0.66 & 14 & 0.41 & 0.66 \\
SDSS J081425.9$+$294116.3  & 19.12 & 0.374 & 8.0 & 43.54 & 0.69 & 0.71 & 1.6 & 0.40 & -0.32 \\
SDSS J081632.1$+$404804.6  & 19.73 & 0.701 & 8.5 & 43.92 & 1.07 & 0.70 & 3.7 & 0.40 & -0.48 \\
SDSS J082033.3$+$382420.4  & 18.58 & 0.648 & 7.6 & 44.29 & 0.52 & 0.44 & 11 & 0.49 & -0.33 \\
SDSS J082930.7$+$272821.9  & 18.29 & 0.321 & 7.5 & 43.72 & 0.53 & 0.91 & 6.3 & 0.57 & -0.40 \\
SDSS J083225.3$+$370736.6  & 15.52 & 0.092 & - & 43.67 & 0.56 & -0.06 & 27 & 0.36 & 1.89 \\
SDSS J083236.3$+$044506.2  & 18.42 & 0.292 & 8.2 & 43.58 & 0.70 & -0.16 & 9.9 & 0.55 & 0.38 \\
SDSS J083533.2$+$494818.8  & 18.49 & 0.198 & 7.7 & 43.19 & 0.51 & -0.31 & 7.9 & 0.89 & 0.75 \\
SDSS J084716.1$+$373218.7  & 17.85 & 0.454 & 7.4 & 44.23 & 0.41 & -0.07 & 11 & 0.31 & 0.30 \\
SDSS J091357.3$+$052229.8  & 19.22 & 0.346 & 5.8 & 43.45 & 0.64 & 0.56 & 4.8 & 0.64 & -0.57 \\
SDSS J092441.1$+$284730.6  & 18.52 & 0.464 & 8.3 & 43.97 & 0.85 & 0.53 & 8.6 & 0.83 & -0.37 \\
SDSS J092736.7$+$153824.3  & 18.25 & 0.555 & 7.9 & 44.26 & 0.59 & -0.91 & 8.4 & 0.45 & -0.32 \\
SDSS J092836.9$+$474245.8  & 19.58 & 0.830 & - & 44.11 & 0.83 &  0.68 & 1.4 & 0.73 & -0.77 \\
SDSS J093017.7$+$470721.7  & 16.52 & 0.160 & 7.3 & 43.75 & 0.47 & 0.28 & 16 & 0.48 & 0.40 \\
SDSS J093329.0$+$291734.1  & 17.90 & 0.262 & 6.8 & 43.67 & 0.54 & 0.60 & 9.6 & 0.76 & -0.48 \\
SDSS J094620.9$+$334746.5  & 16.19 & 0.239 & 7.9 & 44.26 & 0.54 & -0.59 & 19 & 0.86 & 0.50 \\
SDSS J095427.6$+$485638.9  & 18.47 & 0.248 & 8.2 & 43.38 & 0.65 & 1.05 & 8.0 & 0.59 & -0.37 \\
SDSS J095536.8$+$103751.7  & 17.46 & 0.284 & 7.5 & 43.93 & 1.08 & -1.90 & 17 & 0.52 & 2.95 \\
SDSS J095750.0$+$530106.0  & 18.34 & 0.437 & 7.4 & 43.97 & 0.85 & -0.20 & 12 & 0.57 & 1.14 \\
SDSS J100256.2$+$475027.9  & 18.41 & 0.391 & 8.2 & 43.84 & 0.69 & -0.65 & 11 & 0.49 & 0.88 \\
SDSS J100343.3$+$512611.2  & 18.25 & 0.431 & 7.7 & 43.99 & 0.51 & -0.43 & 7.0 & 0.41 & 0.39 \\
SDSS J102614.0$+$523752.0  & 17.68 & 0.259 & 7.4 & 43.74 & 0.69 & -0.37 & 17 & 0.99 & 0.35 \\
SDSS J102752.4$+$421012.5  & 18.42 & 0.933 & 8.4 & 44.68 & 0.64 & 0.93 & 4.3 & 0.60 & 1.13 \\
SDSS J102817.7$+$211508.1  & 18.46 & 0.365 & 8.3 & 43.76 & 0.61 & -0.93 & 6.8 & 0.53 & 0.46 \\
SDSS J103255.9$+$365451.0  & 18.91 & 0.894 & 8.8 & 44.45 & 0.70 & -0.82 & 2.6 & 0.35 & 0.57 \\
SDSS J104254.8$+$253714.2  & 18.30 & 0.603 & - & 44.33 & 0.61 & -0.72 & 8.4 & 0.23 & 0.49 \\
SDSS J110349.2$+$312416.7  & 20.27 & 0.438 & 7.1  & 43.22 &  1.02 & -1.13 & 2.2 & 2.20 & -0.38 \\
SDSS J111617.8$+$251035.0  & 18.36 & 0.534 & 8.3 & 44.16 & 0.92 & 0.63 & 8.1 & 0.50 & -0.45 \\
SDSS J111947.6$+$233539.9  & 17.65 & 0.147 & 7.4 & 43.22 & 0.58 & 0.20 & 11 & 0.62 & -0.49 \\
SDSS J112243.1$+$364141.6  & 18.15 & 0.313 & 7.7 & 43.74 & 0.45 & -0.45 & 8.6 & 0.57 & 0.47 \\
SDSS J113111.1$+$373709.4  & 18.62 & 0.448 & 7.9 & 43.90 & 0.76 & 0.63 &  11 & 0.27 & -0.36 \\
SDSS J113706.9$+$013948.2  & 16.56 & 0.193 & 7.4 & 43.92 & 0.42 & 0.56 & 15 & 0.49 & -0.31 \\
SDSS J113757.7$+$365501.8  & 18.78 & 0.861 & 7.9 & 44.46 & 0.85 & 0.88 & 7.3 & 0.37 & -0.30 \\
SDSS J114408.9$+$424357.5  & 18.08 & 0.272 & 8.6 & 43.63 & 0.62 & 0.88 & 6.1 & 0.29 & -0.39 \\
SDSS J120130.9$+$494049.8  & 18.00 & 0.392 & 7.6 & 44.02 & 0.68 & -0.46 & 12 & 0.74 & 0.39 \\
SDSS J120442.2$+$275411.6  & 16.34 & 0.165 & 8.0 & 43.86 & 0.44 & -0.28 & 18 & 0.51 & 0.58 \\
SDSS J123215.2$+$132032.3  & 17.71 & 0.286 & 8.1 & 43.84 & 0.45 & 0.49 & 7.3 & 0.67 & -0.65 \\
SDSS J123819.6$+$412420.4  & 18.94 & 0.499 & 8.5 & 43.88 & 0.56 & 0.53 & 4.8 & 0.40 & -0.39 \\
SDSS J125757.2$+$322929.6  & 18.20 & 0.806 & - & 44.63 & 1.08 & -1.81 & 18 & 2.00 & 1.58 \\
SDSS J134133.7$+$090356.3  & 16.80 & 0.105 & 7.4 & 43.27 & 0.60 & 0.74 & 17 & 0.81 & -0.70 \\
SDSS J134822.3$+$245650.4  & 17.98 & 0.293 & 7.5 & 43.74 & 0.45 & 0.60 & 4.1 & 0.47 & -0.42 \\
SDSS J135636.6$+$255320.0  & 18.91 & 0.277 & 7.2  & 43.32 & 0.76 & 0.58 & 9.3 & 0.75 & -0.41 \\
SDSS J142852.8$+$271042.9  & 17.64 & 0.445 & 7.7 & 44.28 & 0.55 & 0.25 & 11 & 0.47 & 0.76 \\
SDSS J144118.9$+$485454.8 & 18.60 & 0.289 & 7.0 & 43.52 & 0.89 &-0.47 &  6.6 & 0.78 & 0.73 \\
SDSS J144202.8$+$433709.1 & 17.27 & 0.231 & 7.2 & 43.80 & 0.93 & 1.67 & 19 & 1.10 & -0.70 \\
SDSS J144702.8$+$273747.2 & 18.45 & 0.224 & 7.6 & 43.31 & 0.68 & 0.99 & 9.5 & 0.99 & -0.49 \\
SDSS J145022.7$+$102555.8  & 17.86 & 0.790 & 8.0 & 44.76 & 0.70 & -1.26  & 6.0 & 0.49 & 3.03 \\
SDSS J145755.4$+$435035.5  & 18.50 & 0.528 & 7.5 & 44.10 & 0.51 & -0.67 & 8.1 & 0.66 & 0.56 \\
\hline
\end{tabular}
\end{table*}

\begin{table*}
\centering
\contcaption{}
\begin{tabular}{lcccccccccc}
\hline
Name & $V_m$ & $z$ & $\log(M_{BH})$ & $\log (L_V)$ & $\Delta V$ & $\Delta$BB & $SWV_1$ & $|\Delta W1|$ & $\Delta ($H$\beta$ / [\ion{O}{III}$])$ \\
 & (mag) & & ($M_{\odot}$) & (ergs s$^{-1}$) & (mag) & (mag) & & (mag) & \\
\hline
SDSS J151604.3$+$355025.4  & 18.39 & 0.592 & 7.9 & 44.26 & 0.50 & 0.61  &  6.7 & 0.35 & -0.34 \\
SDSS J152749.9$+$084408.6  & 18.87 & 0.849 & - & 44.44 & 1.24 & -1.93  & 5.0 & 0.28 & 0.37 \\
SDSS J153354.6$+$345504.6  & 18.79 & 0.753 & 8.8 & 44.34 & 0.93 & 1.10 & 9.0 & 0.95 & -0.35 \\
SDSS J153415.4$+$303434.5  & 15.92 & 0.093 & 7.3 & 43.51 & 0.31 & -0.26 & 14 & 0.77 & 0.34 \\
SDSS J155651.4$+$321008.9  & 17.88 & 0.350 & 7.77 & 43.96 & 0.77 & 1.05 & 11 & 1.50 & -0.37 \\
SDSS J155829.4$+$271714.3  & 16.69 & 0.090 & 7.6 & 43.19 & 0.35 & 0.52 & 6.0 & 0.28 & -0.46 \\
SDSS J160743.0$+$432817.1  & 18.46 & 0.596 & 7.8 & 44.23 & 0.69 & 0.94 & 9.8 & 0.92 & -0.54 \\
SDSS J161400.3$-$011005.1  & 17.87 & 0.253 & 7.6 & 43.78 & 0.75 & 0.99 & 15 & 0.99 & -0.31 \\
SDSS J224829.4$+$144418.4  & 18.75 & 0.424 & 7.7 & 43.84 & 0.48 & 0.07 & 6.6 & 0.25 & -0.49 \\
SDSS J230443.6$-$084110.0  & 13.56 & 0.047 & 7.6 & 43.86 & 0.26 & -0.01 & 16 & 0.86 & 0.68 \\
SDSS J231207.6$+$140212.8  & 17.82 & 0.357 & 7.7 & 44.05 & 0.54 & -0.48 & 8.4 & 0.19 & 0.52 \\
SDSS J233136.8$-$105638.0  & 17.79 & 0.373 & 7.4 & 44.06 & 0.44 & -0.65 & 6.9 & 0.21 & 1.16 \\
SDSS J235439.1$+$005751.9  & 18.92 & 0.390 & 8.0 & 43.66 & 1.20 & -1.63 & 14 & 1.50 & 1.43 \\
\hline
\end{tabular}
\end{table*}

The optical variability constraints described above give 65,816 candidates from the initial 1.1 million source data set (rejecting blends) and the MIR variability constraint reduces this to 47,451 sources. Of these, 14,412 have $z < 0.95$ and are therefore suitable for spectroscopic confirmation (i.e., with optical spectroscopy H$\beta$ falls within the wavelength coverage of the optical spectrum). A SDSS DR14 spectrum exists for 7,576 of these and multiepoch SDSS spectra with at least 100 (500) days between epochs are available for 466 (266) objects. For comparison, there are 213,358 SDSS DR14 \citep{abolfathi17} sources classed as QSOs or AGN with $z < 0.95$, of which 8,213 (4,244) have additional spectra taken at least 100 (500) days after the initial epoch.

\subsection{Spectroscopic Selection}
Over the past three years we have obtained second epoch spectra (all at least $>500$ days after the initial SDSS epoch) for an additional 172 candidates (and subsequent epoch spectra for another 35 sources) using either the Double Spectrograph (DBSP) on the Hale 200'' telescope at Palomar Observatory, the Low Resolution Imaging Spectrometer (LRIS) spectrograph on the Keck I telescope at the W. M. Keck Observatory, or the Echellete Spectrograph and Imager (ESI), also on the Keck I telescope (see Table~\ref{tab:observing}). All these spectra were processed using standard procedures and flux calibrated with observations of spectrophotometric standard stars from \cite{massey90} observed on the same night. We note, though, that the photometric quality of the nights was not consistent across multiple observing runs. We fit all spectra with a single power law continuum and measure H$\beta$ and [\ion{O}{III}] emission line profiles relative to this. We assume a two component Gaussian fit for H$\beta$ to model a broad and narrow component and single Gaussians for the $[\ion{O}{III}]$ lines.

There is no objective definition of the spectral variability required to qualify as a CLQ/CSQ in the literature. Some authors \citep{ruan16,yang17} rely on visual inspection to identify sources with obvious broad H$\beta$ emission in one epoch and no detection in another. \cite{macleod16} and \cite{macleod19} compute the flux deviation between two spectra at any given wavelength to determine the significance of a broad emission line (BEL) change and assess the significance of a change relative to the underlying continuum at that wavelength. The earlier work considers a range of significance from $< 2 \sigma$ change for a faint spectrum to $> 8\sigma$, but generally an absence of H$\beta$ at one epoch is still required. The more recent work assumes a significance in the flux deviation of H$\beta$ greater than 3. \cite{yang17} describe two sources transitioning from type 1 to type 1.8 where H$\beta$ does not entirely vanish. We can thus define a measure of this deviation to set a lower limit on the change in H$\beta$ required. 

The narrow [\ion{O}{III}] $\lambda 5007$ emission line is not expected to vary on human timescales and so we can use the change in the flux ratio of H$\beta$ to  [\ion{O}{III}]  between epochs as a robust indicator of H$\beta$ change, i.e., the ratio should not be significantly affected by systematic errors due to observing conditions or spectral reduction. Since there is an expectation (almost by definition) that CSQs are associated with significant spectral variability, we will consider only those sources where the absolute value of the fractional change in H$\beta$/[\ion{O}{III}] $> 0.3$. We also reject all spectra with a signal-to-noise ratio less than 5 as defined in \cite{stoehr08} using: $SNR = 0.605 * \mathrm{median}(f_i) /  \mathrm{median}(|2 f_i - f_{i-2} - f_{i + 2}|)$ where $f_i$ is the flux at pixel $i$.

Fig.~\ref{fig:hbeta} shows the distribution of the ratio for all SDSS sources with multiepoch spectra and H$\beta$ coverage, highlighting known CLQs (bottom panel). Table~\ref{tab:clqs} gives details of the 73 objects that meet our full selection criteria.  

\section{Results}

The light curves and spectra of our final selected sample of 73 CSQs are shown in Fig.~\ref{fig:newclqspec}. We find 36 sources with declining H$\beta$ and 37 sources with increasing H$\beta$. We reject 8 sources which meet the photometric selection criteria but have atmospheric O$_2$ A band absorption coincident with the H$\beta$ - [\ion{O}{III}] complex in the observed frame.

Fig.~\ref{fig:example} shows an example of two sources which pass the photometric selection criteria but only one of which also shows spectroscopic variability. This demonstrates that there are a variety of phenomena which may produce quantitively similar photometric variability but are distinguishable with multiepoch spectra. Note that they might also be differentiated by other observables such as X-ray or radio behavior but this requires further investigation. 

\begin{figure*}
\centering
\includegraphics[width = 6.4in]{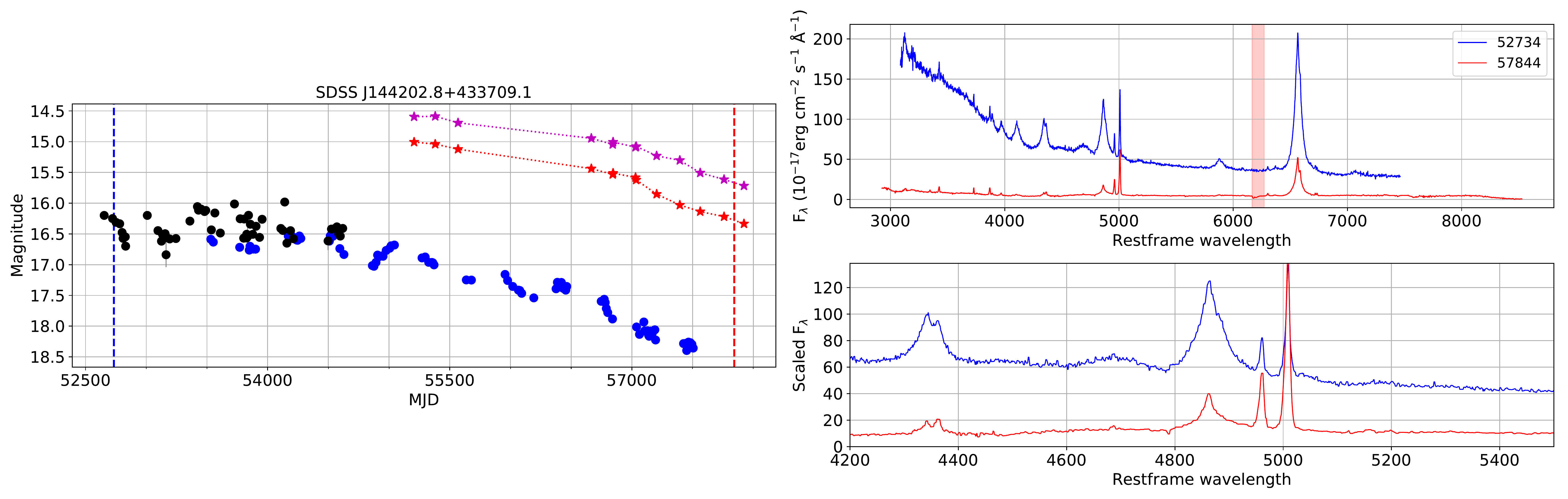}
\includegraphics[width = 6.4in]{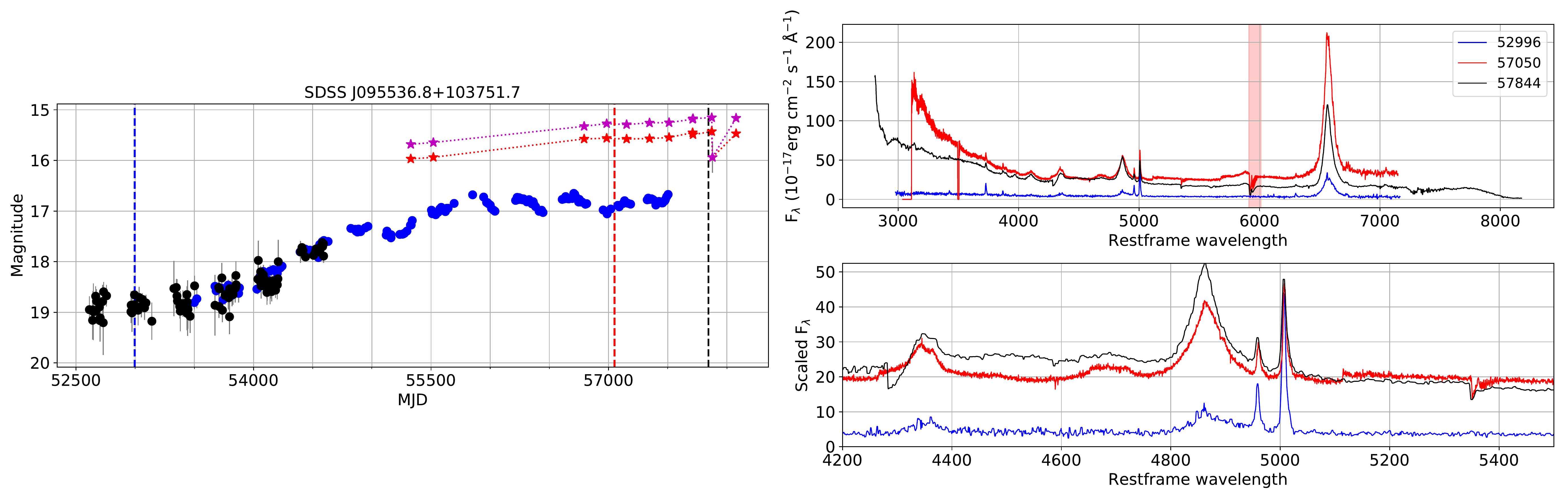}
\includegraphics[width = 6.4in]{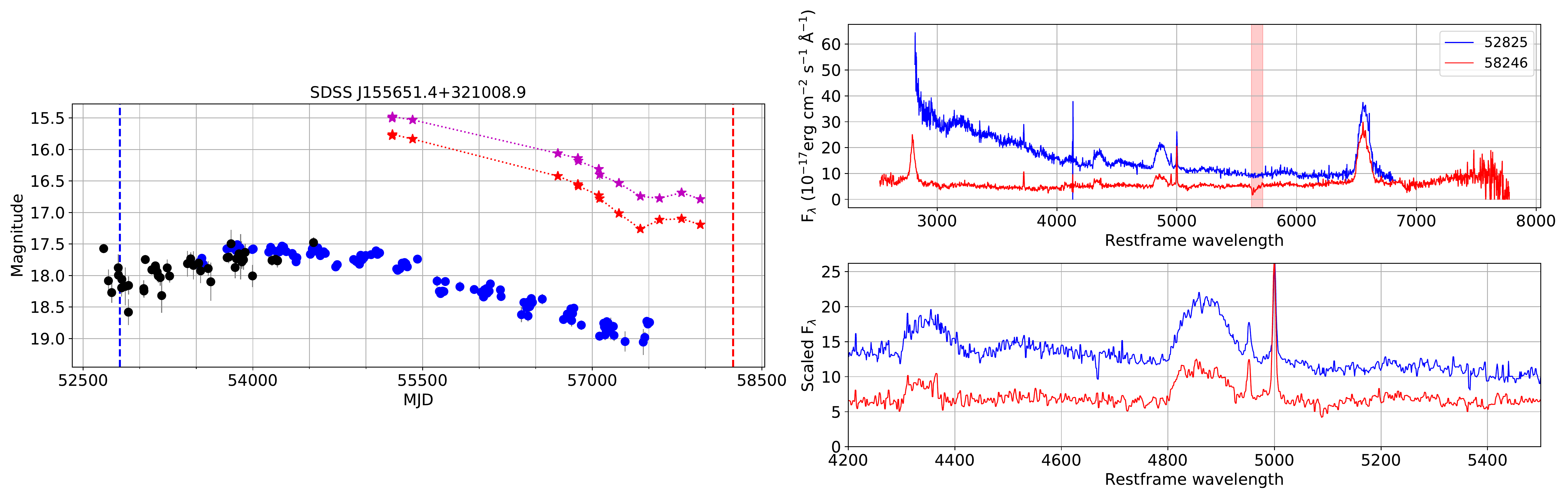}
\includegraphics[width = 6.4in]{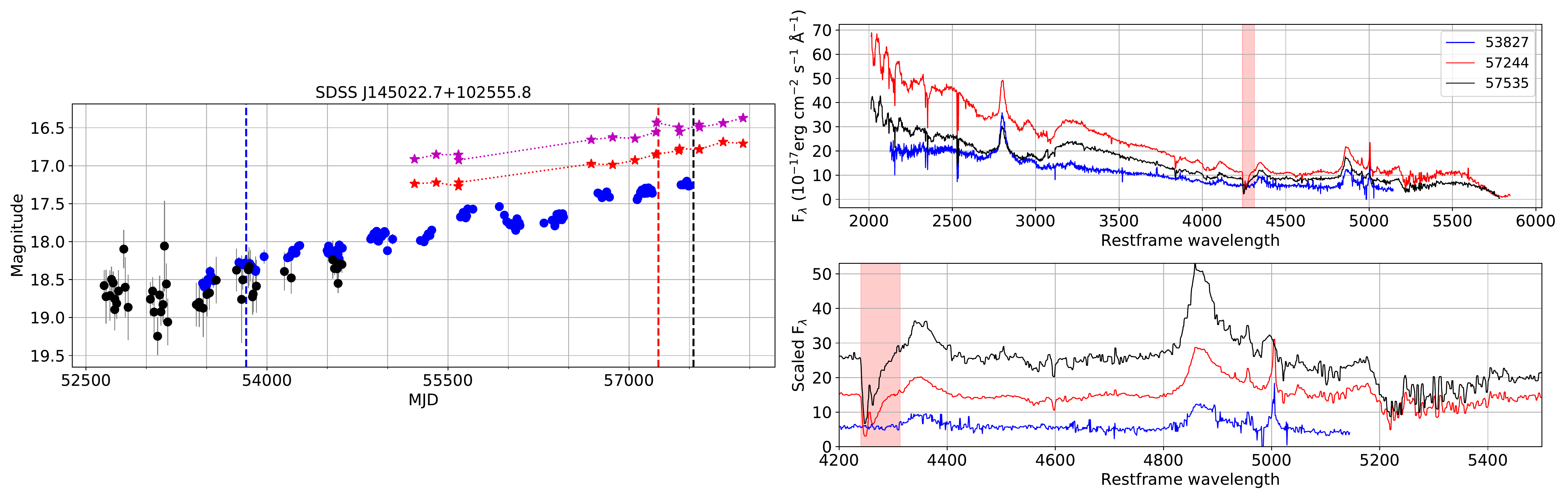}
\caption{Example light curves and spectra of CSQs identified in this paper corresponding to the largest values in each of the selection parameters: $SWV_1$, $\Delta BB$, $|\Delta W|$, and $\Delta($H$\beta$ / [\ion{O}{III}] (from top to bottom).  The left plot for each source shows the CRTS (blue) data and LINEAR (black) data where available. The {\em WISE} $W1$ (maroon) and $W2$ (red) light curves are also shown (binned on a daily basis) and offsets ($W1$ = 2.70, $W2$ = 3.34) have been applied to the {\em WISE} Vega magnitudes for display. The right upper plot for each source shows the SDSS spectra and the spectra obtained in our followup. The lower right plot shows a comparison of the H$\beta$ regions for the spectra scaled to the flux of [\ion{O}{III}] $\lambda \, 5007$ in the earliest spectra. The spectra are smoothed with a 3 \AA\,\. box filter in all cases. The red shaded area indicates the location of the atmospheric O$_2$ A-band absorption feature. The corresponding epochs of the spectra are shown in the left plot by dashed lines.}
\label{fig:extremes}
\end{figure*}

Figure~\ref{fig:extremes} presents the light curves and spectra of CSQs identified in this paper corresponding to the largest values in each of the selection parameters: $SWV_1$, $\Delta BB$, $|\Delta W|$, and $\Delta($H$\beta$ / [\ion{O}{III}]. 
From Figs~\ref{fig:extremes} and \ref{fig:newclqspec}, we find that extreme variable sources are a heterogeneous population reflecting the complexity of the physics of accretion with subsets dominated by particular processes. Different selection techniques drawing from this population may probe different physics and this needs to be borne in mind in any analysis.

\begin{figure}
\includegraphics[width = 3.3in]{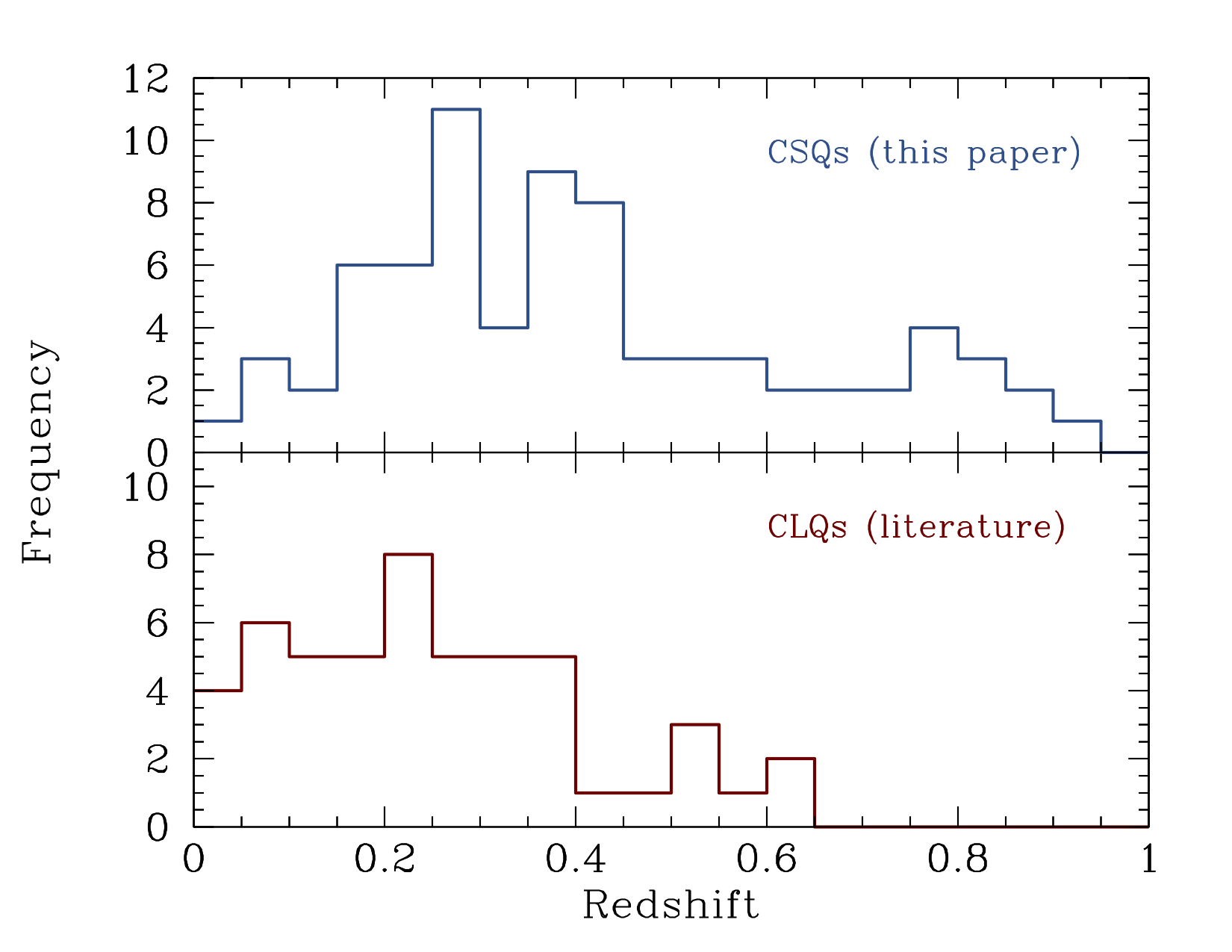} 
\caption{The redshift distribution of known CLQs from the literature (red) and the CSQs selected here (blue).}
\label{fig:redshift}
\end{figure}

We have identified a sample of CSQs on the basis of excess optical variability over specific timescales associated with a distinct change in levels of activity over a decade, strong MIR variability, and changes in the strength of H$\beta$ emission. We are keen to compare this study to previous and contemporary analyses but note that this is not a straightforward exercise. Table~\ref{tab:knownclqs} and Fig.~\ref{fig:knownclq} describe the collective set of CLQs from the literature with CRTS light curves and publicly available spectra. Since these are the result of a variety of selection techniques, we will consider in this section the two samples - CSQs and CLQs - in terms of the selection criteria used in this work.

\begin{figure}
\includegraphics[width=3.5in]{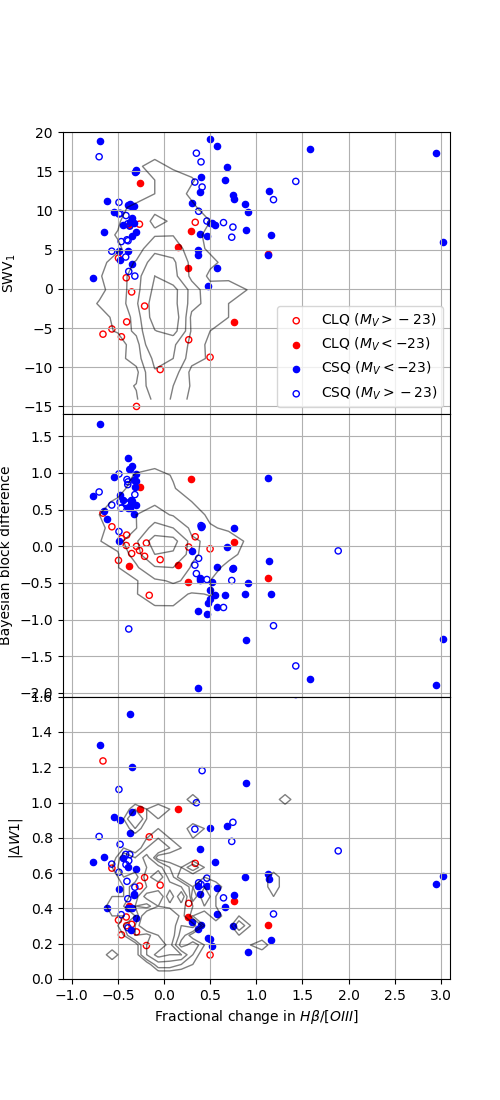}
\caption{The distributions of the photometric selection criteria -- the distance to the median Slepian wavelet variance curve, $SWV_1$,  the difference between initial and final Bayesian block (BB) states, and the maximum $W_1$ difference -- against the spectroscopic variability for the CSQs identified in this paper (blue), known CLQs in the literature (red) where the multiepoch spectra are available, and the general population of quasars with CRTS lightcurves (gray contours). Open circles indicate low luminosity sources with $M_V > -23$. A positive BB difference indicates a transition from a higher (brighter) to a lower (fainter) state.}
\label{fig:hbphot}
\end{figure}

Fig.~\ref{fig:redshift} shows that the CSQ sample spans a wider redshift range than the known CLQs. The distributions of these two samples as well as the general quasar population in terms of the photometric criteria compared to the spectroscopic are shown in Fig.~\ref{fig:hbphot}. We also distinguish between low and high luminosity sources (taking $M_V < -23$ as the fiducial boundary). The absolute magnitude of a source is given by:
\[ M_V = m_V - A_V - DM - K_V \]

\noindent
where $A_V$ is the Galactic extinction, $DM$ is the distance modulus, $K_V$ is the K-correction, and $m_V$ is the median magnitude from the CRTS lightcurve. We obtain\footnote{http://irsa.ipac.caltech.edu/applications/DUST/} extinction values at the source position from the \cite{schlafly11} recalibration of the \cite{schlegel98} reddening maps. We assume a K-correction of: $K = -2.5 (\alpha + 1) \log (1 + z)$ for a power law SED of $F_{\nu} \propto \nu^\alpha$ with $\alpha = -0.5$.

It is clear that the CSQ sample represents a more extreme level of photometric variability than the known CLQ set and that the lower luminosity CLQs have a less temporally characterizable variability (smaller SWV$_1$ values) than their brighter counterparts (Fig.~\ref{fig:bbdiff}). This is could be likely due to an overall stronger host contribution to the optical light curve from the generally lower redshift, lower luminosity CLQs as compared to CSQs. The spectroscopic variability also seems marginally stronger in CSQs with objects transitioning from a lower to a higher state of activity, although there is no distinction between high and low luminosity sources within this group.

The difference between the initial and final states in the Bayesian block representation of the light curve correlates with the fractional change in the H$\beta$/[\ion{O}{III}] ratio, although there is no difference between low and high luminosity sources. However, the flux change associated with just H$\beta$ (dis)appearing is not sufficient to account for the full scale of the magnitude change seen in the light curves. Using a composite quasar spectrum, such as \cite{selsing16}, the magnitude difference for a source with $V \sim 18$ at $z = 0.25$ (which places H$\beta$ roughly at the peak of the CRTS equivalent filter) with and without H$\beta$ is only $\Delta m \sim0.07$ mag. An accompanying reduction in continuum flux of a third is also required to give a magnitude change of $\Delta m \sim 0.5$, a typical value seen in CSQs. This suggests that there will be quasars experiencing the same physical changes as CLQ/CSQs but which may only exhibit a significant photometric change (corresponding to a change in continuum flux) without showing a corresponding spectral line change, i.e., a population of CLQs without the broad emission line variability.

\begin{figure}
\includegraphics[width = 3.4in]{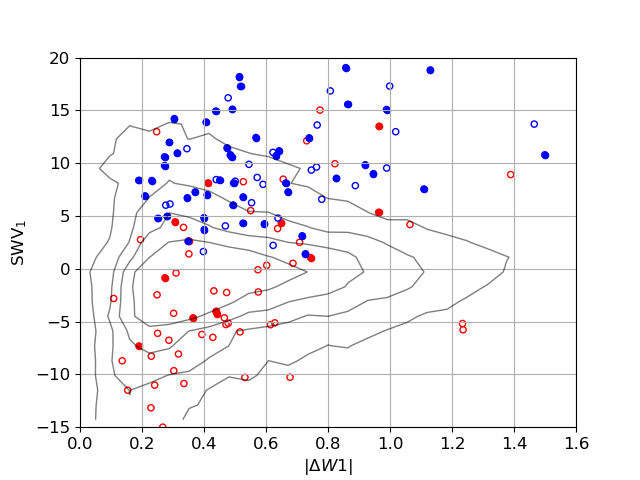}
\caption{The distribution of optical photometric variability strength against MIR photometric variability. Points are colored as in Fig.~\ref{fig:hbphot}.}
\label{fig:wise}
\end{figure}

There is no correlation, though, between the amplitudes of the MIR photometric variability and the optical spectral variability of these sources (or the general population of quasars). This argues that the physical mechanism underpinning the change of activity manifests differently at the AGN dust torus than at the broad emission line region, although it is also possible that a correlation does exist but that the different physical locations of the emitting regions within the quasar impose a several year temporal delay between MIR and spectral variability, \citep[e.g.,][]{jun15} that simple amplitude measures or the available data do not capture. Fig.~\ref{fig:wise} shows a trend between the MIR and optical photometric variability for CLQ/CSQs but not for the general quasar population. Since the former are known to show spectral variability, this also supports the idea that continuum variability is the stronger component overall as compared to just spectral line changes. 

\begin{figure*}
\begin{tabular}{cc}
\includegraphics[width = 3.4in]{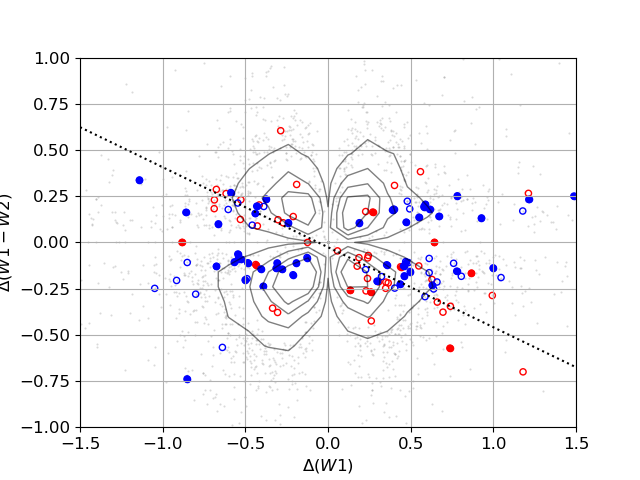} &
\includegraphics[width = 3.4in]{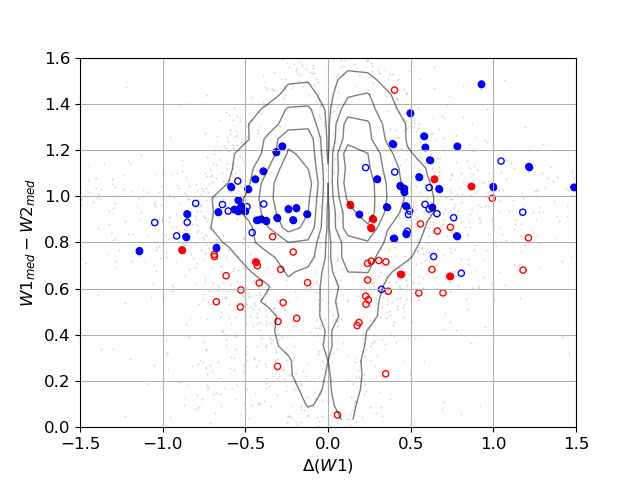}
\end{tabular}
\caption{The distribution of MIR color variability (left) and MIR color (right) in terms of MIR magnitude variability. The dotted line in the left plot is the brighter-when-redder relationship reported in Yang et al. (2017). The points are colored as in Fig.~\ref{fig:hbphot}. The contour lines show the 50th - 90th percentiles of the distributions respectively for the general population of quasars with CRTS lightcurves. }
\label{fig:wisecol}
\end{figure*}

\cite{yang17} reported a ``redder when brighter" correlation between MIR color and magnitude amplitudes, indicative of a stronger contribution from the AGN dust torus when the AGN turns on. Fig.~\ref{fig:wisecol} shows that this correlation holds only for the known CLQs and possibly for the lower luminosity CSQs but not for the higher luminosity objects. CSQs are also a redder population at MIR wavelengths than CLQs (a large proportion of which would not pass the MIR selection criteria used in this paper). The higher luminosity, more variable CSQs may thus already have a strong contribution in their MIR flux from the AGN dust torus that the relative change associated with the changing-state mechanism is not so significant an effect as with the lower luminosity sources.

\begin{figure}
\includegraphics[width = 3.4in]{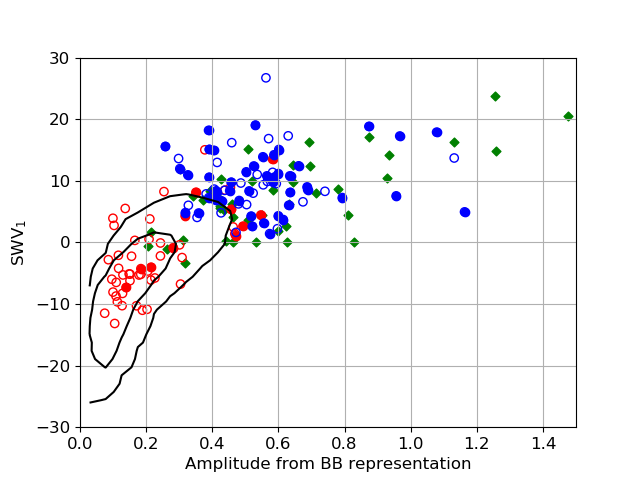}
\caption{The distribution of the magnitude amplitude from the Bayesian block components and the similarity (aggregate distance) to the median Slepian wavelet variance for the quasars in our data set. The points and contours are as in Fig.~\ref{fig:bbdiff}.}
\label{fig:bbamp}
\end{figure}

Our selection criteria are somewhat broader by design than those employed in other works so it is also worthwhile considering CSQs in the context of other criteria used. \cite{macleod19} and other CLQ searches have employed a simple variability amplitude criterion (typically $|\Delta g| > 1$) to select sources exhibiting strong photometric variability over any of the available time baselines probed by surveys, such as SDSS and Pan-STARRS 1. We used the difference between the initial and final states of the Bayesian block representation of a light curve rather than its amplitude (i.e, the difference between its maximum and minimum states) to identify sources showing significant photometric change but avoiding objects showing flaring activity such as reported in \cite{graham17} and \cite{lawrence16}. Fig.~\ref{fig:bbamp} shows that almost all CSQs lie outside the 95th percentile contour in the SWV$_1$ - BB amplitude plane but that flaring quasars from \cite{graham17} are the majority source at higher amplitudes (compare with Fig.~\ref{fig:bbdiff}). Known CLQs still form a predominantly less strongly variable population. 

\begin{figure*}
\includegraphics[width = 6.in]{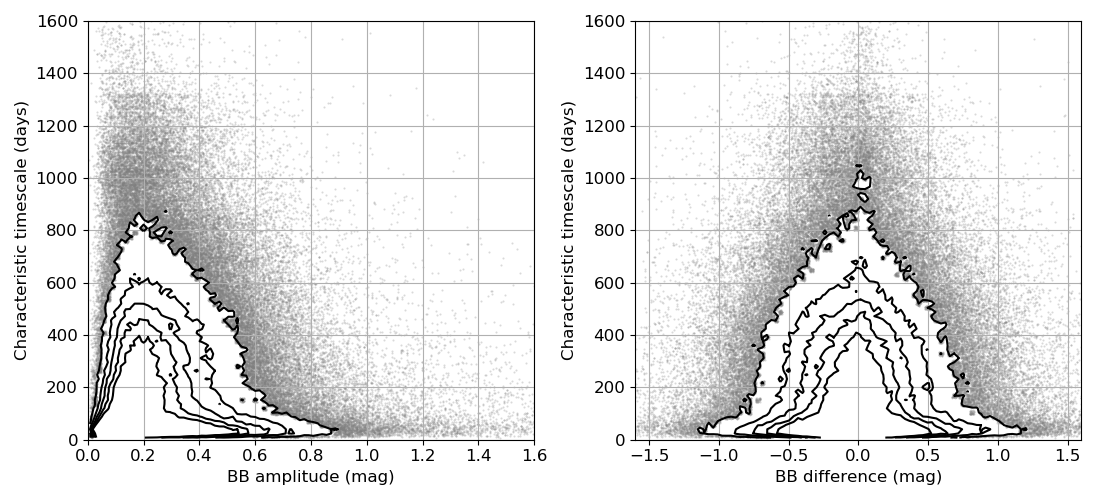}
\caption{The distributions of the Bayesian block amplitude and difference against the characteristic timescale from the Slepian wavelet variance for the initial 1.1 million sources in the MQ superset. The contours show the 50th - 90th percentiles of the distributions.}
\label{fig:bbtemp}
\end{figure*}

It is possible, however, that the long baseline of typically 4000 days between the initial and final states in the observed frame could bias our selection toward quasars with longer timescale variability and miss shorter timescale variability potentially associated with CLQs. The Slepian wavelet variance measure, $SWV_1$, has a characteristic timescale associated with it (see Sec.~\ref{sec:timescale}) and Fig.~\ref{fig:bbtemp} shows the distribution of BB amplitude and difference in relation to this timescale for the 1.1 million initial sources in MQ. It is clear that with increasing variability, measured either through the amplitude or the difference, there are fewer objects with longer characteristic timescales and no indication of a selection bias towards them.

\begin{figure*}
\includegraphics[width = 3.4in]{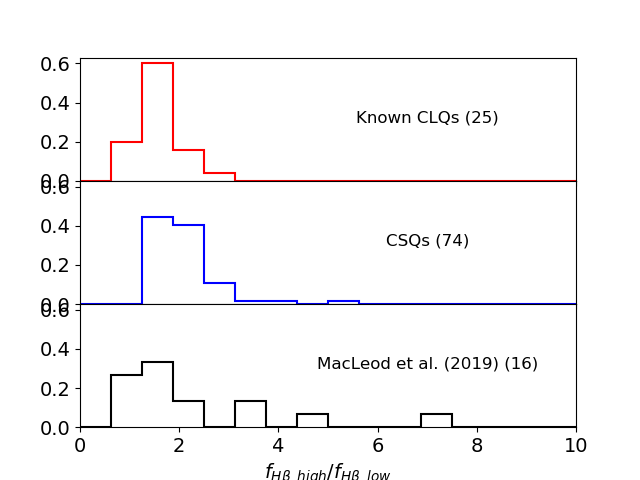}
\includegraphics[width = 3.4in]{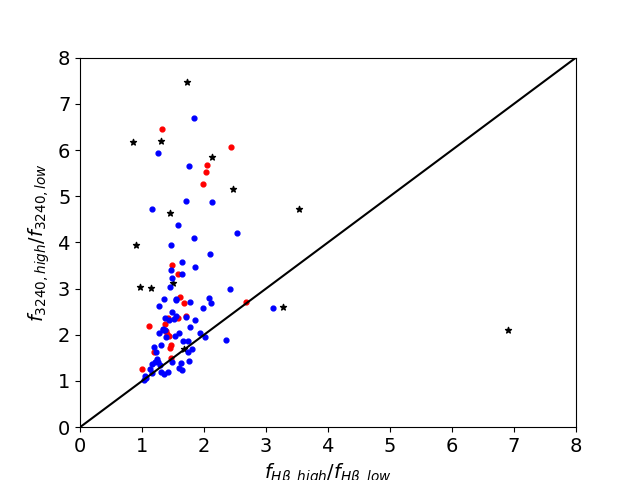}
\caption{The distributions of the H$\beta$ flux ratios (left) and H$\beta$ and continuum (at restframe 3240\AA) flux ratios (right) between the high (bright) and low (dim) states for CSQs (blue), known CLQs with available spectra (red), and the CLQs reported by M19 (black stars). The numbers in the parentheses in the left plot give the number of objects in each data set. The black line with unit slope in the right panel is the expectation for a linear response of BEL flux to the continuum variability (see M19, Fig.~4).}
\label{fig:ratios}
\end{figure*}

The spectral variability constraint (fractional change in H$\beta$ / [\ion{O}{III}] > 30\%) that is employed here also differs from the ad hoc visual criteria that have defined CLQs in other searches. As M19 note, the application of such a quantitative definition could lead to a different set of ambiguities and make comparison with other CLQ samples difficult. Fig.~\ref{fig:hbeta} shows that our spectral criterion is not met by about 45\% of the known CLQs for which we have multiepoch spectra. CSQs are associated with large continuum luminosity changes and MIR variability over long timescales but not necessarily with a complete (visual) absence of H$\beta$ flux in their lower activity (fainter) state. M19 uses spectral flux ratios between high (bright) and low (dim) states to determine more clearly how the H$\beta$ line varies relative to the continuum (measured at restframe 3460 \AA). Fig.~\ref{fig:ratios} shows that the distribution of H$\beta$  and continuum (measured at restframe 3240\AA) flux ratios for CSQs is consistent with the known CLQs for which multiepoch spectra are available and the values reported by M19. Note that flux ratios measured for sources where H$\beta$ has largely disappeared in the low state will be a lower limit.

A further complication can arise due to the timing of, e.g., spectroscopic follow-up observations for candidates. Given the data-taking, it is not always possible to obtain a spectrum and sample the light curve, say, at the largest amplitude of variability (in flux, BB, or $SWV_1$ parameter space). This could present a bias for/against a particular class of object depending on when the spectra are obtained relative to the light curve. We do not claim completeness in our sample: for example, we note that the changing source SDSS J2232-0806 \citep{kynoch19} is a photometric candidate in this work but lacks a second epoch spectrum in our data set. Similarly 57 of the 262 CLQ candidates (22\%) from \cite{macleod19} also appear in our sample of CSQ candidates with single epoch spectra.

More generally, we can consider our sources in the context of the extreme variability quasars (EVQs) of R17. In a comparison with a control sample of SDSS quasars matched in redshift and $g$ magnitude to their EVQ sample, R17 find that EVQs have a larger variability amplitude (from the structure function) than control quasars at all timescales from days to years. This is equivalent to the $SWV_1$ selection criteria we have employed with the source $SWV_1$ greater than the median $SWV_1$ at the source magnitude across timescales. However, we compare CSQ quasars to the general population of MQ sources with CRTS light curves rather than a magnitude and redshift-matched sample.

\section{Discussion}

\subsection{Luminosity}
\label{sec:luminosity}

\begin{figure}
\centering
\includegraphics[width = 3.65in]{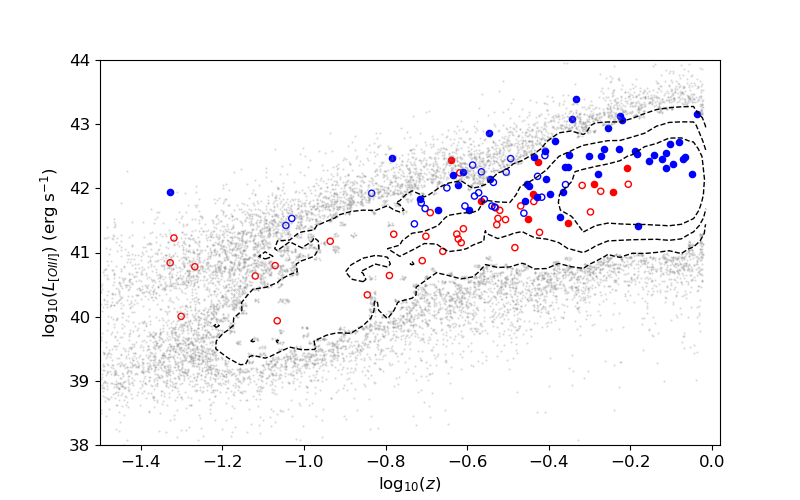}
\caption{The distribution of the peak [OIII] luminosity as a function of redshift. The points are colored as in Fig.~\ref{fig:hbphot}. The contour lines indicate the 50th, 70th, and 90th percentiles respectively.}
\label{fig:oiii}
\end{figure}

Quasar variability is known to be anticorrelated with luminosity in that low luminosity quasars have a larger probability of showing large amplitude variation over multi-year timescales \citep[e.g.,][R17]{hook94}. Using the 5100 \AA \, continuum luminosity as a proxy for the intrinsic AGN luminosity, \cite{macleod16} have shown that CLQs seem to preferentially be low-luminosity AGN. Fig.~\ref{fig:oiii} shows the distribution of the peak [\ion{O}{III}] luminosity from multiepoch spectra as a function of redshift for the CSQs reported here, the known CLQs, and values measured for 30,000 SDSS quasar at $z < 0.95$. Our CSQs expands the range of changing AGN to more luminous sources and higher redshifts. Furthermore, this type of variability is not dependent on the luminosity of the object.

R17 find that there is a trend of decreasing Eddington ratio with variability. To determine the population distribution, we have calculated the bolometric luminosity for the 4,244 SDSS quasars with possible H$\beta$ coverage $(z < 0.95)$ and at least 500 days between multiepoch spectra using:
\[ L_{bol,V} = b_V L_{\odot,V}10^{(M_{\odot,V} - M_V) / 2.5} \]

\noindent
where the solar constants for $V$-band are $M_{\odot,V} = 4.83$ and $L_{\odot,V} = 4.64 \times 10^{32}$ erg s$^{-1}$ and $b_V$ is the bolometric correction. A comparison with the bolometric luminosities calculated for DR12 quasars from SDSS by \cite{kozlowski17} (hereafter K17) with $L_{bol,V}$ gives a mean bolometric correction of $\log_{10} b_V = 1.46$, which we use as a fiducial value hereafter. We estimate the black hole virial mass using:
\[ \log \left( \frac{M}{M_\odot} \right) = a + b \log \left( \frac{L_{5100}}{10^{44} \,\, \mathrm{erg \cdot s^{-1}}} \right) + 2 \log \left( \frac{\mathrm{FWHM(H\beta)}}{1000 \,\,\mathrm{km \cdot s^{-1}}} \right) \]

\noindent
with $a = 6.91$ and $b = 0.533$ for H$\beta$ \citep{ho15}. We find that estimates for the same source from multiepoch spectra are typically consistent within 0.4 dex.

\begin{figure}
\includegraphics[width = 3.45in]{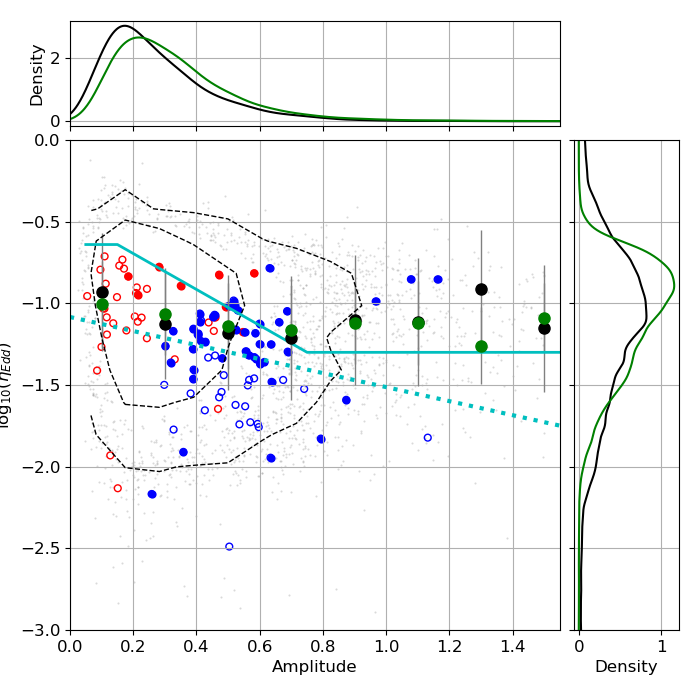}
\caption{The distribution of Eddington ratio as a function of variability amplitude for SDSS quasars with $z < 0.95$ and multiple spectra separated by more than 500 days (contours and grey points; contours show 68th and 95th percentiles). The black points show the median value for this data set in each bin of variability and the green show the same quantity for all SDSS DR12 quasars from Kozlowski (2017) with $z < 0.95$. The values of known CLQs (red) and the CSQs (blue) reported here are also shown. The solid cyan line indicates the Eddington ratio trend from R17 Fig.~10 and the dashed cyan line shows a linear fit to the CLQ/CSQ sources.}
\label{fig:leddamp}
\end{figure}

The Eddington ratio is defined as $\eta_{\rm Edd} = L _{bol} / L_{\rm Edd}$ where $L_{\rm Edd}$ is the Eddington luminosity. Fig.~\ref{fig:leddamp} shows the distribution of the Eddington ratio for this sample as a function of variability amplitude as measured from the Bayesian block fits to the respective time series (note that here amplitude = 0.5 $\times$ (max value - min value)). It also shows this distribution for SDSS quasars with potential H$\beta$ coverage ($z < 0.95$) with Eddington ratio estimates derived as above and also from K17 as a consistency check. The general population does not show any strong indication of decreasing Eddington ratio with increasing variability, as had been previously reported (e.g., R17). Instead, we find that low-amplitude (amp < 0.2) low-luminosity sources have a fractionally higher Eddington ratio ($\eta_{\rm Edd} = 0.2$) relative to an otherwise flat relationship between Eddington ratio and variability amplitude.  We note that R17 used Eddington ratio values from \cite{shen11} whereas K17 derived their values from their MgII and CIV-based black hole virial mass estimates. Our values are based on our own H$\beta$ virial mass estimates and a bolometric luminosity correction derived from K17 data. Although virial mass estimators based on different lines can systematically disagree, this is not sufficient to explain this difference.

The discrepancy may lie in the R17 control sample having no upper redshift bound (the majority of their comparison sample is at $z > 1$) whereas both our CSQs and the control sample data is constrained to $z < 0.95$. There is a correlation between bolometric luminosity (Eddington ratio) and redshift in the R17 sample so that low variability amplitude bins in R17 will be biased toward higher Eddington ratio as a large fraction of objects will have $z > 1$. Higher amplitude bins have comparatively more sources with $z < 1$ and so are more consistent with our results. However, we agree that CLQ/CSQs do show the reported anticorrelation between Eddington ratio and amplitude of variability and that it is stronger for low luminosity ($M_V < -23)$ sources. This is consistent with attributing the changes seen to accretion physics occurring preferentially in lower activity systems but not necessarily just in low luminosity sources. We also note, though, that although the CSQ (this paper), CLQ (M19), and EVQ (R17) samples all have $\log(\eta_{\rm Edd}) = -2$ to $-0.5$, this does not imply the same physical mechanisms are necessarily involved across these samples.

\subsection{Timescales}
\label{sec:timescale}
\begin{figure}
\centering
\includegraphics[width = 3.4in]{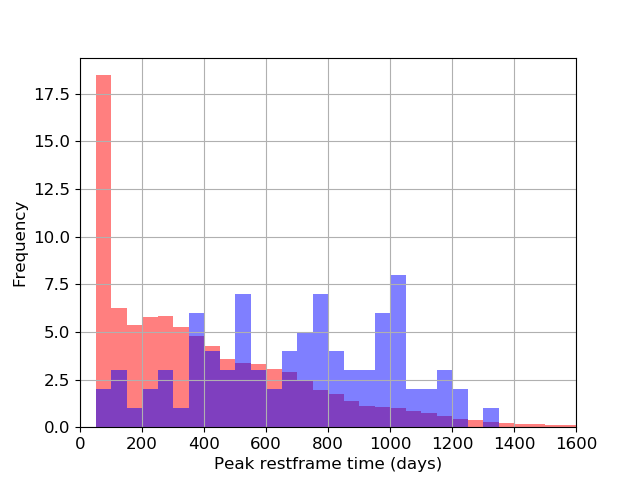}
\caption{The distribution of the peak restframe timescale contributing to the source variability as measured from Slepian wavelet variance. The blue bars indicate the peak timescale from CSQs/CLQs whilst the red bars show the distribution of peak timescale for 136,000 quasars with $z < 1.1$ normalized to the CSQ/CLQ sample size.}
\label{fig:peak_timescales}
\end{figure}

One of the selection criteria in Sec.~\ref{sec:swv} was an excess of variability relative to a median level for a magnitude range as measured by Slepian wavelet variance (SWV). We employed a cumulative measure looking for an overall significant signal rather than one at any specific timescale. However, CSQs display a particular pattern of activity over the period covered by their light curve and therefore the timescale which contributes most to the variability of the source should be associated with (or even characteristic of) the physical mechanism driving the change. We have determined the restframe timescale for each CSQ in our data (and CLQ in the literature) at which the SWV of the source has its largest value relative to the median value (see Fig.~\ref{fig:peak_timescales}) as well as the distribution of such timescales for 137,000 quasars with $z < 1.1$. The distribution of peak time values for the CSQ/CLQ sample is significantly different than the population distribution ($>> 5 \sigma$ according to the Anderson-Darling test)  and so the timescales are indicative of process(es) associated with the observed variability.

As given in \cite{stern18}, the relevant disk timescales for a black hole of mass $M_{BH}$ at $R \sim 150r_g$ can be parameterized as:

\begin{equation}
\label{eq:t_th}
 t_{\mathrm{th}} \sim \mathrm{1 yr} \left( \frac{\alpha}{0.03} \right) ^{-1} \left( \frac{M_{BH}}{10^8 M_{\sun}} \right) \left(\frac{R}{150 r_g} \right)^{3/2} 
\end{equation}
\begin{equation}
\label{eq:t_front} 
 t_{\mathrm{front}} \sim \mathrm{20 yr} \left( \frac{h/R}{0.05} \right)^{-1} \left( \frac{\alpha}{0.03} \right) ^{-1} \left( \frac{M_{BH}}{10^8 M_{\sun}} \right) \left(\frac{R}{150 r_g} \right)^{3/2} 
 \end{equation}
 \begin{equation}
 \label{eq:t_v}
 t_{\mathrm{v}} \sim \mathrm{400 yr} \left( \frac{h/R}{0.05} \right)^{-2} \left( \frac{\alpha}{0.03} \right) ^{-1} \left( \frac{M_{BH}}{10^8 M_{\sun}} \right) \left(\frac{R}{150 r_g} \right)^{3/2} 
\end{equation}

\noindent 
where $\alpha$ is the disk viscosity parameter, $h/R$ is the disk aspect ratio, $R$ is the disk radius, and $r_g = GM_{BH}/c^2$ is the gravitational radius. The thermal timescale $t_{\mathrm{th}}$ corresponds to the timescale on which the disk heats or cools with cooling and heating fronts crossing the disk of timescales of $t_{\mathrm{front}}$. The viscous disk timescale, $t_{\mathrm{v}}$, gives the characteristic timescale of mass flow. 

\begin{figure}
\centering
\includegraphics[width = 4.5in, trim=2.8cm 2.8cm -2.4cm 2.8cm, clip]{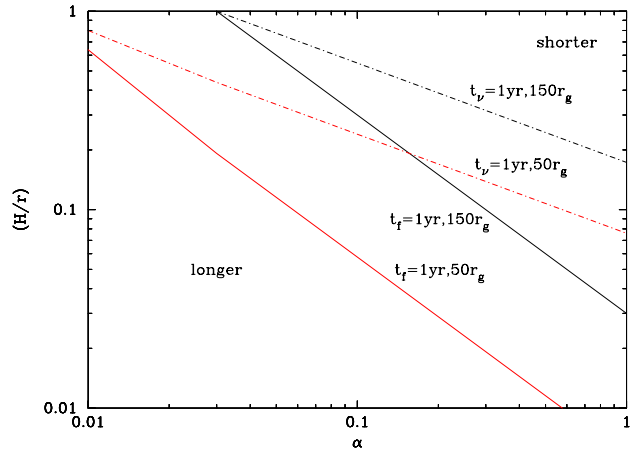}
\caption{Loci of timescales from eqns.~(\ref{eq:t_front}) (thermal front timescale; solid lines) and~(\ref{eq:t_v}) (viscous timescale; dash-dot lines) plotted as a function of both $\alpha$ and $(h/R)$. We assumed a fiducial timescale of one year in each case, at disk radii  $50r_{g}$ (red) and $150r_{g}$ (black) around a black hole of mass $M_{\rm BH}=10^{8}M_{\odot}$. If timescales are shorter, the curves will shift in parallel towards the top right of the Figure. If the timescales are longer, the curves will shift in parallel towards the bottom left of the Figure. As the disk approaches a spherical configuration ($H/r \rightarrow 1$), $t_{\rm front} \rightarrow t_{\rm v}$ as we see from eqns.~(\ref{eq:t_front}) and (\ref{eq:t_v}) and the curves meet. Note that the thermal timescale in eqn.~(\ref{eq:t_th}) is independent of $H/r$ and would therefore correspond to a vertical line at $\alpha=(0.16) \,\, 0.03$ at $(50) \, 150r_{g}$.}
\label{fig:timescales}
\end{figure}

From eqns.~(\ref{eq:t_th})-(\ref{eq:t_v}), timescales associated with processes in AGN disks are expected to be functions of the disk aspect ratio ($h/R$), disk viscosity parameter ($\alpha$), black hole mass $M_{\rm BH}$ and disk radius ($R(r_{g})$). In Fig.~\ref{fig:timescales} we show the loci of timescales from eqns. (\ref{eq:t_front}) and (\ref{eq:t_v}) as a function of both $\alpha$ and $(h/R)$. Solid lines correspond to $t_{\rm front}=1{\rm yr}$ and dash-dot lines correspond to $t_{\rm v}=1\rm{yr}$ located at $50r_{g}$ (red) and $150r_{g}$ (black) respectively in a disk around a $M_{\rm BH}=10^{8}M_{\odot}$. We can read Fig.~\ref{fig:timescales} as follows: if a CSQ is observed to change state on a one year timescale and we model the associated spectral change with the propagation of a front from the ISCO to $150r_{g}$, then disk properties ($h/R, \alpha$) must live on the black solid line (e.g., both $h/R \sim 0.1, \alpha \sim 0.3$ and $h/R \sim 0.3, \alpha \sim 0.1$ are possible solutions for the model. Likewise if a CSQ varying on a timescale of one year is modeled in terms of a viscous change at $50r_{g}$, then the disk parameters ($h/R,\alpha$) must live on the red dash-dot line (e.g., $h/R \sim 0.5, \alpha \sim 0.04$ is a possible solution).

\begin{figure*}
\centering
\includegraphics[width = 6.8in]{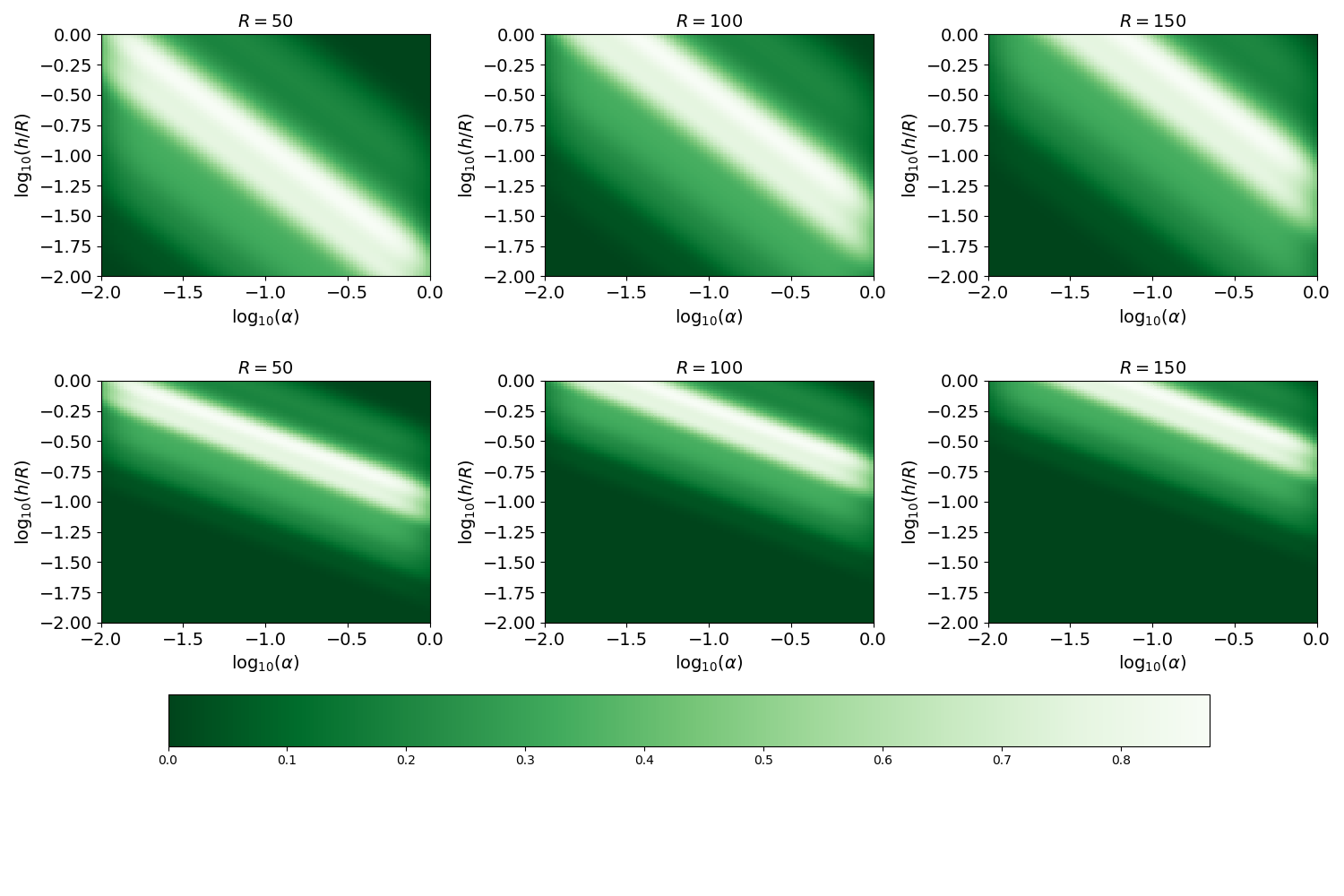}
\caption{The distribution of $t_{\mathrm{front}}$ (top) and $t_{\mathrm{v}}$ (bottom) timescale curves in the $h/R - \alpha$ plane for the CSQ/CLQ sample at values of $R = 50, 100,$ and $150 r_g$ respectively. Brighter color indicates denser distribution. Observations and simulations broadly support values of $\alpha$ in the range $0.03 \lesssim \alpha \lesssim 0.5$.}
\label{fig:bestfittimes}
\end{figure*}

For a given value of $R$, each quasar will define a curve in the $(h/R, \alpha)$ plane and Fig.~\ref{fig:bestfittimes} shows the distribution of these for $t_{\mathrm{front}}$ and $t_{\mathrm{v}}$ for the CLQ/CSQ sample. There are clearly different preferred regions of the parameter space depending on whether the timescales are interpreted as front crossing or viscous. For example, \cite{king07} argue that observations favor a typical range of $\alpha \sim 0.1 - 0.4$ and this would suggest disk scale ratios of $\sim 0.03 - 0.1$ at $R = 50 r_g$ for front crossing timescales but ratios of $\sim0.2 - 0.3$ for viscous timescales. Numerical simulations, however, favor $\alpha \sim 0.03$ \citep[e.g.,][]{hirose09, davis10} and thus very thick disks. Although there are degeneracies, broad constraints can be placed on viable disk geometries: a viscous timescale process favors a thicker disk and less change in disk thickness with increasing radius whereas a front crossing process can support a thinner disk but also one that expands more in height with increasing radius.

\subsection{Physical mechanisms}

Cooling fronts have been proposed as the mechanism for CLQ/CSQs, either as a result of a sudden change in torque applied by the magnetic field at the innermost stable circular orbit (ISCO) \citep{ross18}, or a drop in mass accretion rate causing an advection-dominated accretion flow (ADAF) in the inner disk and spectral state transition by disc evaporation \citep{noda18}. \cite{sniegowska19} also propose that for sources operating at a few percent of the Eddington limit, there is a radiation pressure instability in a narrow zone between the outer cold gas-dominated disk and an inner hot ADAF flow which can lead to outbursts producing changing look behavior. \cite{noda18} suggest that CLQ/CSQ sources can be placed in one of three groups, depending on which particular aspect of the process they exhibit: (1) a factor of two to four decrease in luminosity associated with disc evaporation/condensation; (2) large mass accretion rate change due to thermal front propagation; and (3) a variability amplitude of more than ten indicative of both phenomena. Most objects should be in either the first or the third group whereas the second group will contain sources showing large variability but not showing any significant spectral changes (although given the results of Sec.~4, MIR variability would be expected to be shown in addition to optical). 

\begin{figure}
\includegraphics[width = 3.45in]{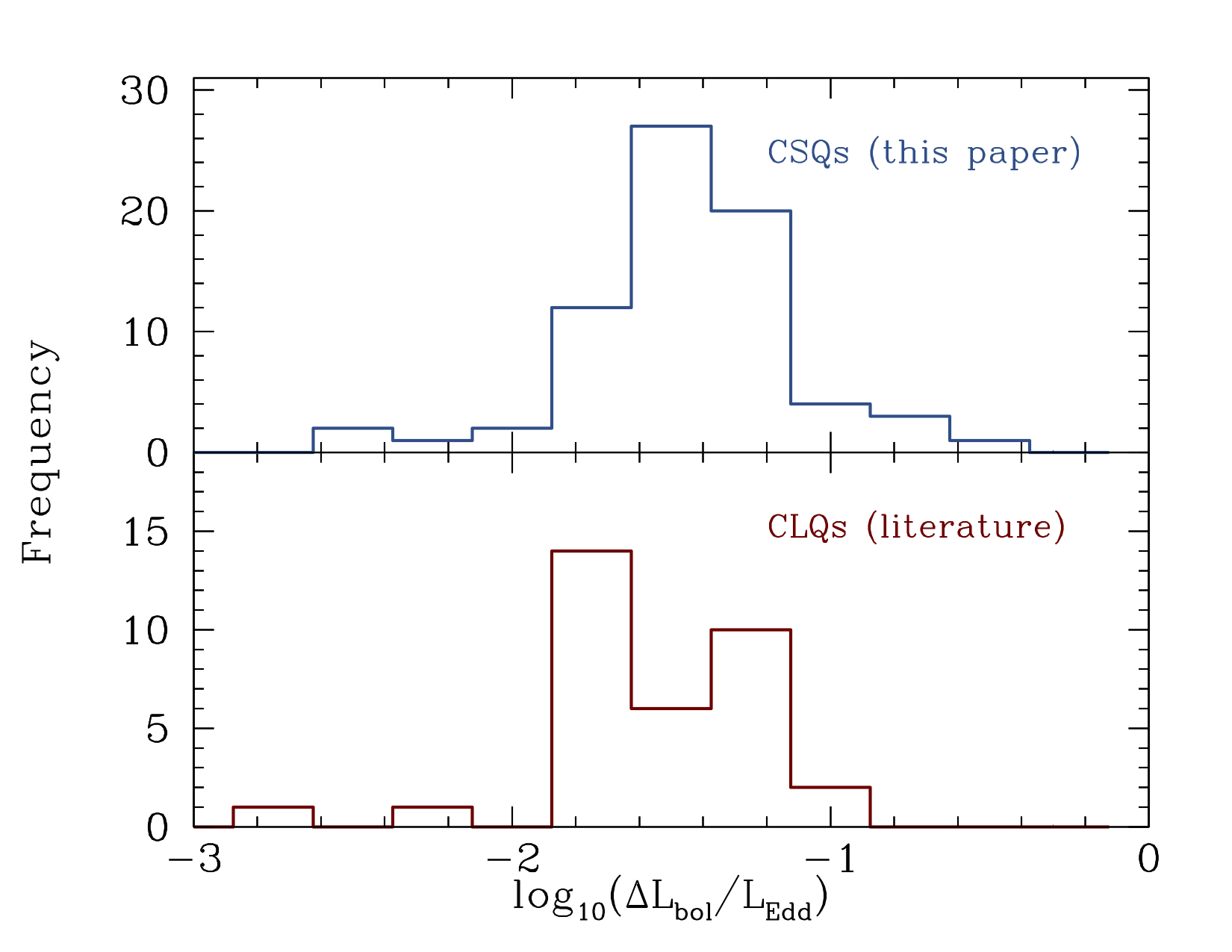} 
\caption{The difference in Eddington ratio associated with the highest and lowest states in the Bayesian block representation of a light curve. Known CLQs are shown in red and CSQs in blue.}
\label{fig:leddiff}
\end{figure}

\begin{figure*}
\includegraphics[width = 3.4in]{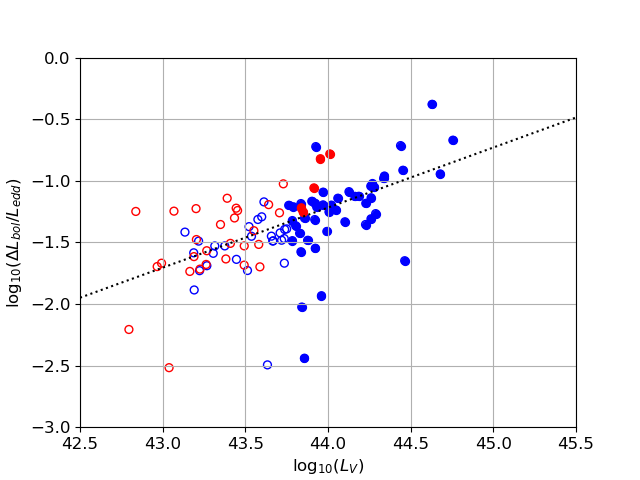}
\includegraphics[width = 3.4in]{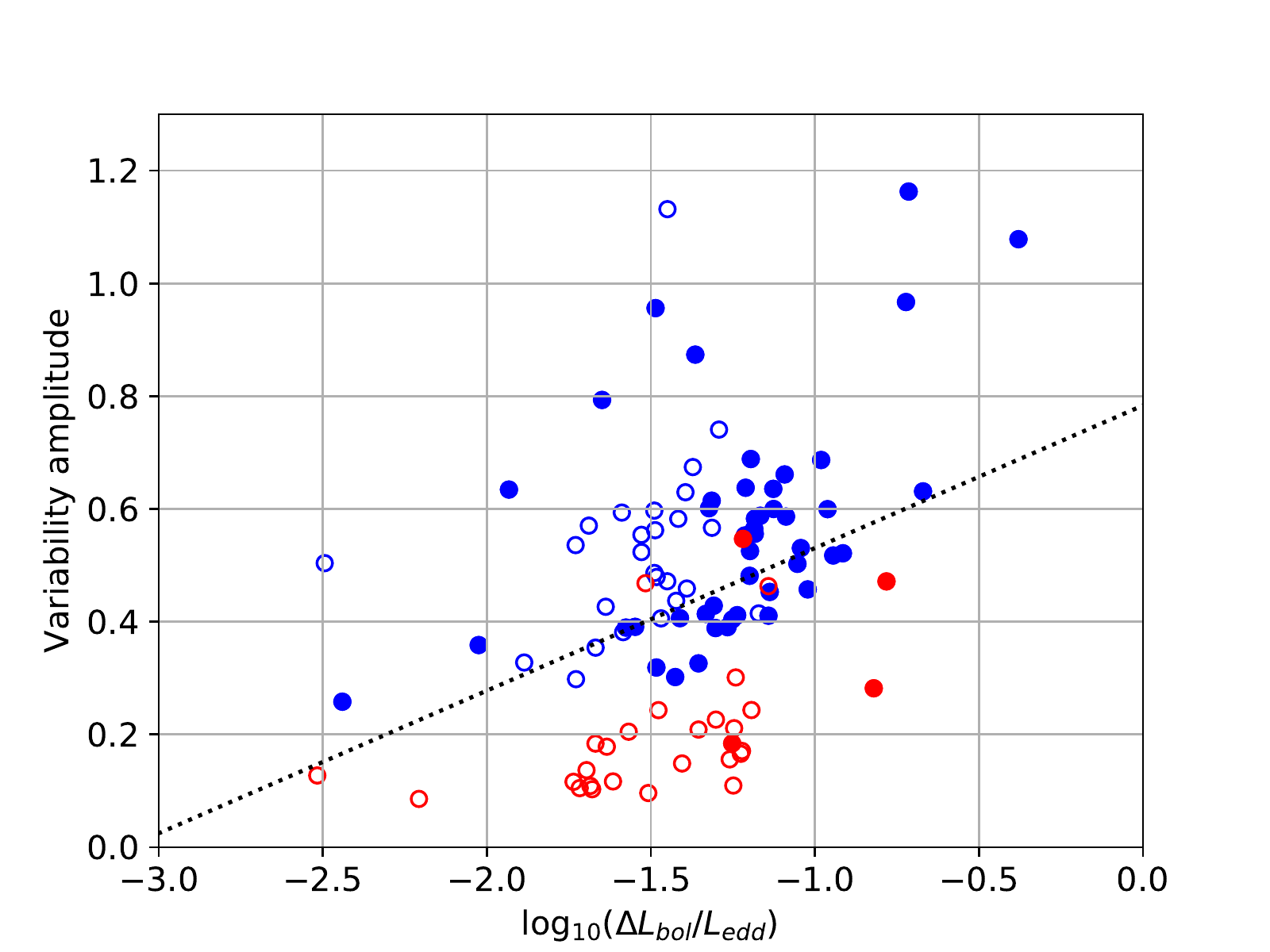}
\caption{The change of accretion rate (difference in Eddington ratio) associated with the photometric variability as (left) a function of median luminosity and (right) variability amplitude. The same color scheme is used as in Fig.~\ref{fig:hbphot}. The black dotted line is the Theil-Sen linear fit to the data. Both distributions favor a linear model over a constant model (no trend) with $p < 10^{-4}$ from the F-test.}
\label{fig:dled}
\end{figure*}

Fig.~\ref{fig:leddiff} shows that most CSQs/CLQs are associated with a change of Eddington ratio (accretion rate) of between 1\% and 10\% $L_{\rm edd}$, consistent with the observational predictions from \cite{noda18} and placing these sources in their third group. We note as well (see Fig.~\ref{fig:dled} and Table~\ref{tab:relations}) that the magnitude of the change in $\eta_{\rm Edd}$ shows a trend with (median) luminosity and also with the amplitude of variability: $amp \propto \log(\Delta \eta_{\rm Edd})$ and $\Delta \eta_{\rm Edd} \propto L / \log (\eta_{\rm Edd})$, where the inverse relationship comes from Sec.~\ref{sec:luminosity}. In other words, more extreme variability is associated with larger changes of $\eta_{\rm Edd}$ in higher luminous systems but also with lower actual  $\eta_{\rm Edd}$ or, conversely, lower luminosity systems with higher $\eta_{\rm Edd}$ but only able to support a smaller change in $\eta_{\rm Edd}$. If the magnitude of the change in $\eta_{\rm Edd}$ correlates with either a change in torque at the ISCO or a change in mass accretion rate then larger systems show stronger fluctuations. This suggests that disk instabilities, e.g., magneto-hydrodynamical, may be a more likely cause than local perturbative events in the disk, e.g., an embedded supernova, since the latter should not scale with the size of the accreting system. Such instabilities may be driven by the larger environment: \cite{charlton19} reported that four CLQs are associated with galaxy mergers and \cite{kim18} have proposed that changing look activity in Mrk 1018 is due to a recoiling SMBH perturbing the accretion flow on a 29-year period. Alternatively, both disk instabilities and local perturbative events may be present but with a bias for the latter in lower luminosity systems.

\begin{figure}
\includegraphics[width=3.4in]{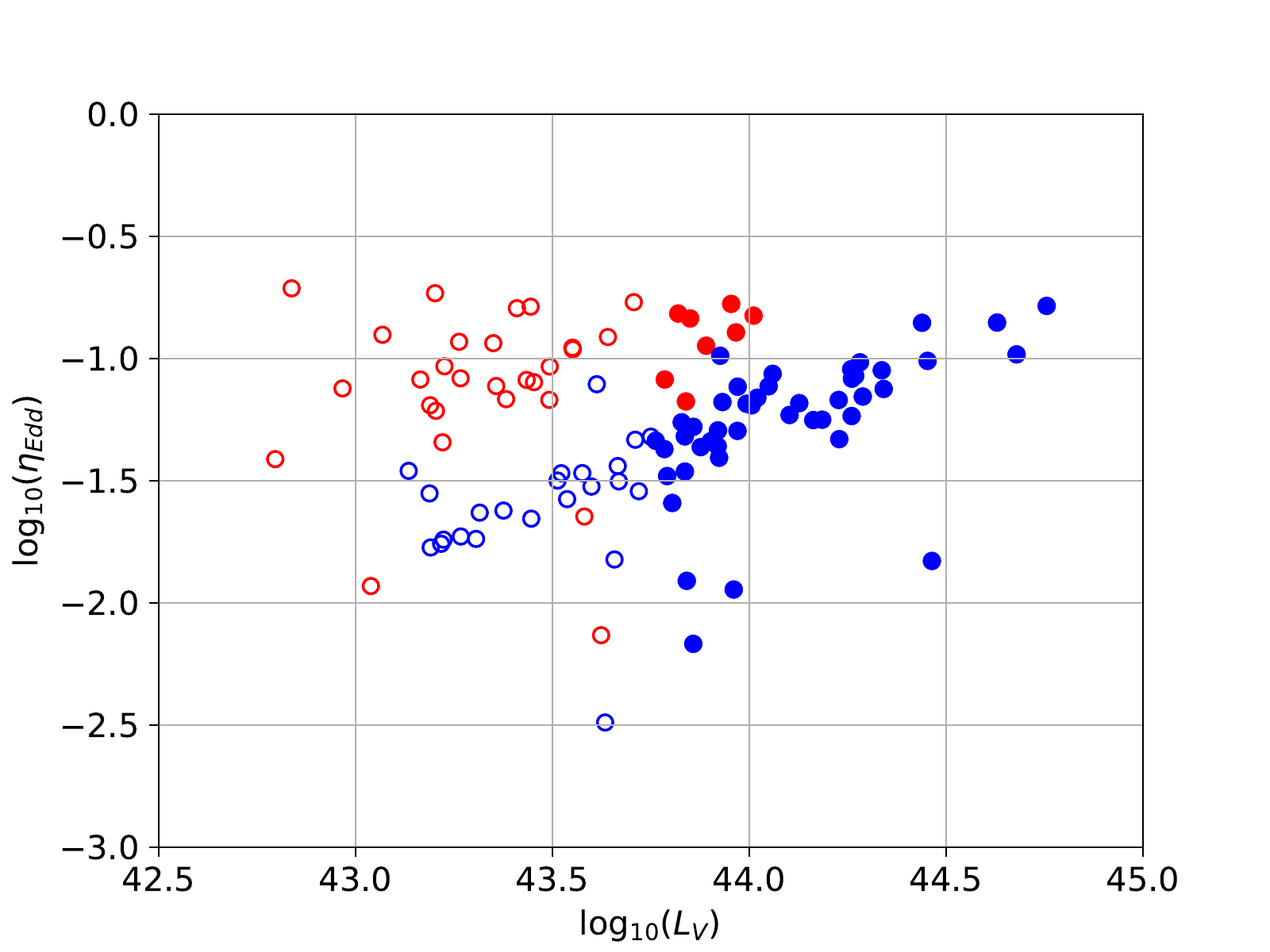}
\caption{Eddington ratio as a function of luminosity. The same color scheme is used as in Fig.~\ref{fig:hbphot}.}
\label{fig:eddlum}
\end{figure}

We expect that low luminosity AGN (LLAGN, open  circles in Figs.~\ref{fig:hbphot}-\ref{fig:eddlum}) should be more heterogeneous in origin than the high luminosity AGN population (filled circles in Figs.~\ref{fig:hbphot}-\ref{fig:eddlum}). This is because either the intrinsic SMBH mass is lower than for the higher luminosity population even if $\eta_{\rm Edd}$ is comparable, or because $\eta_{\rm Edd}$ is intrinsically lower than for the higher luminosity population, or some combination of these. So we should anticipate a lack of correlation between luminosity among the LLAGN and $\eta_{\rm Edd}$. This is confirmed in Fig.~\ref{fig:eddlum} where the LLAGN (open circles) form a scatterplot. Conversely, for the high luminosity AGN, there is an apparent correlation with $\eta_{\rm Edd}$ above $\sim 10^{44}$ erg s$^{-1}$ in Fig.~\ref{fig:eddlum}. While we expect LLAGN are more heterogeneous than the high luminosity AGN, a change in the accretion rate in LLAGN (effectively a change in $\eta_{\rm Edd}$) should correlate with luminosity. This is indeed apparent in the left panel of Fig.~\ref{fig:dled}. The LLAGN variability amplitude anti-correlates with the accretion rate (Fig.~\ref{fig:leddamp}) which suggests that it is harder to significantly change the accretion rate in LLAGN.

\begin{table*}
\centering
\caption{A summary of the relationships shown for high and low luminosity subsets.}
\label{tab:relations}
\begin{tabular}{lcccc}
\hline
Sample & $amp \propto \Delta \eta_{\rm Edd}$ & $\log(\Delta \eta_{\rm Edd}) \propto L$ & $\log(\eta_{\rm Edd}) \propto 1 / amp$ & $L \propto 1 / \eta_{\rm Edd}$ \\
 & (Fig.~\ref{fig:dled}R) & (Fig.~\ref{fig:dled}L) & (Fig.~\ref{fig:leddamp}) & (Fig.~\ref{fig:eddlum})\\
\hline
Low luminosity & No & Yes & Yes & No \\
High luminosity & Yes & Yes & No & $L \propto \eta_{\rm Edd}$ \\
\hline
\end{tabular}
\end{table*}

\section{Conclusions}

We have identified 73 quasars exhibiting strong coherent optical and MIR photometric variability with significant contemporaneous spectroscopic variability that are comparable to the $\sim$60 existing changing-look quasars reported in the literature. Our sample, however, forms a higher luminosity (and higher redshift) counterpart to the known CLQs showing that this phenomenon is not restricted to low luminosity systems. The characterizing preference is rather for systems with low Eddington ratios and with the amplitude of the associated variability correlated with a change of Eddington ratio. Characteristic timescales of the photometric variability suggest that it most closely matches the timescale associated with a cooling/heating front propagating through the disk as has been proposed for individual sources \citep{stern18, ross18, noda18}. The lack of large variability in smaller systems may also indicate disk instabilities associated with magnetic phenomena as the more likely physical cause for the fronts, particularly in larger systems.

The next generation of sky surveys, such as the Zwicky Transient Facility (ZTF; Bellm et al. 2019; Graham et al. 2019) and LSST, will effectively monitor all AGN in the sky ever few nights. Generative models of quasar variability can be learnt from archival data and predicted behavior compared to that observed with unexpected changes identified far more quickly -- in weeks or months rather than years -- than waiting for a significant magnitude change to be detected. In this way, these changes of the state of the accretion disk can be tracked with follow-up resources as they happen rather than serendipitously or after the fact. This will allow us to test more easily the theoretical explanations for this phenomena.

Future work will employ machine learning to identify potential further sources - the combination of known CLQs and the CSQs reported here ensures that there is now adequate coverage of the parameter space and a suitable training set can be defined. Further characterization of the sources will also aid this activity. We have also undertaken a program to find CSQs with $z > 0.95$, i.e, where H$\beta$ does not fall into the optical spectral range. Candidates sharing the same photometric variability as their lower redshift counterparts have been identified and optical and near-IR spectra are being obtained, the latter to capture H$\beta$. Although multiepoch near-IR spectra are unlikely to exist, we will explore the possible correlations between \ion{Mg}{ii} and H$\beta$ variability for these sources relative to a more expected lack of correlation in the general population. We are interested as well in those objects which meet the photometric selection criteria but not the spectroscopic to understand if we are probing the same population but missing the spectral variability due to delayed discovery and followup or whether such objects are associated with a different phenomenon. Similar efforts are underway for sources with mid-IR variability but no associated optical change. 

\section*{Acknowledgements}

We thank Chelsea MacLeod for useful discussions. 

This work was supported in part by the NSF grants AST-1313422, AST-1413600, AST-1518308, and AST-1815034, and the NASA grant 16-ADAP16-0232. The work of DS was carried out at Jet Propulsion Laboratory, California Institute of Technology, under a contract with NASA.

NPR acknowledges support from the STFC and the Ernest Rutherford Fellowship scheme. 

This work made use of the Million Quasars Catalogue.

Funding for SDSS-III has been provided by the Alfred P. Sloan Foundation, the Participating Institutions, the National Science Foundation, and the U.S. Department of Energy Office of Science. The SDSS-III web site is http://www.sdss3.org/.

This publication makes use of data products from the Near-Earth Object Wide-field Infrared Survey Explorer (NEOWISE), which is a project of the Jet Propulsion Laboratory/California Institute of Technology. NEOWISE is funded by the National Aeronautics and Space Administration.

SDSS-III is managed by the Astrophysical Research Consortium for the Participating Institutions of the SDSS-III Collaboration including the University of Arizona, the Brazilian Participation Group, Brookhaven National Laboratory, Carnegie Mellon University, University of Florida, the French Participation Group, the German Participation Group, Harvard University, the Instituto de Astrofisica de Canarias, the Michigan State/Notre Dame/JINA Participation Group, Johns Hopkins University, Lawrence Berkeley National Laboratory, Max Planck Institute for Astrophysics, Max Planck Institute for Extraterrestrial Physics, New Mexico State University, New York University, Ohio State University, Pennsylvania State University, University of Portsmouth, Princeton University, the Spanish Participation Group, University of Tokyo, University of Utah, Vanderbilt University, University of Virginia, University of Washington, and Yale University.








\appendix

\section{Known CLQs}

\begin{table*}
\centering
\caption{CLQs reported in the literature with CRTS coverage and associated features including the median CRTS magnitude ($V_m$), the optical amplitude (Amp), absolute change in $W1$, Bayesian block change ($\Delta$BB), and Slepian wavelet variance measure ($SWV_1$). Sources marked with an asterisk have multiepoch spectra in the public domain. SMBH virial mass estimates are calculated as described in Sec.~\ref{sec:luminosity} except for sources where no spectra are available.}
\label{tab:knownclqs}
\begin{tabular}{lccccccccc}
\hline
Name  & Transition & $V_m$ & $z$ & $\log (M_{BH})$ & Amp & $|\Delta W1|$ & $\Delta$BB & $SWV_1$ & Ref. \\
 & & (mag) && $(M_{\sun})$  & (mag) & (mag) & (mag) &  & \\
\hline
SDSS J000904.5$-$103428 &  Disappear & 18.18 & 0.241 & 8.0 & 0.56 & 0.70 & 0.61 & 8.9 &(12) \\
SDSS J002311.0$+$003517* &  Appear & 18.53 & 0.422 & 8.2 & 0.57 & 0.42  & 0.92 & 7.3 & (1) \\
SDSS J004339.3$+$134437 & Disappear & 19.92 & 0.527 & - & 0.81 & 0.25 & -0.18 & 2.3 & (12) \\
SDSS J012648.0$-$083948 &  Disappear & 18.00 & 0.198 & 7.8 & 0.21 & 0.10 & 0.03 & 1.3 & (2) \\
SDSS J013458.3$-$091435* &  Disappear & 18.64 & 0.443 & 8.2 & 0.32 & 0.44 & 0.43 & -4.0 & (12) \\
SDSS J015957.6$+$003310* &  Disappear & 18.83 & 0.312 & 7.8 & 0.23 &  0.27 & 0 & -15 & (3) \\
SDSS J022556.0$+$003026* &  Both & 19.68 & 0.504 & 8.2 & 0.59 & 0.33 & -0.19 & 3.9 & (1) \\
SDSS J022652.2$-$003916* &  Disappear & 20.25 & 0.625 & 8.6 & 0.89 &  0.35 & 0.01 & 1.4 & (1) \\
SDSS J035301.0$-$062326 &  Appear & 16.35 & 0.076 & 7.6 & 0.10 &   0.23 & -0.04 & -13 &(7) \\
SDSS J074511.9$+$380911 &  Disappear & 17.84 & 0.237  & 9.1 & 0.19 & 0.20 & 0.13 & -6.2 & (12) \\
SDSS J081319.3$+$460849* & Appear & 15.64 & 0.054 & 7.6 & 0.13 &   0.43 & -0.05 & -2.1 & (7) \\
SDSS J083132.2$+$364617 &  Appear & 17.40 & 0.195 & - & 0.21 &  0.48 & -0.13 & -5.1 & (6) \\
SDSS J084748.2$+$182439 &  Disappear & 16.35 & 0.085 & 7.7 & 0.18 &   0.47 & 0.18 & -4.6 & (7) \\
SDSS J084957.7$+$274728 &  Disappear & 18.36 & 0.299 & 7.9 & 0.32 &   0.57 & 0.38 & -0.1 & (6) \\
SDSS J090902.3$+$133019 &  Appear & 15.39 & 0.050 & 7.3 &  0.17 & 0.60 & 0.13 & 0.3 & (7) \\
SDSS J090932.0$+$474730 &  Appear & 19.14 & 0.117 & - & 0.19 & 0.82  & 0.08 & 9.9 & (6) \\
SDSS J092702.3$+$043308 & Disappear & 17.93 & 0.322 & - & 0.29 & 0.49 & 0.18 & 7.7 & (12) \\
SDSS J093730.3$+$260232* &  Appear & 17.33 & 0.162 & 7.6 & 0.18 &   0.53 & -0.18 & 10 & (6) \\
SDSS J093812.3$+$074340 &  Disappear & 14.56 &  0.022 & 7.5 & 0.08 &   0.11 & 0.03 & -2.8 & (7) \\
SDSS J094838.4$+$403043 &  Disappear & 14.99 &  0.047 & 7.5 & 0.10 &  0.20 & 0.06 & 0.1 & (7) \\
SDSS J100220.1$+$450927* &  Disappear & 18.66 & 0.400 & 6.3 & 0.42 & 0.52   & -0.06 & 8.3 & (1) \\
SDSS J100323.4$+$352503 &  Appear & 17.54 & 0.119 & 7.3 & 0.25 &  0.64 & -0.06 & 3.8 &  (6) \\
SDSS J101152.9$+$544206* &  Disappear & 18.30 & 0.246 & 7.9 & 0.37 &     1.24 & 0.45 & -5.8 & (4) \\
SDSS J102152.3$+$464515* &  Disappear & 17.58 & 0.204 & 7.8 & 0.19 &    0.63 & 0.27 & -5.1 & (1) \\
WISE J105203.5$+$151929 &  Disappear & 18.95 & 0.302 & 7.9 & 0.73 & 0.72 & 0.93 & 2.5 & (9) \\
SDSS J110057.7$-$005304* &  Dis/appear & 18.09 & 0.378 & 8.2 & 0.51 & 0.40  & -0.27 & 8.1 & (11) \\
SDSS J110423.2$+$634305 &  Disappear & 19.10 & 0.164 & 6.8 & 0.21 & 0.67  & -0.14 & 12 & (6) \\
SDSS J110455.1$+$011856 &  Disappear & 19.29 & 0.575 & 8.2 & 0.70 &    0.74 & 0.94 & 1.1 & (6) \\
SDSS J111025.4$-$000334* & Appear & 18.31 & 0.219 & 7.6 & 0.31 &    0.69 & -0.13 & 0.5 & (6) \\
SDSS J111329.6$+$531338* &  Disappear & 18.44 & 0.239 & 7.8 & 0.29 & 0.42 & 0.11 & 6.1 & (12) \\
SDSS J111536.5$+$054449* &  Appear & 17.01 & 0.090 & 7.7 & 0.22 & 1.06  & 0.17 & 4.2 & (6)  \\
SDSS J111829.6$+$320359* &  Disappear & 19.85 & 0.365 & 7.7 & 0.69 &    0.58 & -0.14 & -2.2 & (6) \\
SDSS J113229.1$+$035729 &  Appear & 17.04 & 0.091 & 8.3 & 0.13 &   0.68 & -0.06 &  -10 &  (6) \\
SDSS J115039.3$+$363258* &  Disappear & 19.06 & 0.340 & 7.9 & 0.49  &  0.31 & -0.1 & -0.4 & (6) \\
SDSS J115227.4$+$320959 &  Disappear & 18.19 & 0.374 & 8.1 & 0.34 &  0.27 & 0.56 & -0.9 & (6) \\
SDSS J123359.1$+$084211 &  Disappear & 17.92 & 0.255 & 8.7 & 0.25 & 0.85 & 0.30 & -11 & (12) \\
SDSS J125916.7$+$551507 &  Appear & 17.96 & 0.198 & 7.9 & 0.22 &  0.61 & -0.18 & -5.3 &  (6) \\
SDSS J131930.7$+$675355* &  Appear & 17.26 & 0.166 & 7.7 & 0.13 &    0.30 & -0.12 & -9.7 & (6) \\
SDSS J132457.2$+$480241* & Disappear & 17.51 & 0.272 & 8.0 & 0.21 &  0.44 & 0.06 & -4.3 & (1) \\
SDSS J135855.8$+$493414*  & Appear & 18.06 & 0.116 & 6.9 & 0.26 &   0.43 & -0.01 & -6.5 & (6) \\
SDSS J141324.7$+$530527*  & Appear & 18.91 & 0.456 & 8.2 & 0.60 & 0.96 & -0.26 & 5.3 & (10) \\
WISEA J142846.7$+$172353  & Disappear & 17.44 & 0.104 & 7.6 & 0.19 & 1.24 & 0.06 & -5.2 & (8) \\
SDSS J144754.2$+$283324  & Appear & 16.69 & 0.163 & 7.8 & 0.18 &    0.47 & -0.31 & -2.2 & (6) \\
SDSS J153355.9$+$011029*  & Appear & 17.02 & 0.143 & 7.9 & 0.12 &    0.14 & -0.03 & -8.7 & (6) \\
SDSS J153612.8$+$034245  & Disappear & 18.08 & 0.365 & 8.2 & 0.44 & 0.65 & 0.50 & 4.3 & (12) \\
SDSS J153734.0$+$461358  & Disappear & 18.92 & 0.378 & 8.0 & 0.31 & 0.24 & 0.25 & -11 & (12) \\
SDSS J154507.5$+$170951  & Appear & 15.84 & 0.048  & 7.4 & 0.14 &   0.55 & -0.02 & 5.5 & (7) \\
SDSS J154529.6$+$251127  & Appear & 16.72 & 0.117 & 7.5 & 0.13 &   0.52 & -0.04 & -6.0 & (6) \\
SDSS J155017.2$+$413902  & Appear & 19.98 & 0.220 & - & 0.70 &   0.25 & -0.62 & -2.5 & (6) \\
SDSS J155258.3$+$273728  & Appear & 17.10 & 0.086 & - & 0.13 &  0.32 & -0.07 & 8.1 & (6) \\
SDSS J155440.2$+$362952  & Appear & 18.17 & 0.237 & - & 0.52 &  0.77  & -0.76 & 15 & (5) \\
SDSS J160111.2$+$474509  & Disappear & 18.29 & 0.297 & 7.9 & 0.19 & 0.15 & 0.03 & -12 & (12) \\
SDSS J161711.4$+$063833  & Disappear & 17.32 & 0.229 & 8.0 & 0.65 & 0.48 & 0.80 & 14 & (12) \\
SDSS J162415.0$+$455130  & Disappear & 19.31 & 0.481 & 8.1 & 0.45 & 0.29 & 0.63 & -6.8 & (12) \\
\hline
\end{tabular}
\end{table*}

\begin{table*}
\centering
\contcaption{}
\begin{tabular}{lccccccccc}
\hline
Name  & Transition & $V_m$ & $z$ & $\log (M_{BH})$ & Amp & $|\Delta W1|$ & $\Delta$BB & $SWV_1$ & Ref. \\
 & & (mag) && $(M_{\sun})$  & (mag) & (mag) & (mag) &  & \\
\hline
SDSS J210200.4$+$000501  & Disappear & 18.41 & 0.329 & 8.0 & 0.26 & 0.47 & 0.26 & -5.3 & (12) \\
SDSS J214613.3$+$000930*  & Appear & 19.55 & 0.621 & 8.3 & 0.84 &   0.35 & -0.48 & 2.6 & (1) \\
SDSS J220537.7$-$071114 & Disappear & 18.18 & 0.295 & 8.0 & 0.21 & 0.23 & 0.15 & -8.3 & (12) \\
SDSS J225240.3$+$010958*  & Appear & 19.62 & 0.534 & 8.2 & 0.89 &   0.66 & 0.13 & 8.5 & (1) \\
SDSS J233317.3$-$002303*  & Appear & 19.12 & 0.513 & 8.3 & 0.82 &   0.31 & -0.44 & 4.4 & (1) \\
SDSS J233602.9$+$001728*  & Disappear & 18.95 & 0.243 & 7.7 & 0.32 &   0.30 & 0.15 & -4.2 &  (2) \\
SDSS J235107.4$-$091318  & Disappear & 17.95 & 0.355 & 7.9 & 0.20 & 0.20 & 0.16 & -7.3 & (12) \\
\hline
\end{tabular}

References: (1) MacLeod et al. 2016; (2) Ruan et al. 2016; (3) LaMassa et al. 2015; (4) Runnoe et al. 2016; (5) Gezari et al. 2017; (6) Yang et al. 2017; (7) Runco et al. 2016; (8) Assef et al. 2018; (9) Stern et al. 2018; (10) Wang, Xu \& Wei 2018; (11) Ross et al. 2018; (12) MacLeod et al. 2019
\end{table*}

\begin{figure*}
\centering
\includegraphics[width = 6.5in] {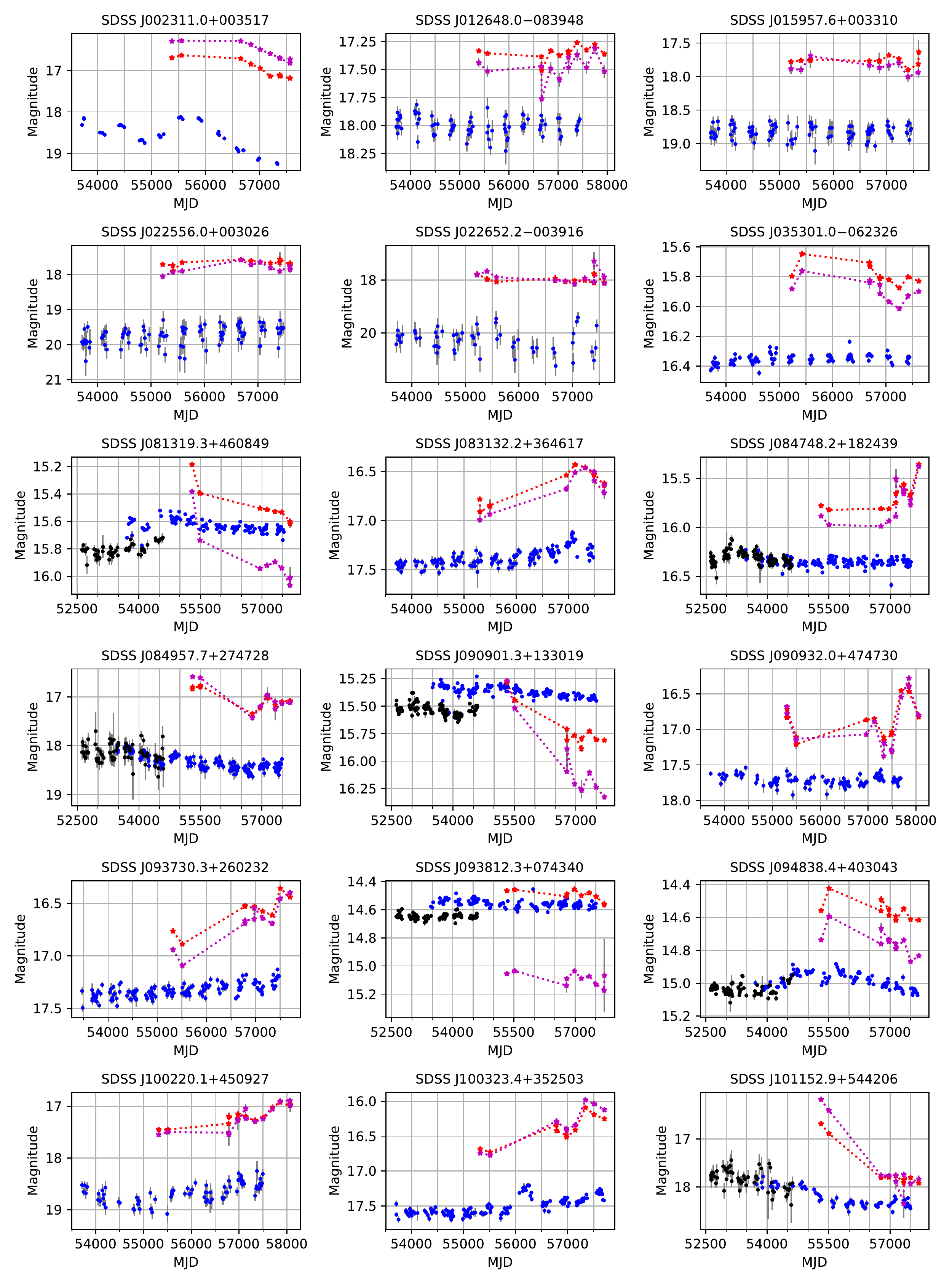}
\caption{Light curves of CLQs identified in the literature showing CRTS (blue) data and LINEAR (black) data where available. The {\em WISE} W1 (maroon) and W2 (red) light curves are also shown (binned on a daily basis) and offsets ($W1$ = 2.70, $W2$ = 3.34) have been applied to the {\em WISE} Vega magnitudes for display.}
\label{fig:knownclq}
\end{figure*}

\begin{figure*}
\centering
\includegraphics[width = 6.7in] {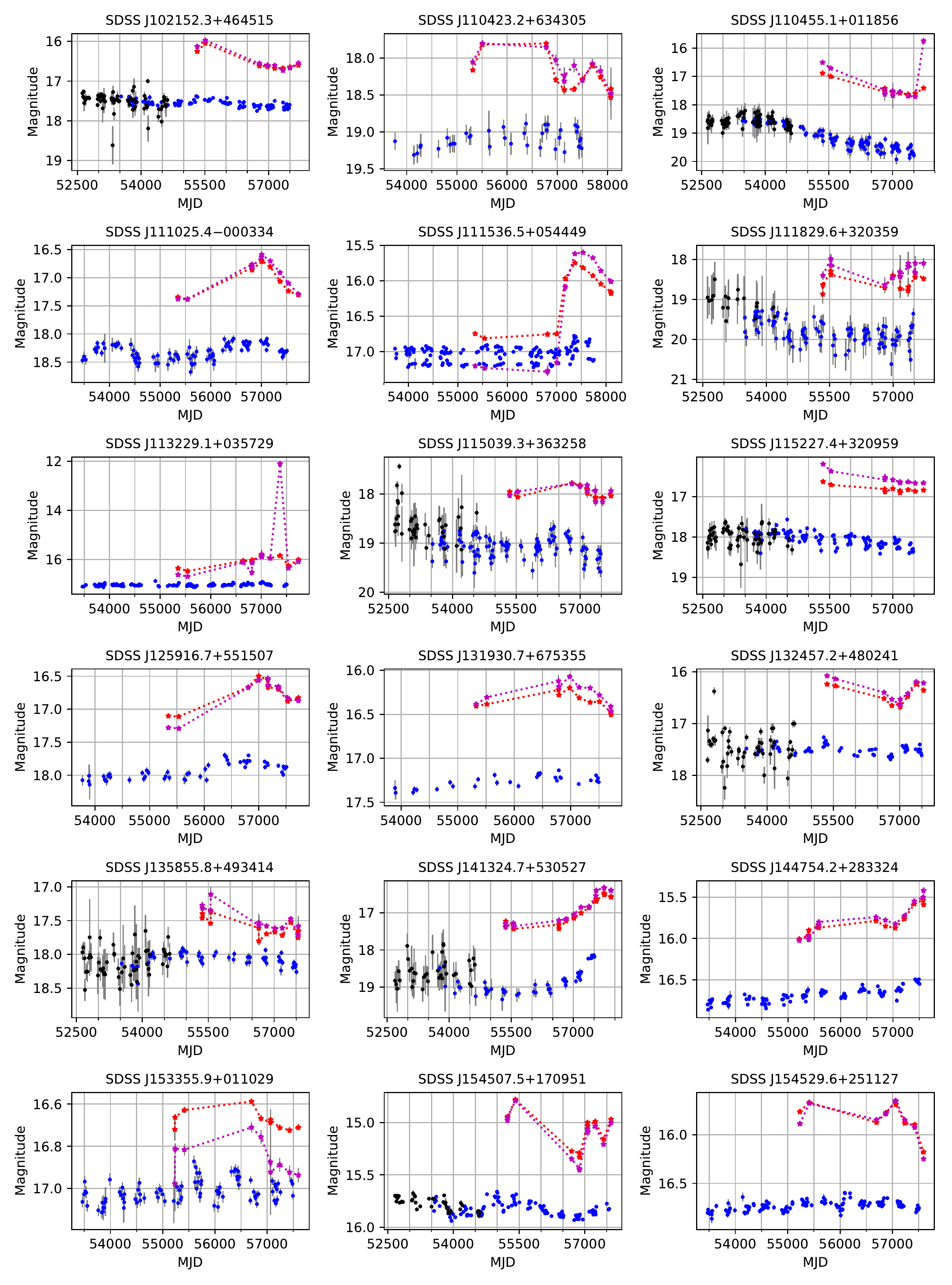}
\contcaption{}
\end{figure*}

\begin{figure*}
\centering
\includegraphics[width = 6.7in] {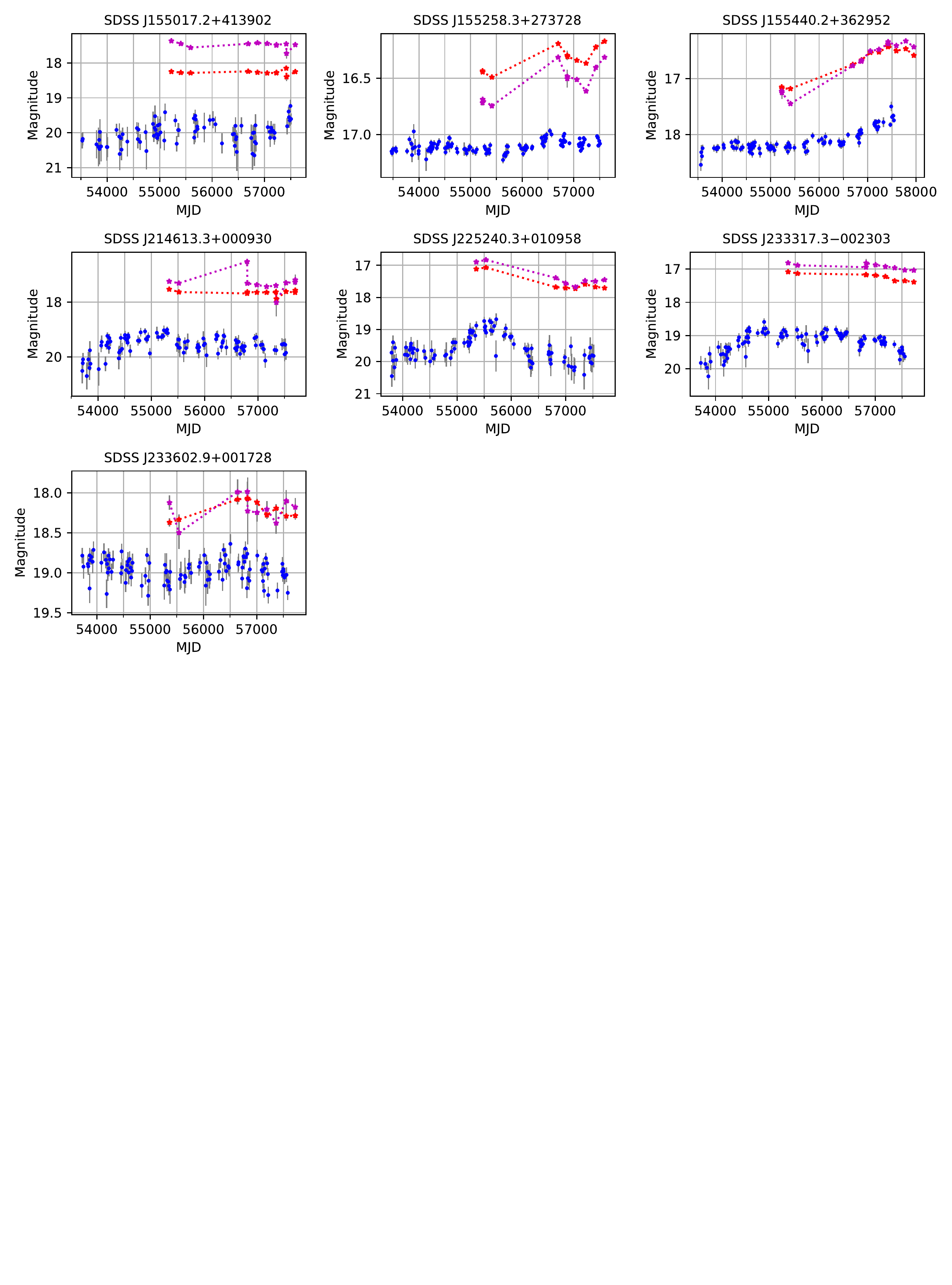}
\contcaption{}
\end{figure*}

\section{CLQ candidates}

\begin{table*}
\centering
\caption{Spectroscopic observations of CSQ candidates. First epoch spectra are from SDSS; subsequent epoch spectra are from SDSS, Palomar (DBSP), and Keck (LRIS, ESI) as described in the text.}
\label{tab:observing}
\begin{tabular}{lll}
\hline
Name & First epoch & Subsequent epochs  \\
  & (MJD) & (MJD) \\
\hline
SDSS J002353.5$-$025159.1 & 57362 & 58054 (DBSP) \\
SDSS J011919.3$-$093721.6 & 52163 & 58013  (DBSP) \\
SDSS J012937.2$+$004609.3  & 52226 & 57373 (SDSS) \\
SDSS J022014.6$-$072859.3  & 52162 & 57360 (DBSP) \\
SDSS J024533.6$+$000745.2  & 52177 & 52946 (SDSS), 56273 (SDSS), 56572/6 (SDSS), 56602 (SDSS), 58013 (DBSP) \\
SDSS J025003.0$+$010930.7  & 52177& 52295 (SDSS), 57690/8 (DBSP), 57779 (DBSP) \\
SDSS J025410.1$+$034912.5  & 55477 & 57399 (SDSS) \\
SDSS J025505.7$+$002523.5  & 51816 & 51877 (SDSS), 52175 (SDSS), 58013 (DBSP) \\
SDSS J025619.0$+$004501.0  & 51816 & 51877 (SDSS), 52175 (SDSS), 56984 (SDSS), 58054 (DBSP), 58070 (DBSP) \\
SDSS J074542.3$+$421404.5  & 51885 & 58054 (DBSP)  \\
SDSS J074908.7$+$453009.0  & 55208 & 56328 (SDSS)  \\
SDSS J075440.3$+$324105.1  & 52583 & 57461 (ESI) \\
SDSS J075728.3$+$245510.1  & 52669 & 57698 (DBSP) \\
SDSS J080138.7$+$423355.2  & 55178 & 55245 (SDSS), 57073 (SDSS) \\
SDSS J080500.3$+$340225.6  & 52584 & 57698 (DBSP) \\
SDSS J081425.9$+$294116.3  & 52618 & 55542 (SDSS)  \\
SDSS J081632.1$+$404804.6  & 52264 & 57361 (SDSS)  \\
SDSS J082033.3$+$382420.4  & 52589 & 58212 (DBSP)  \\
SDSS J082930.7$+$272821.9  & 52932 & 57698 (DBSP)  \\
SDSS J083225.3$+$370736.6  & 52312 & 57050 (ESI)  \\
SDSS J083236.3$+$044506.2  & 52646 & 57698 (DBSP)  \\
SDSS J083533.2$+$494818.8  & 55290 & 58212 (DBSP)  \\
SDSS J084716.1$+$373218.7  & 52323 & 57452 (SDSS) \\
SDSS J091357.3$+$052229.8  & 52652 & 57844 (DBSP), 58154 (DBSP)  \\
SDSS J092441.1$+$284730.6  & 53389 & 57050 (ESI) \\
SDSS J092736.7$+$153824.3  & 54068 & 58131 (DBSP)  \\
SDSS J092836.9$+$474245.8  & 52637 & 56740 (SDSS)  \\
SDSS J093017.7$+$470721.7  & 52316 & 56685 (SDSS), 58131 (DBSP)  \\
SDSS J093329.0$+$291734.1  & 53389 & 58131 (DBSP)  \\
SDSS J094620.9$+$334746.5  & 53387 & 57461 (ESI), 57844 (DBSP)  \\
SDSS J095427.6$+$485638.9  & 52703/8 & 58131 (DBSP), 58154 (DBSP)  \\
SDSS J095536.8$+$103751.7  & 52996 & 57050 (ESI), 57844 (DBSP)  \\
SDSS J095750.0$+$530106.0  & 52385 & 52400 (SDSS), 56993 (SDSS), 57844 (DBSP) \\
SDSS J100256.2$+$475027.9  & 52339 & 56338 (SDSS), 58154 (DBSP) \\
SDSS J100343.3$+$512611.2  & 52385 & 52400 (SDSS), 58246 (DBSP) \\
SDSS J101857.9$+$103625.8  & 52999 & 55957 (SDSS) \\
SDSS J102614.0$+$523752.0  & 52644 & 58212 (DBSP) \\
SDSS J102752.4$+$421012.5  & 55588 & 58154 (DBSP) \\
SDSS J102817.7$+$211508.1  & 53741 & 58154 (DBSP) \\
SDSS J103255.9$+$365451.0  & 55575 & 58246 (DBSP) \\
SDSS J104254.8$+$253714.2  & 53792 & 56358 (SDSS) \\
SDSS J110249.9$+$525013.5  & 52652 & 57374 (SDSS) \\
SDSS J110349.2$+$312416.7  & 53472 & 56367 (SDSS) \\
SDSS J110438.7$+$333059.7  & 55626 & 56369 (SDSS) \\
SDSS J111334.9$+$322527.2  & 53786 & 57898 (DBSP) \\
SDSS J111617.8$+$251035.0  & 54115 & 58246 (DBSP) \\
SDSS J111947.6$+$233539.9  & 54154 & 56304 (SDSS), 58212 (DBSP) \\
SDSS J112243.1$+$364141.6  & 53467 & 58212 (DBSP) \\
SDSS J113111.1$+$373709.4  & 53446 & 57426 (SDSS), 57844 (DBSP) \\
SDSS J113615.2$+$103431.1 & 52765 & 57461 (ESI) \\
SDSS J113706.9$+$013948.2  & 51989 & 57461 (ESI) \\
SDSS J113757.7$+$365501.8  & 55673 & 57427 (SDSS) \\
SDSS J114408.9$+$424357.5  & 53062 & 57520 (SDSS) \\
SDSS J120130.9$+$494049.8  & 52442 & 54849 (SDSS), 58212 (DBSP) \\
SDSS J120442.2$+$275411.6  & 53819 & 56337 (SDSS), 58212 (DBSP) \\
SDSS J123215.2$+$132032.3  & 53166 & 58131 (DBSP), 58154 (DBSP) \\
SDSS J123819.6$+$412420.4  & 53090 & 57511 (SDSS) \\
SDSS J125257.0$+$001052.6  & 51689 &  51994 (SDSS), 55575 (SDSS) \\
SDSS J130323.5$+$011103.0  & 51986 & 58154 (DBSP) \\
SDSS J134133.7$+$090356.3  & 53886 & 58154 (DBSP) \\
SDSS J134822.3$+$245650.4  & 53535 & 58212 (DBSP) \\
SDSS J135636.6$+$255320.0  & 53792 & 58245 (DBSP), 58249 (DBSP) \\
\hline
\end{tabular}
\end{table*}

\begin{table*}
\centering
\contcaption{}
\begin{tabular}{llll}
\hline
Name & First epoch & Subsequent epochs  \\
  & (MJD) & (MJD) \\
\hline
SDSS J142852.8$+$271042.9  & 56067 & 57844 (DBSP), 58339 (DBSP)  \\
SDSS J144118.9$+$485454.8  & 52733 & 56370 (SDSS), 58246 (DBSP)  \\
SDSS J144202.8$+$433709.1  & 52734 & 57844 (DBSP)  \\
SDSS J144702.8$+$273747.2  & 54208 & 57844 (DBSP), 57930 (DBSP)  \\
SDSS J145022.7$+$102555.8  & 53827 & 57244 (DBSP), 57535 (DBSP)  \\
SDSS J145755.4$+$435035.5  & 52734 & 58249 (LRIS) \\
SDSS J151604.3$+$355025.4  & 53083 & 57930 (DBSP)  \\
SDSS J152749.9$+$084408.6  & 56002 & 57570 (DBSP)  \\
SDSS J153354.6$+$345504.6  & 53144 & 57570 (DBSP) \\
SDSS J153415.4$+$303434.5  & 53119 & 57930 (DBSP)  \\
SDSS J155651.4$+$321008.9  & 52825 & 58246 (DBSP)  \\
SDSS J155829.4$+$271714.3  & 52817 & 57926 (LRIS) \\
SDSS J160743.0$+$432817.1  & 52756 & 57244 (DBSP), 57902 (DBSP), 57930 (DBSP)  \\
SDSS J161400.3$-$011005.1  & 51693 & 57535 (DBSP)  \\
SDSS J224829.4$+$144418.4  & 52263 & 57188 (DBSP), 57570 (DBSP), 58013 (DBSP), 57926 (LRIS), 58285 (LRIS)  \\
SDSS J230443.6$-$084110.0  & 52258 & 58013 (DBSP)  \\
SDSS J231207.6$+$140212.8  & 52251 & 57660 (DBSP) \\
SDSS J233136.8$-$105638.0  & 52523 & 57660 (DBSP) \\
SDSS J235439.1$+$005751.9  & 51788 & 52523 (SDSS), 56959 (SDSS), 58054 (DBSP)\\
\hline
\end{tabular}
\end{table*}

\begin{figure*}
\centering
\includegraphics[width = 7.0in]{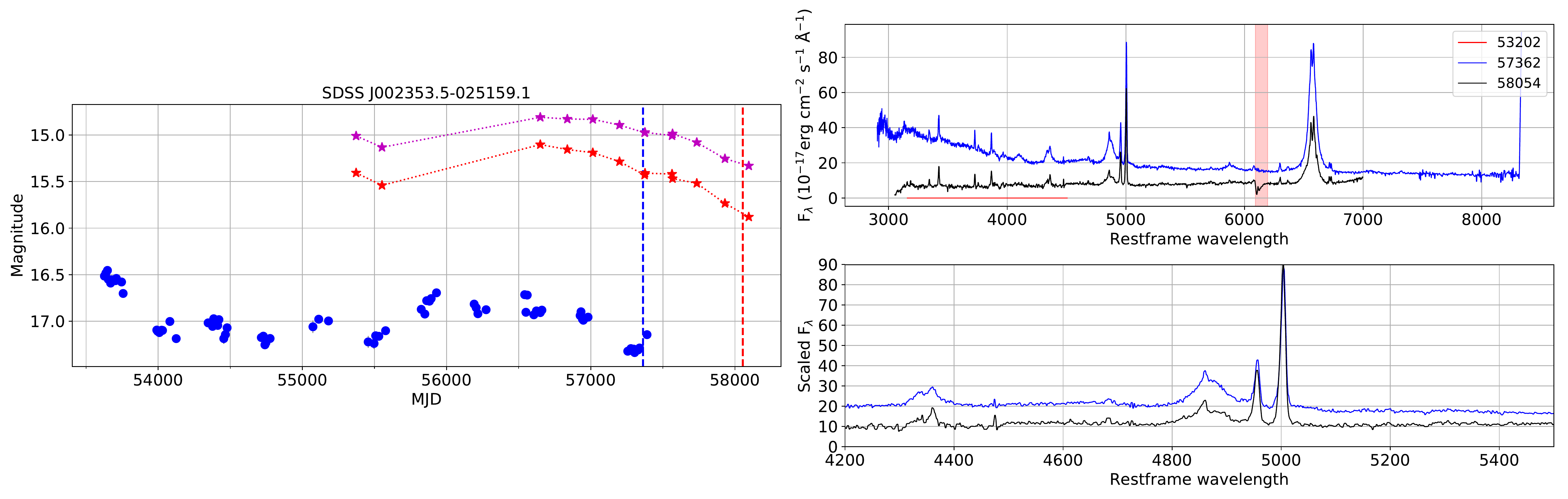}
\includegraphics[width = 7.0in]{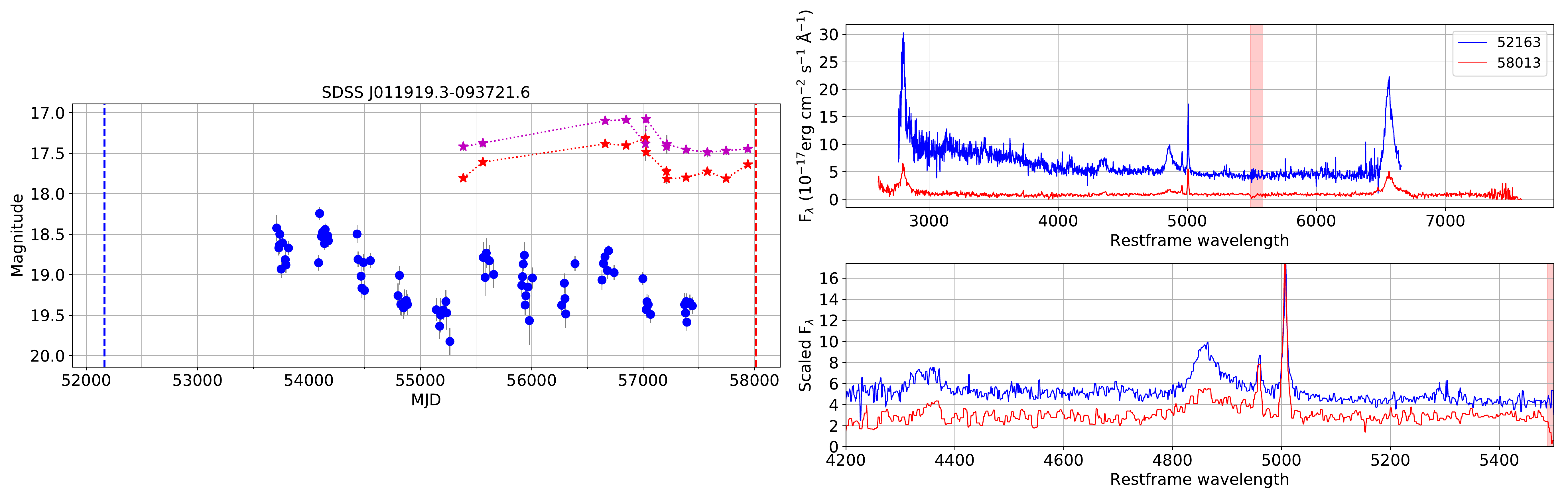}
\includegraphics[width = 7.0in]{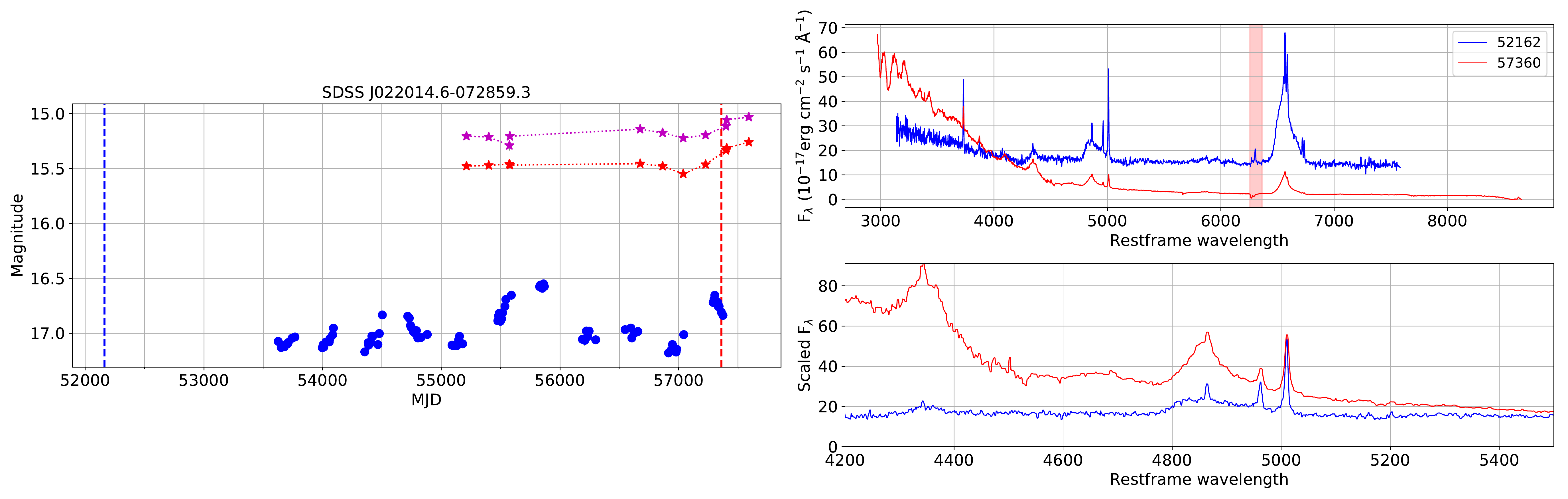}
\caption{Light curves and spectra of CSQs identified in this paper. The left plot for each source shows the CRTS (blue) data and LINEAR (black) data where available. The {\em WISE} W1 (maroon) and W2 (red) light curves are also shown (binned on a daily basis) and offsets ($W1$ = 2.70, $W2$ = 3.34) have been applied to the {\em WISE} Vega magnitudes for display. The right upper plot for each source shows the SDSS spectra and the spectra obtained in our followup. The lower right plot shows a comparison of the H$\beta$ regions for the spectra scaled to the flux of [\ion{O}{III}] $\lambda \, 5007$ in the earliest spectra. The spectra are smoothed with a 3 \AA\,\. box filter in all cases. The red shaded area indicates the location of the atmospheric O$_2$ A-band absorption feature. The corresponding epochs of the spectra are shown in the left plot by dashed lines. The full set is available online.}
\label{fig:newclqspec}
\end{figure*}

\begin{figure*}
\centering
\includegraphics[width = 7.0in]{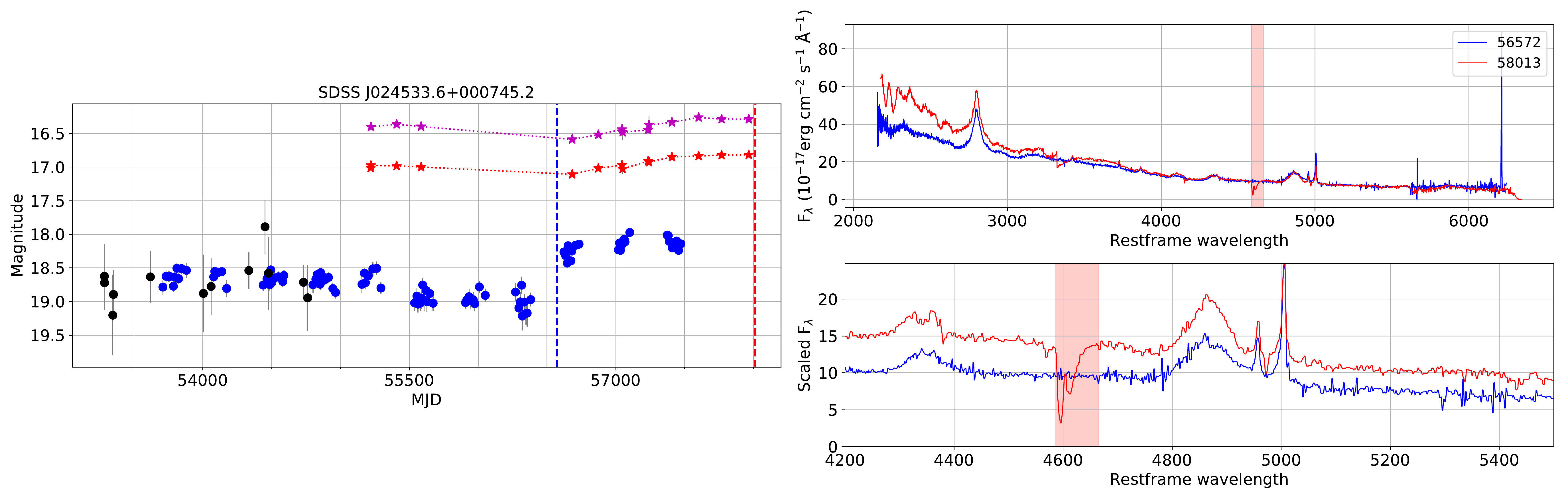}
\includegraphics[width = 7.0in]{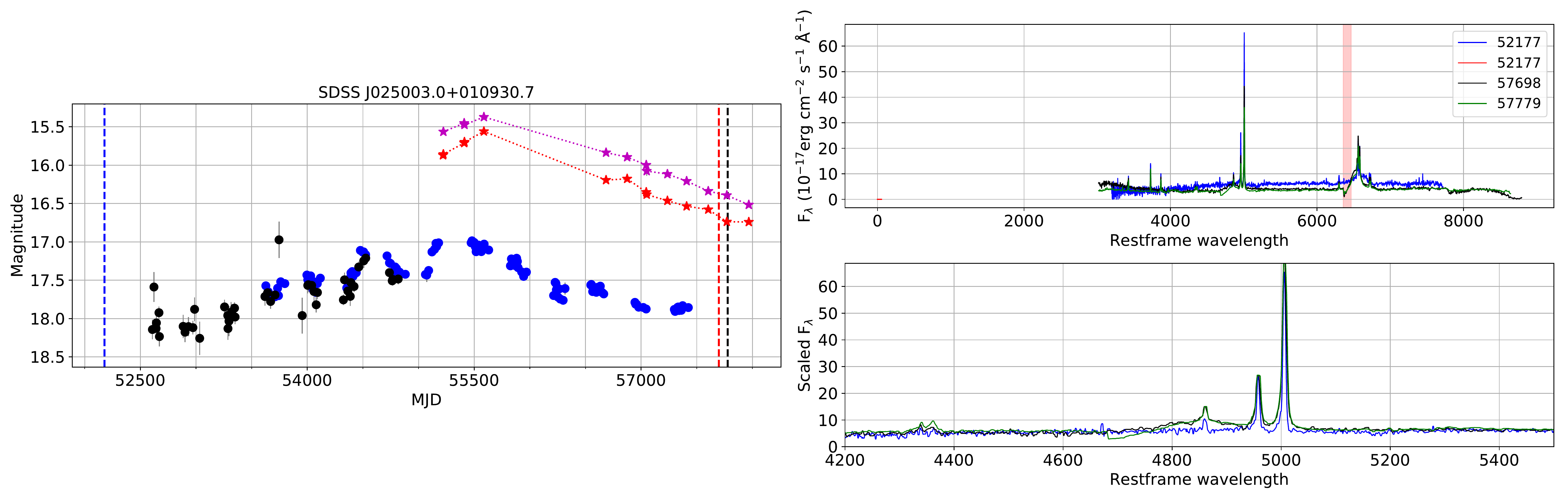}
\includegraphics[width = 7.0in]{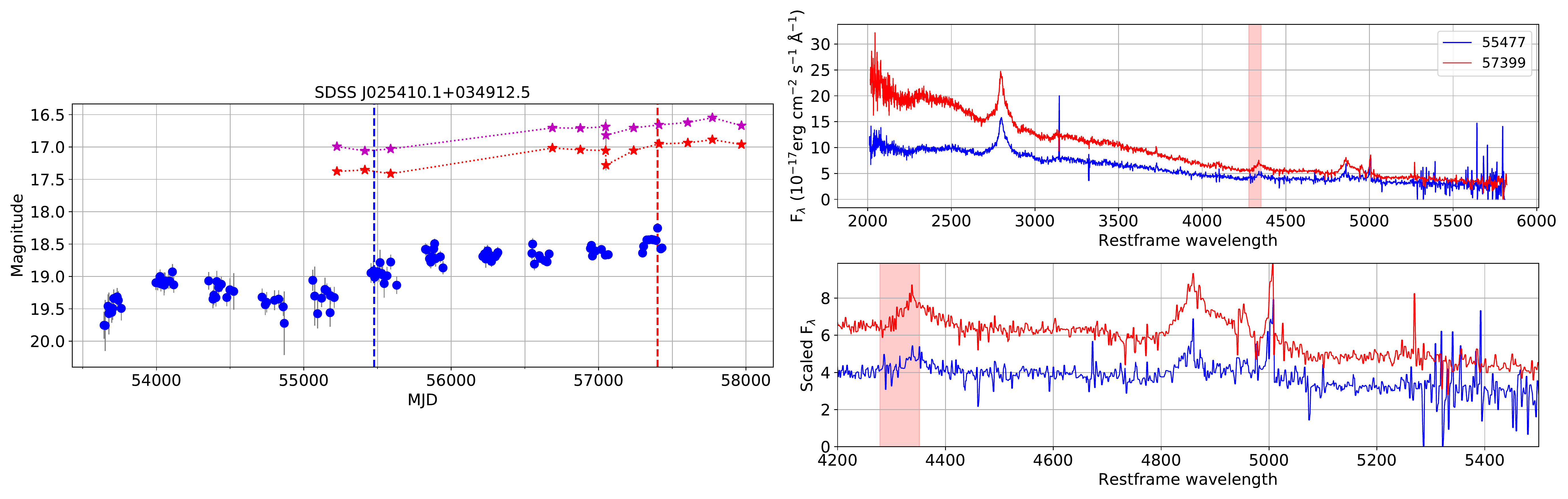}
\includegraphics[width = 7.0in]{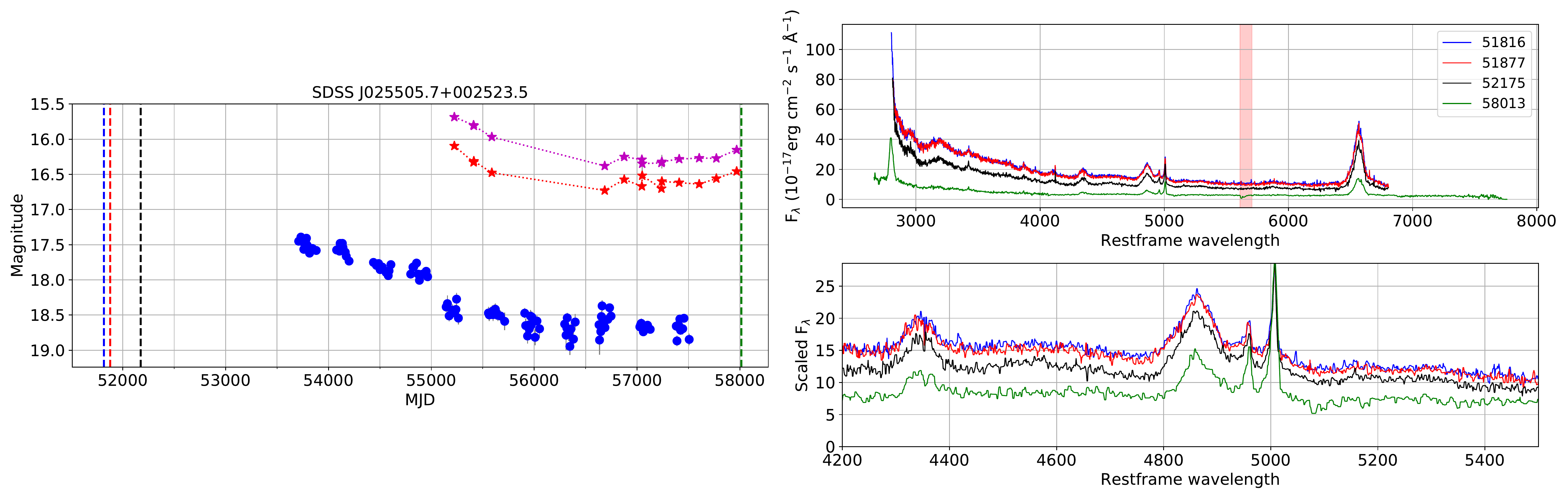}
\contcaption{}
\end{figure*}

\begin{figure*}
\centering
\includegraphics[width = 7.0in]{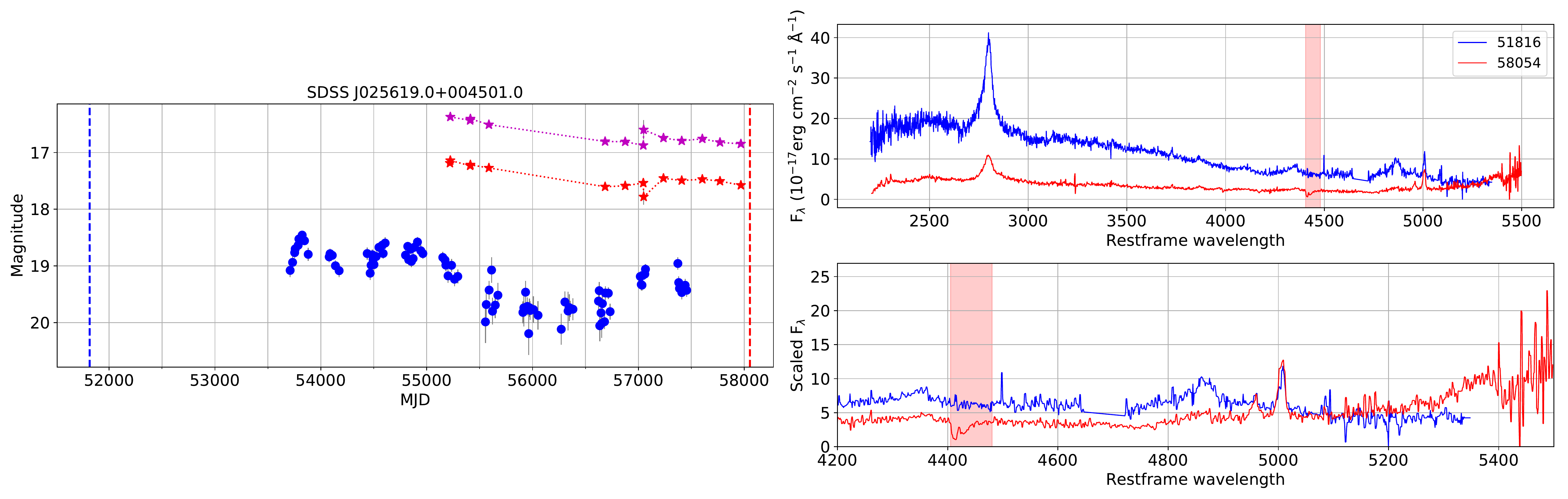}
\includegraphics[width = 7.0in]{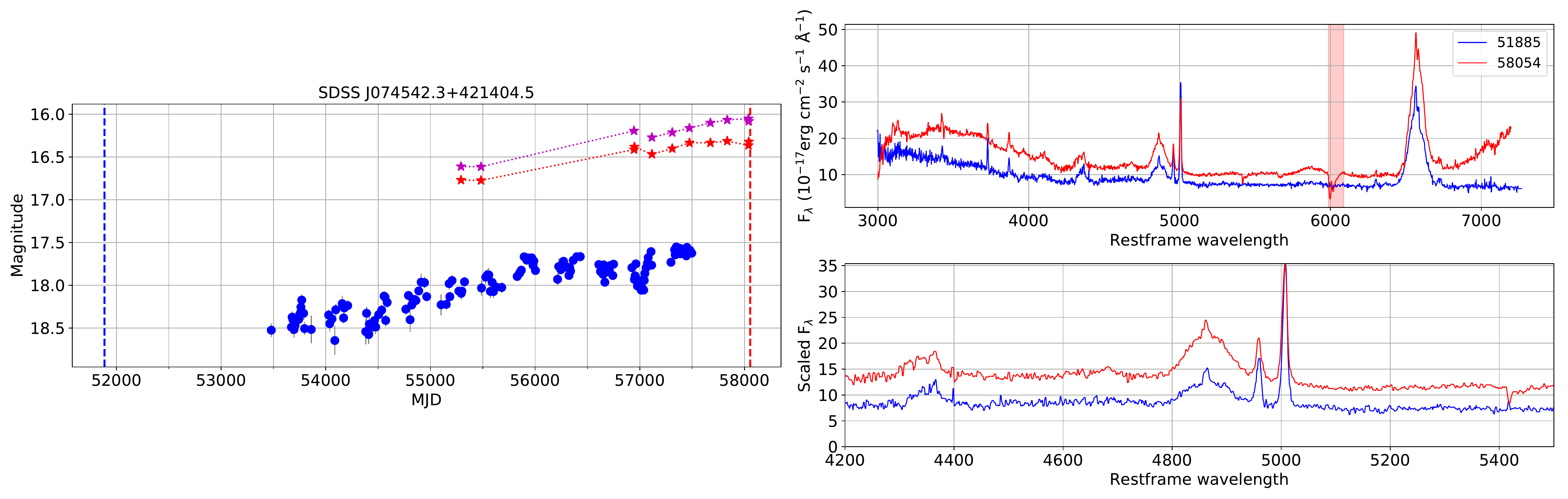}
\includegraphics[width = 7.0in]{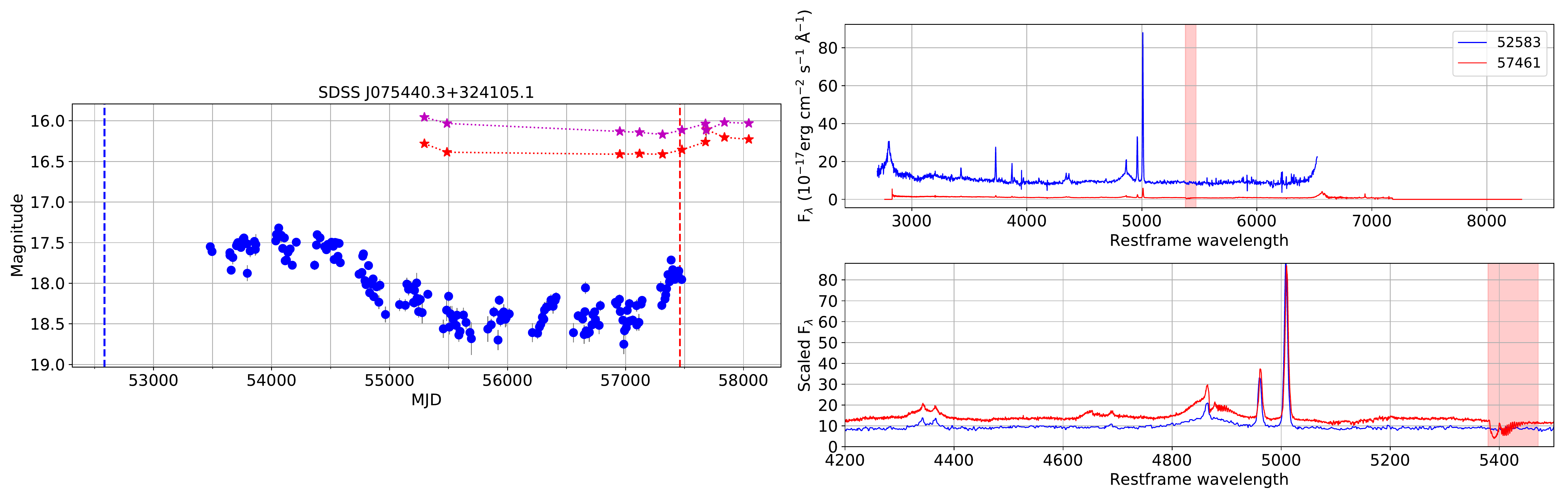}
\includegraphics[width = 7.0in]{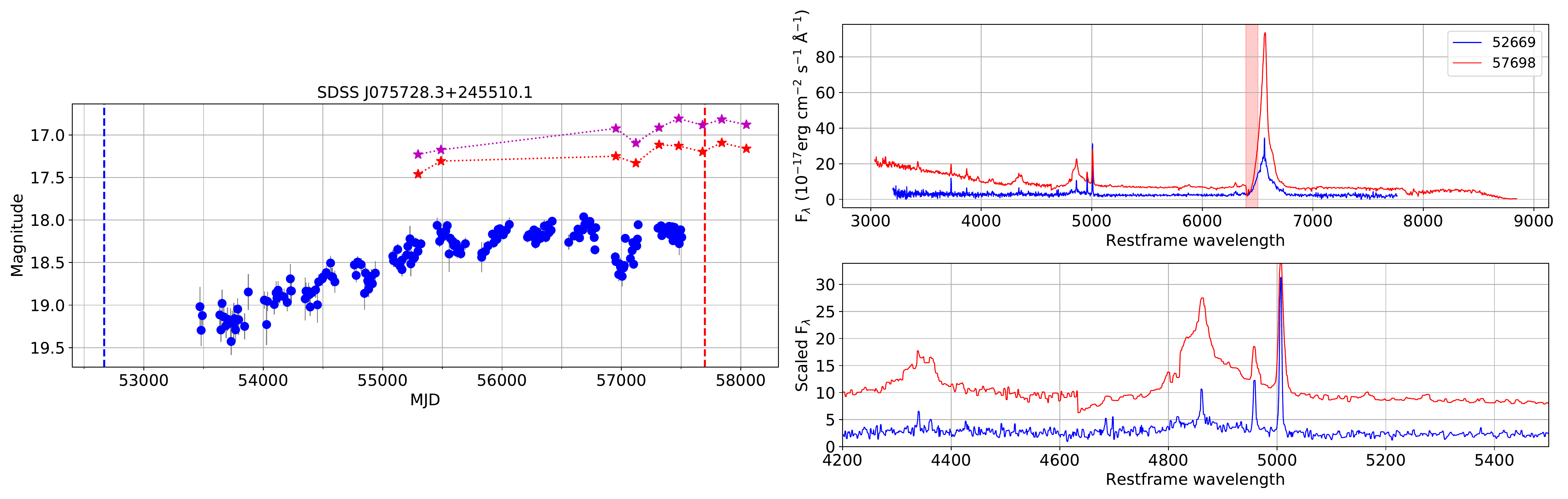}
\contcaption{}
\end{figure*}

\begin{figure*}
\centering
\includegraphics[width = 7.0in]{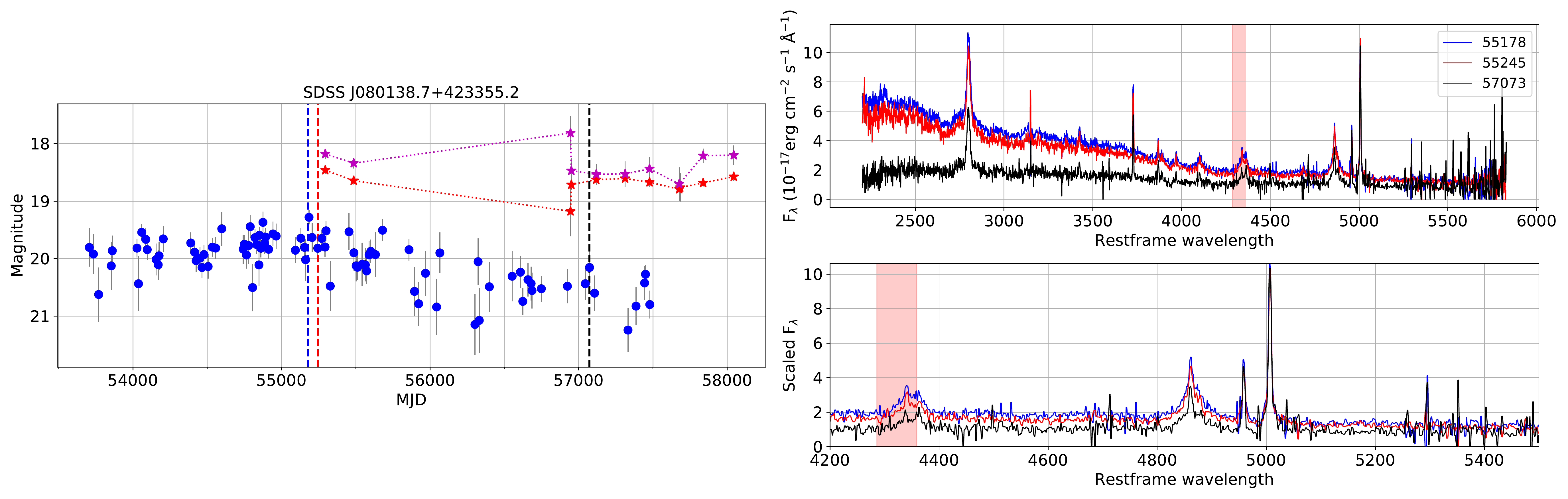}
\includegraphics[width = 7.0in]{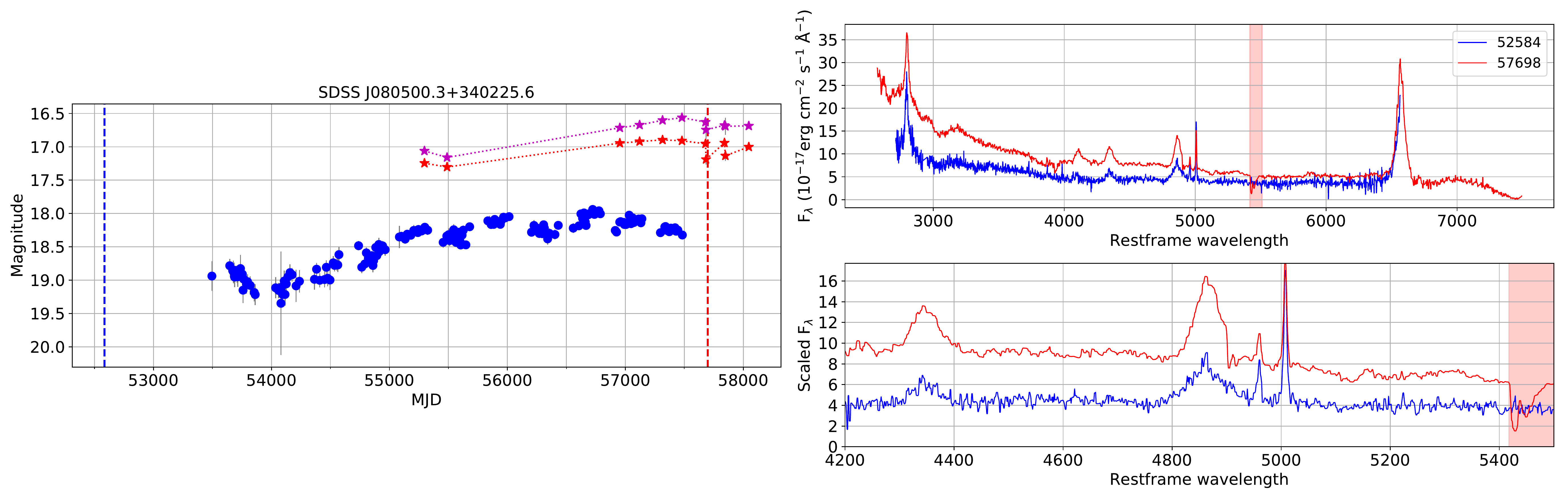}
\includegraphics[width = 7.0in]{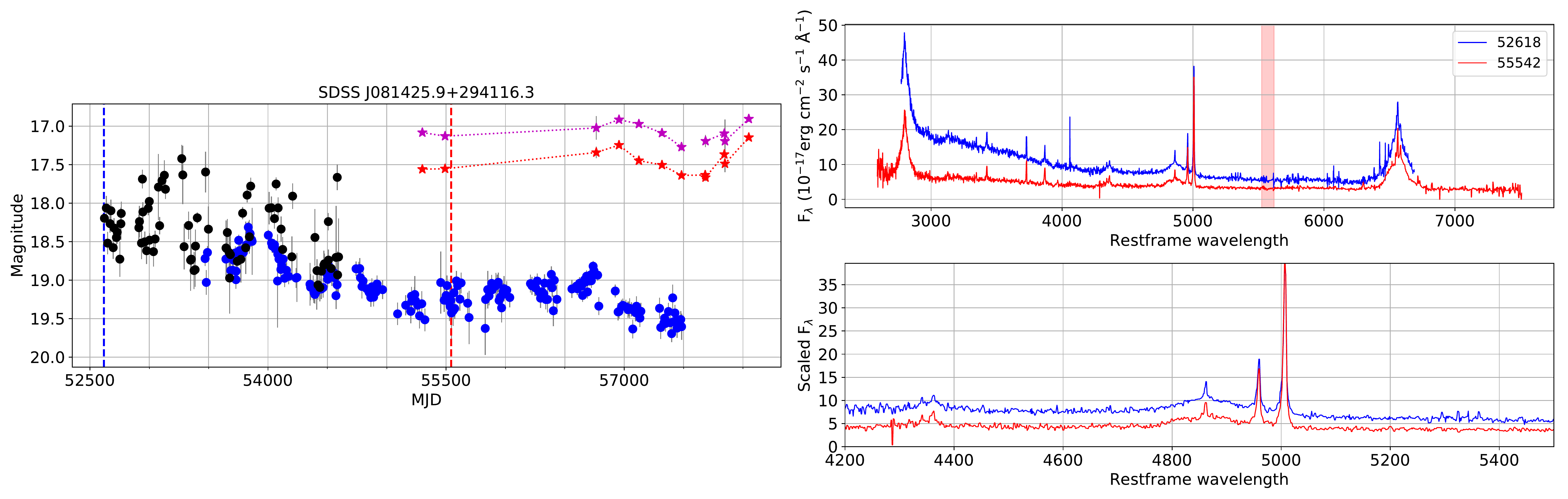}
\includegraphics[width = 7.0in]{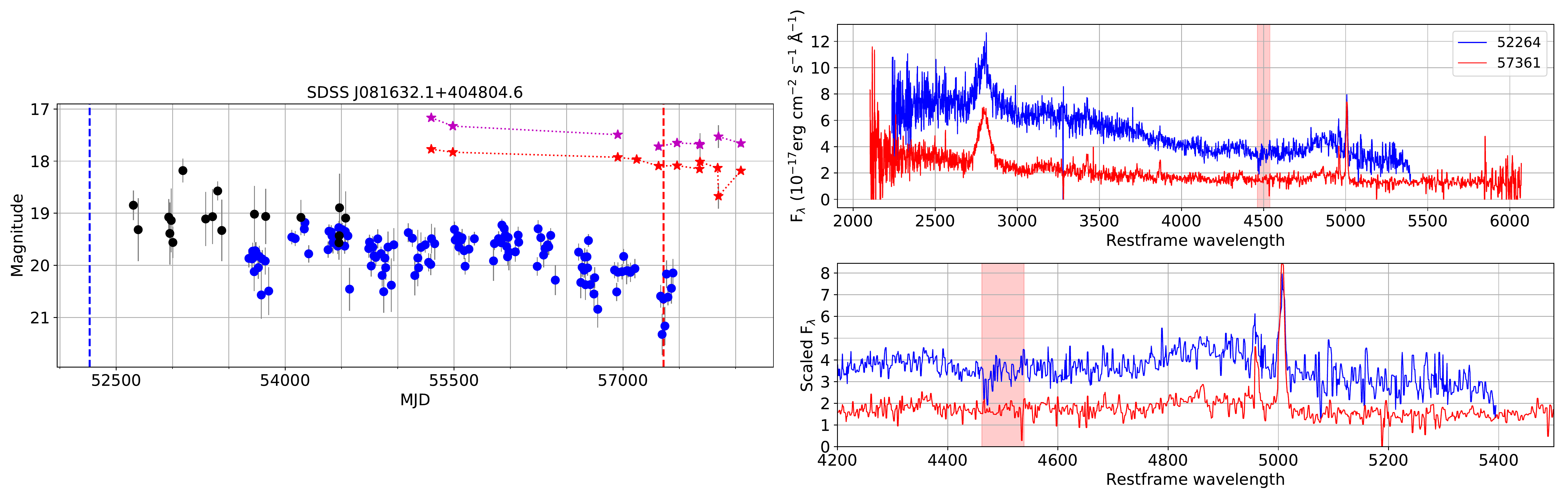}
\contcaption{}
\end{figure*}

\begin{figure*}
\centering
\includegraphics[width = 7.0in]{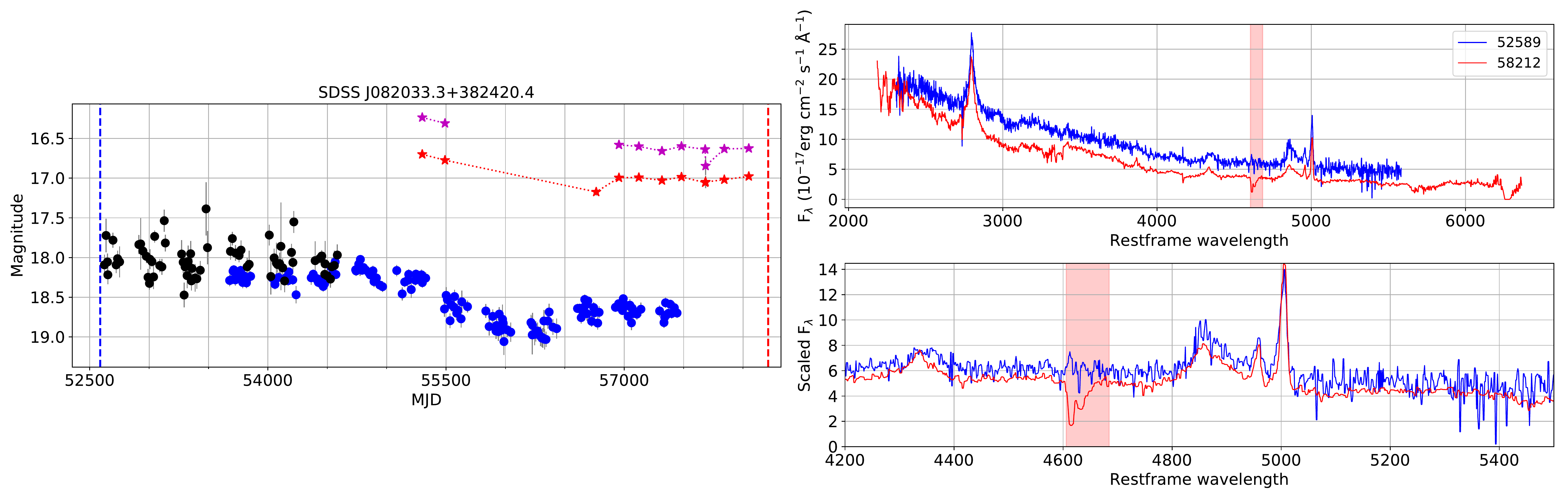}
\includegraphics[width = 7.0in]{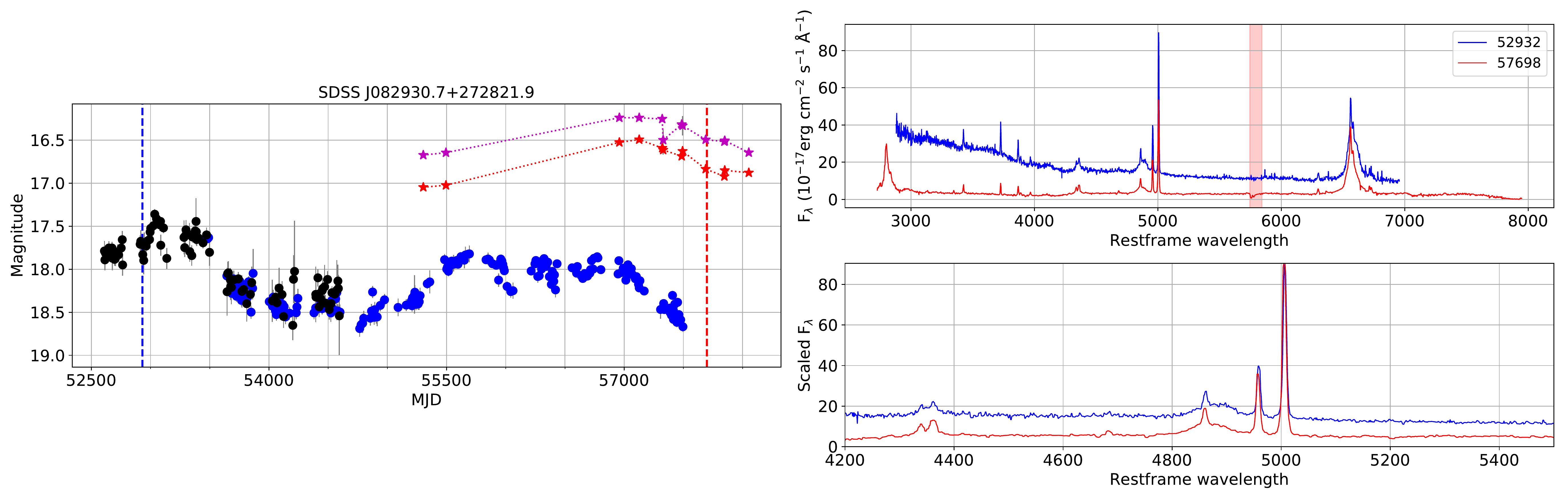}
\includegraphics[width = 7.0in]{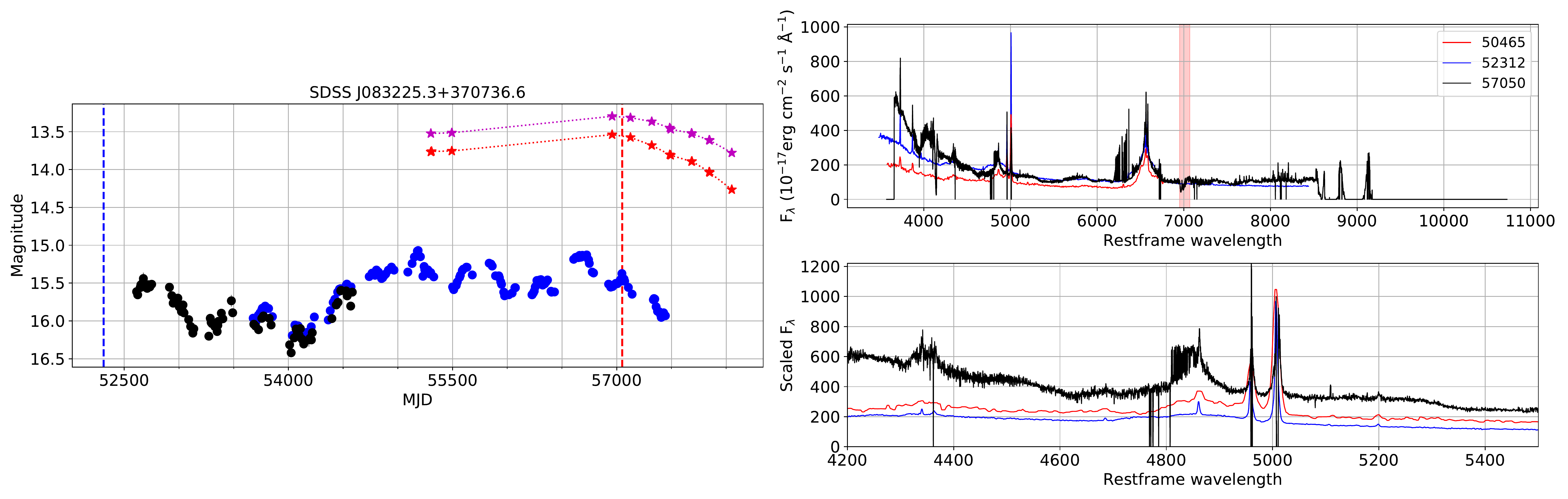}
\includegraphics[width = 7.0in]{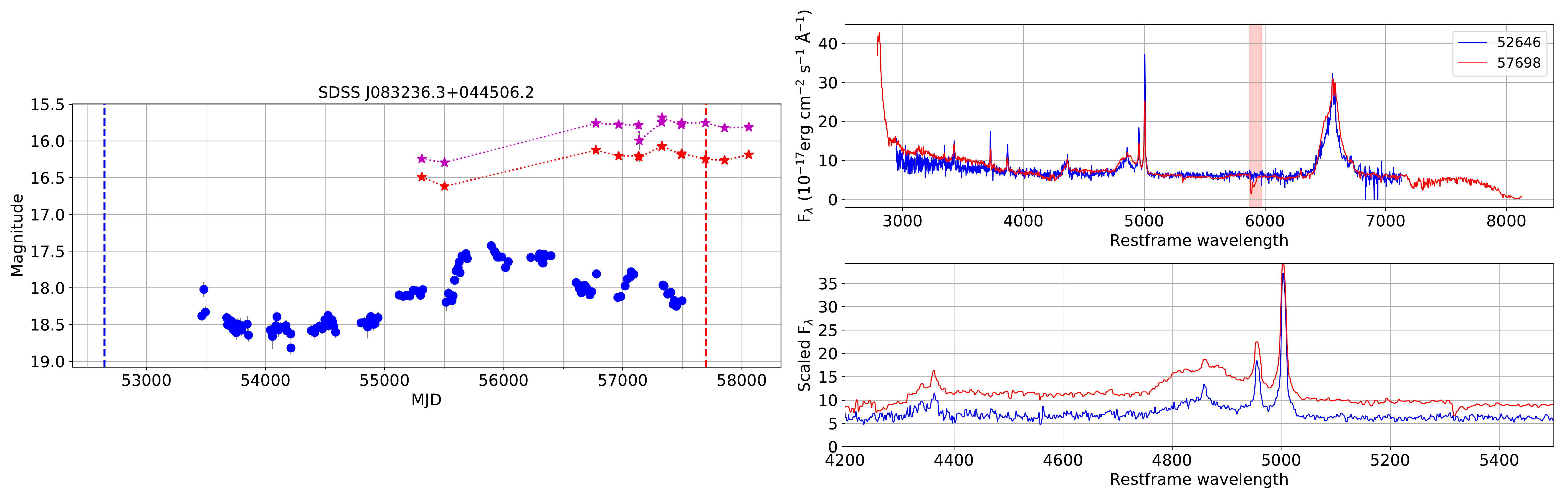}
\contcaption{}
\end{figure*}

\begin{figure*}
\centering
\includegraphics[width = 7.0in]{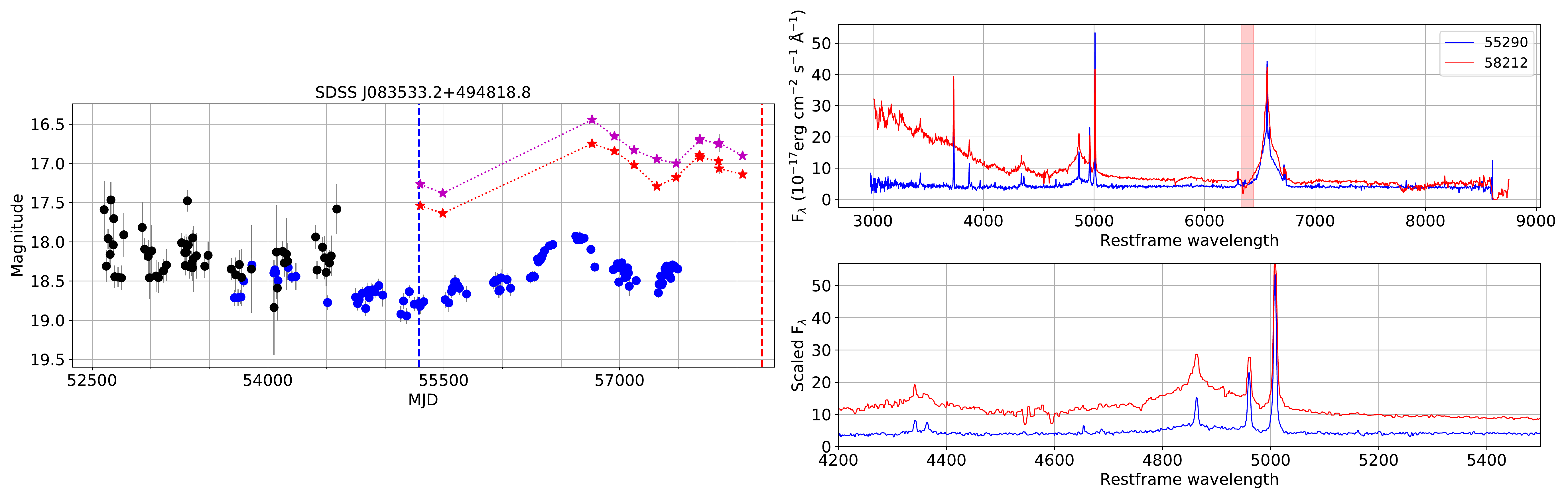}
\includegraphics[width = 7.0in]{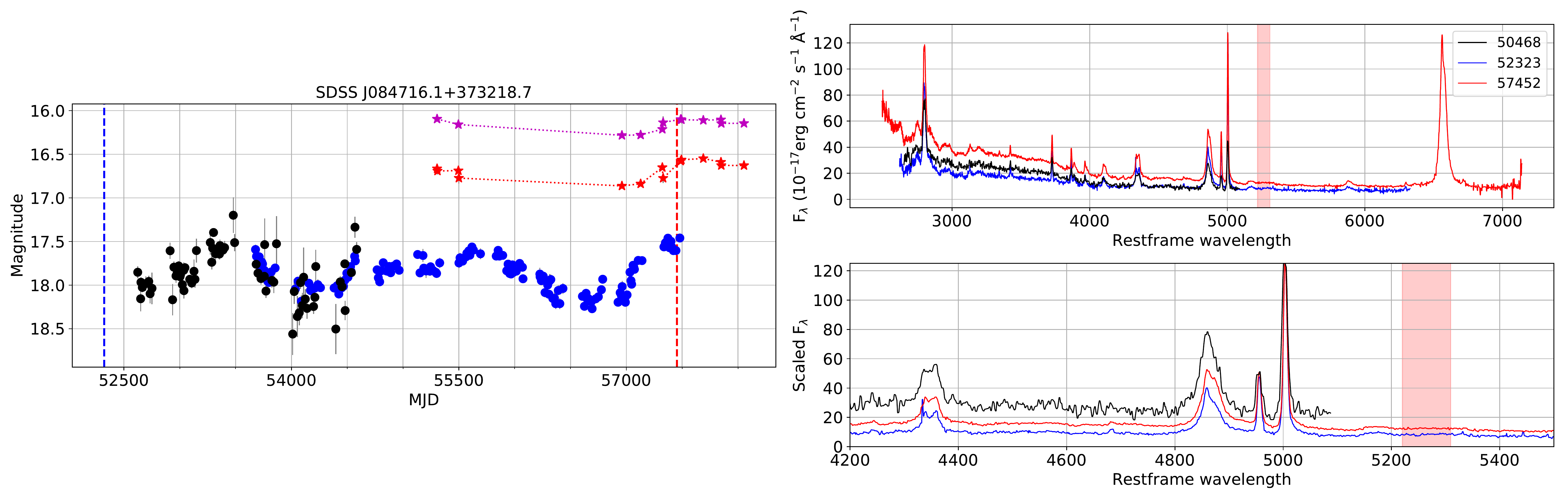}
\includegraphics[width = 7.0in]{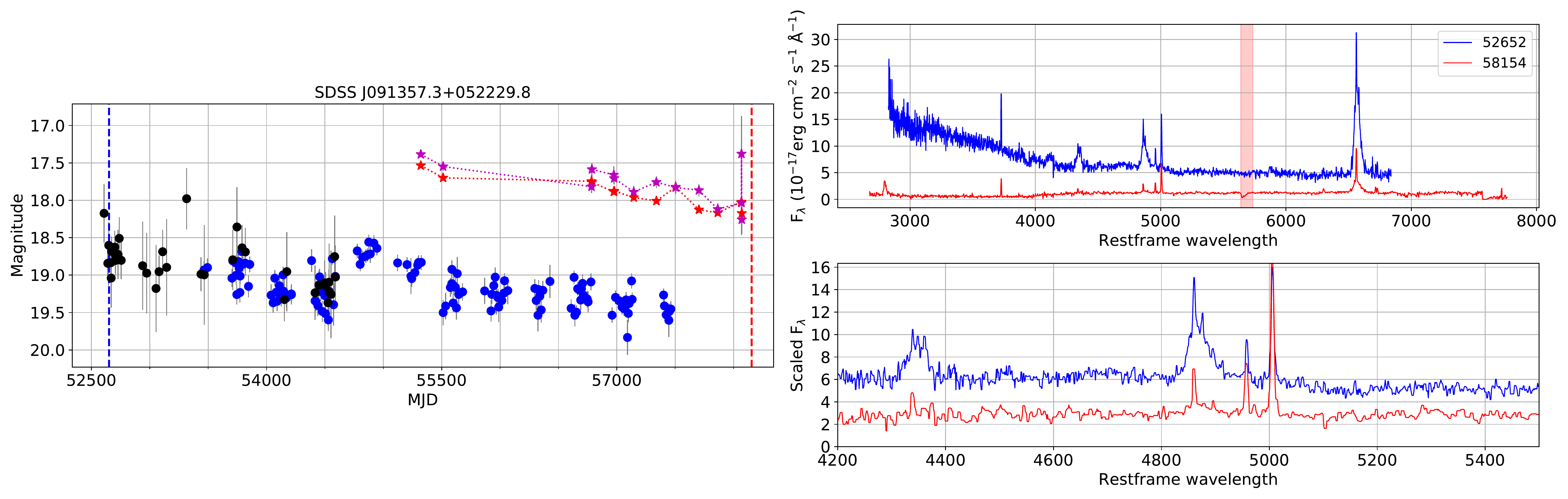}
\includegraphics[width = 7.0in]{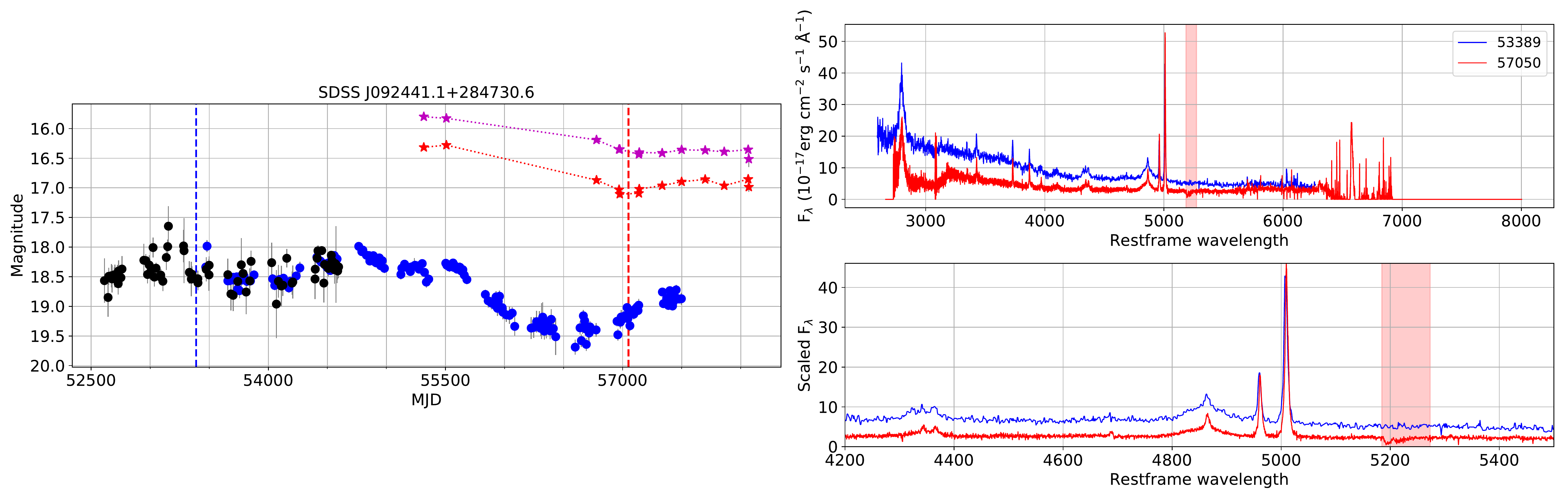}
\contcaption{}
\end{figure*}

\begin{figure*}
\centering
\includegraphics[width = 7.0in]{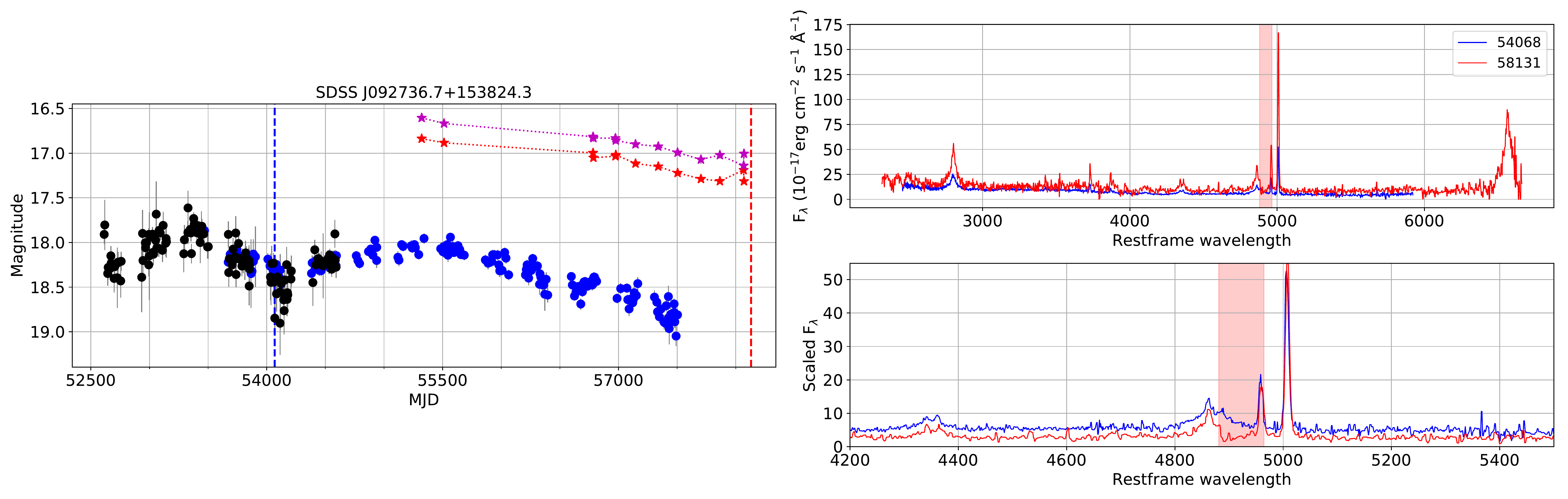}
\includegraphics[width = 7.0in]{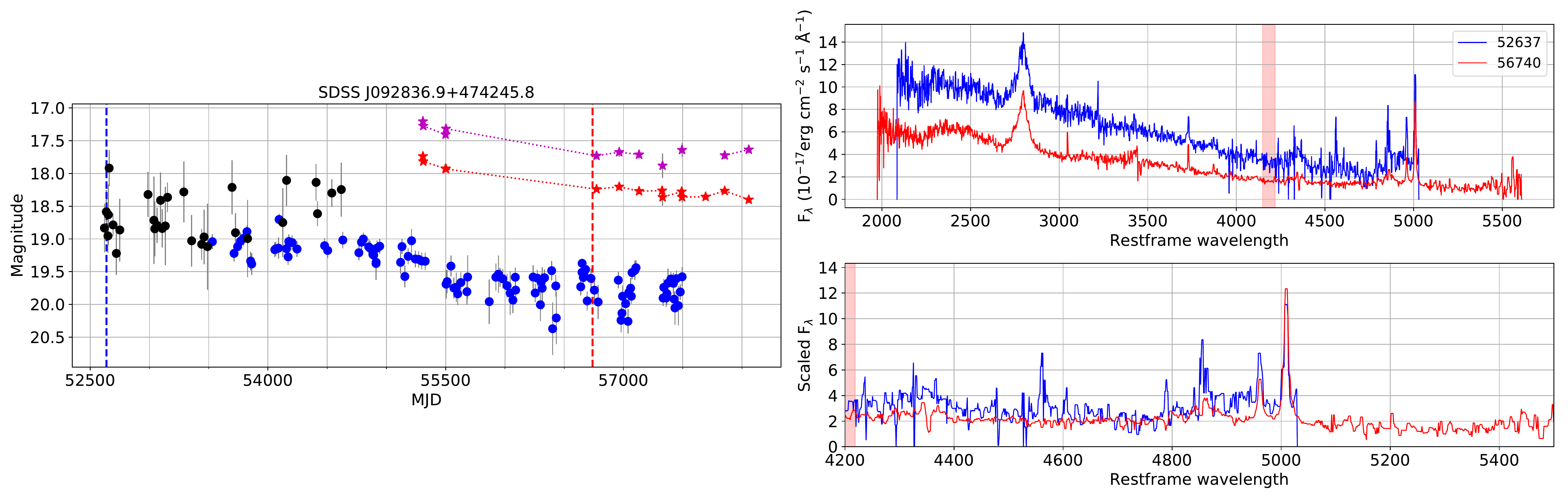}
\includegraphics[width = 7.0in]{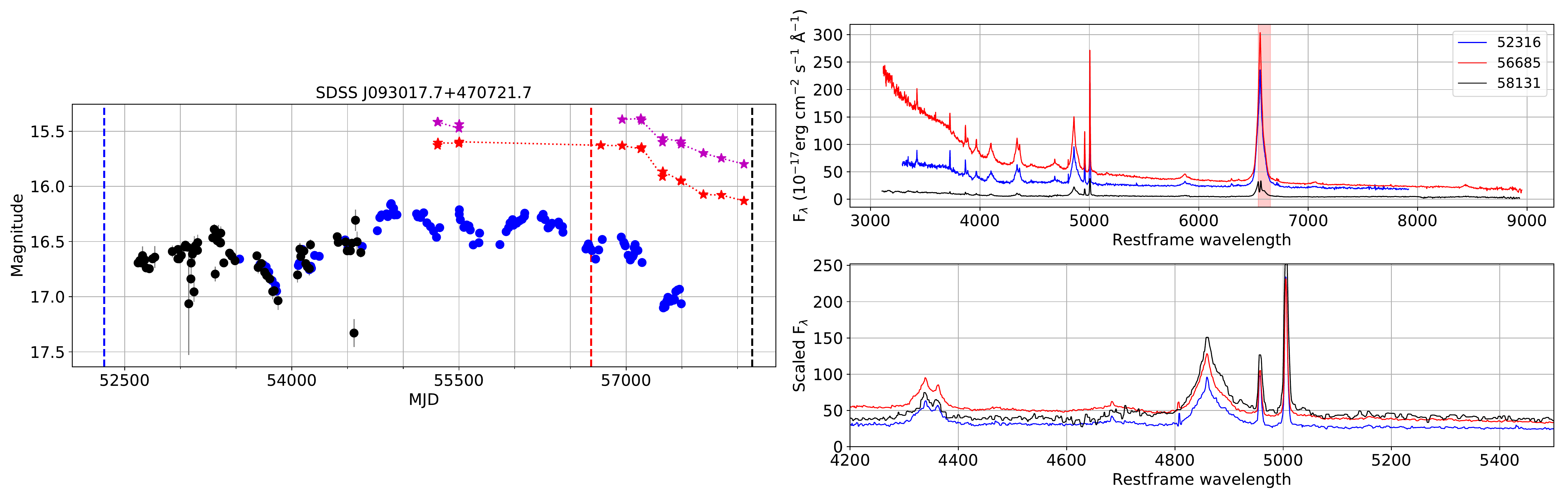}
\includegraphics[width = 7.0in]{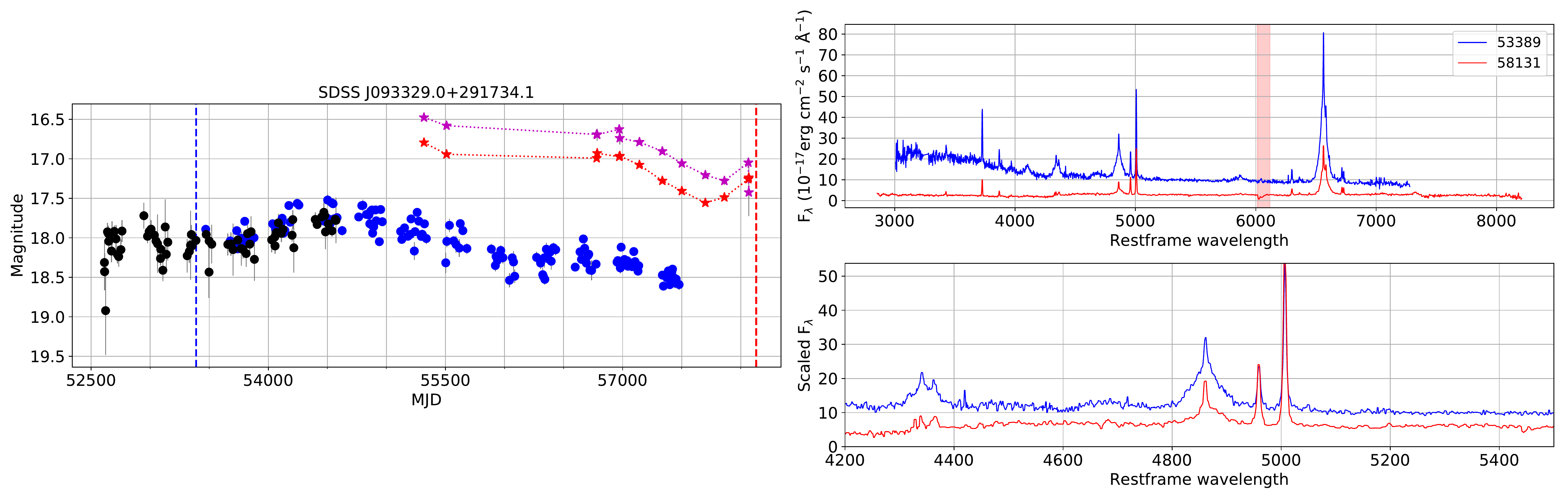}
\contcaption{}
\end{figure*}

\begin{figure*}
\centering
\includegraphics[width = 7.0in]{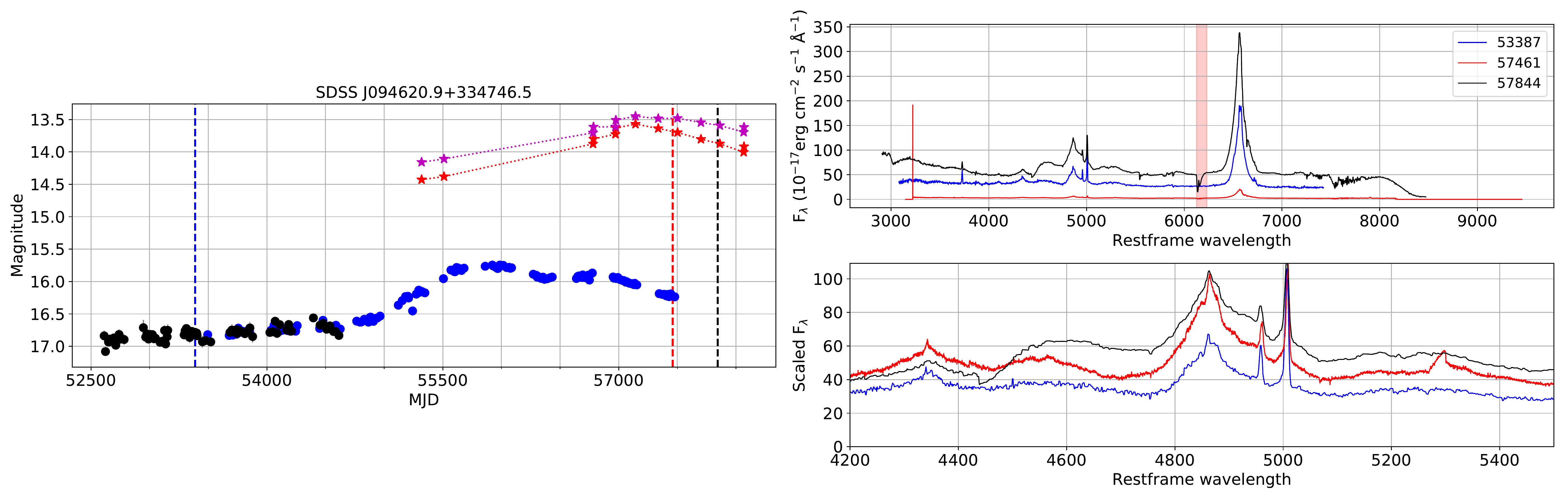}
\includegraphics[width = 7.0in]{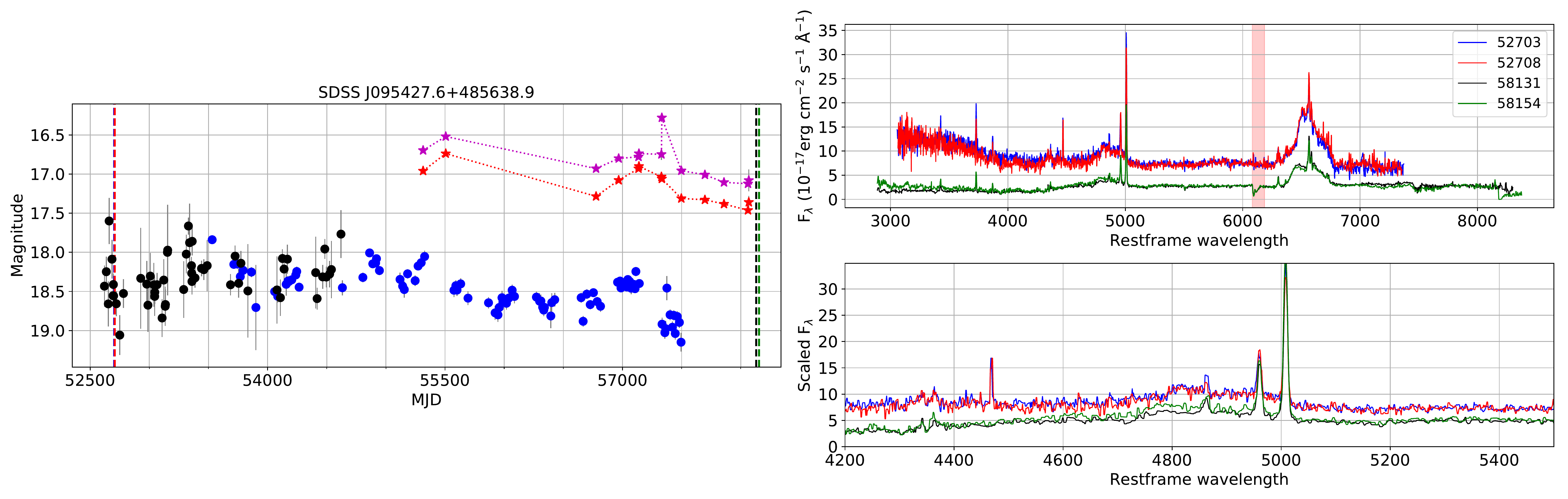}
\includegraphics[width = 7.0in]{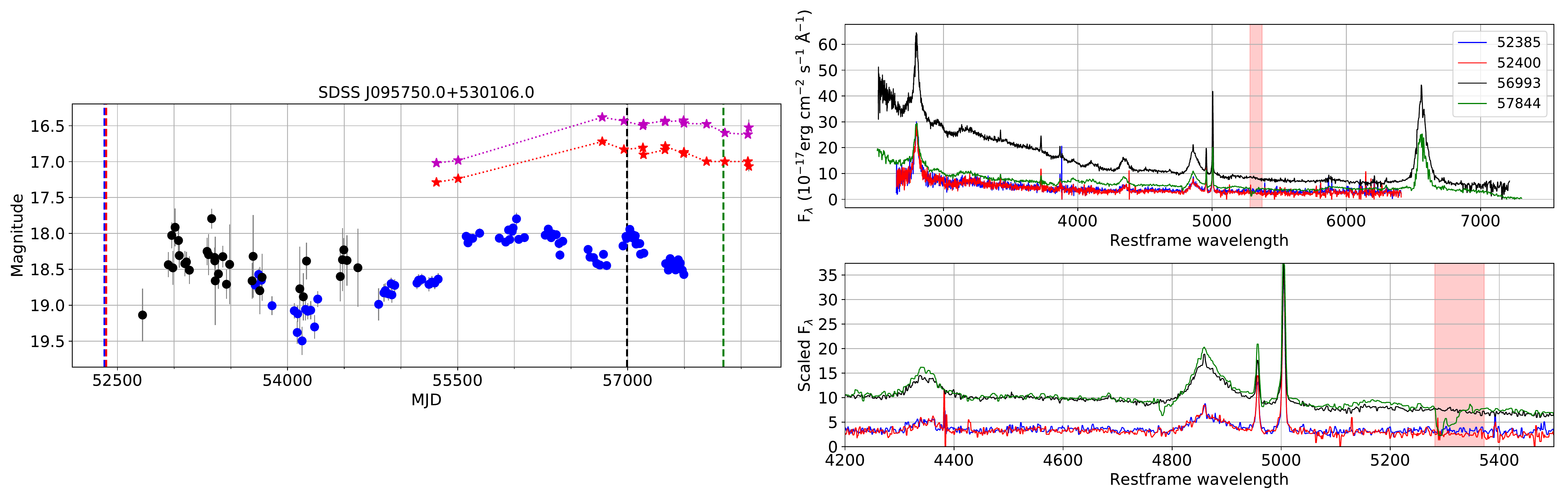}
\includegraphics[width = 7.0in]{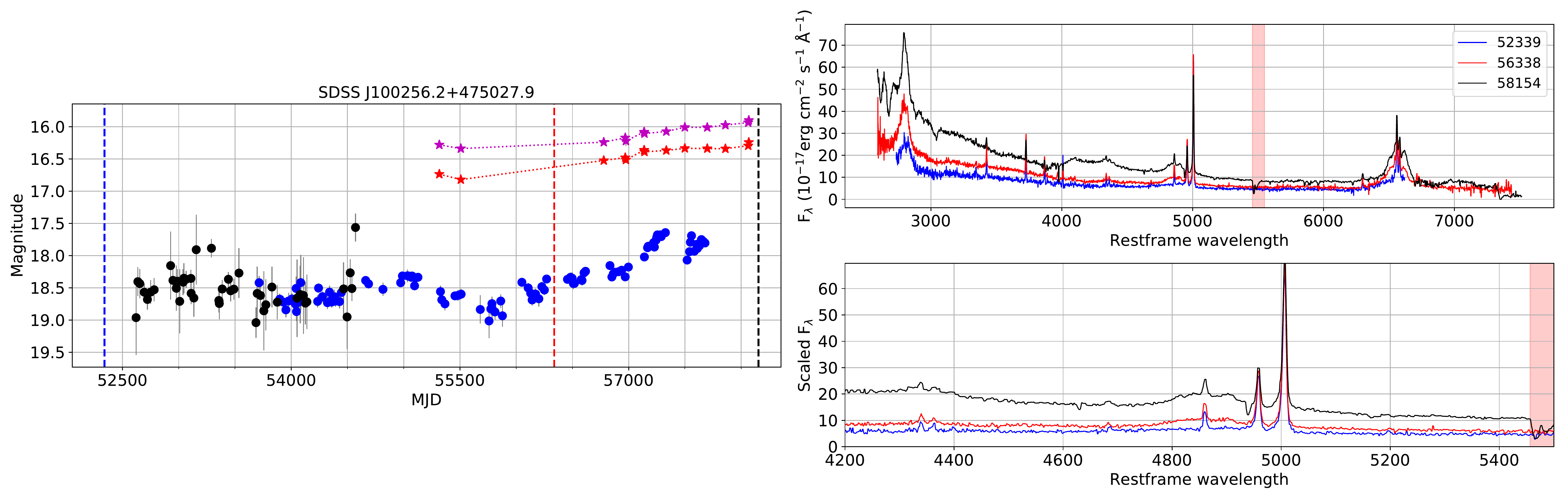}
\contcaption{}
\end{figure*}

\begin{figure*}
\centering
\includegraphics[width = 7.0in]{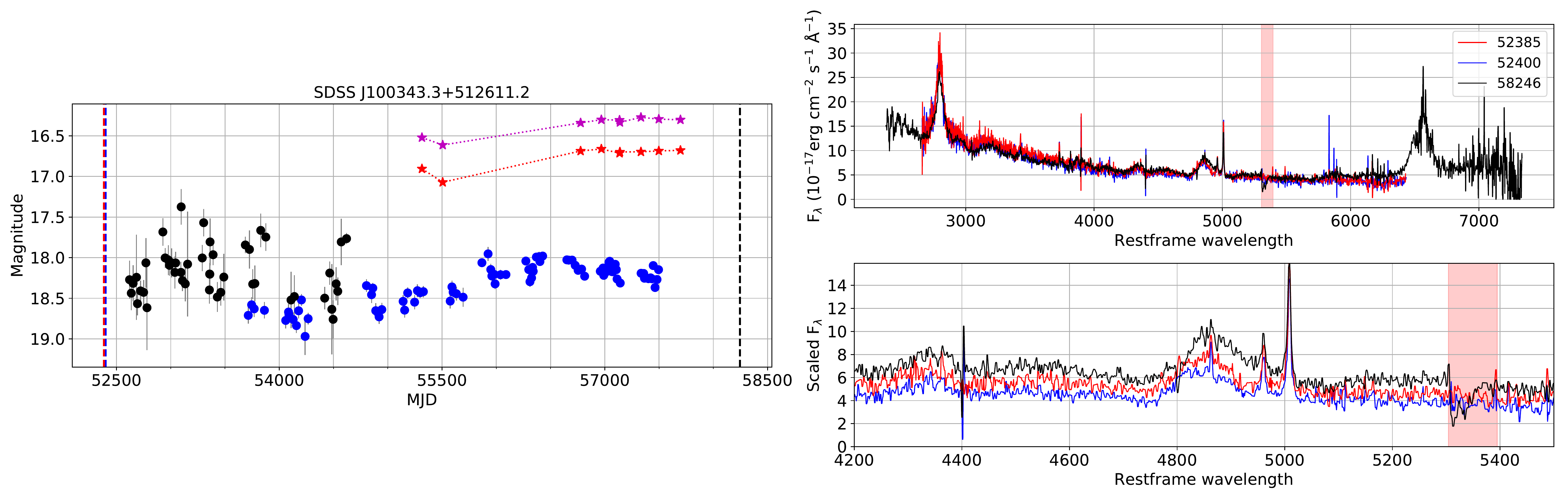}
\includegraphics[width = 7.0in]{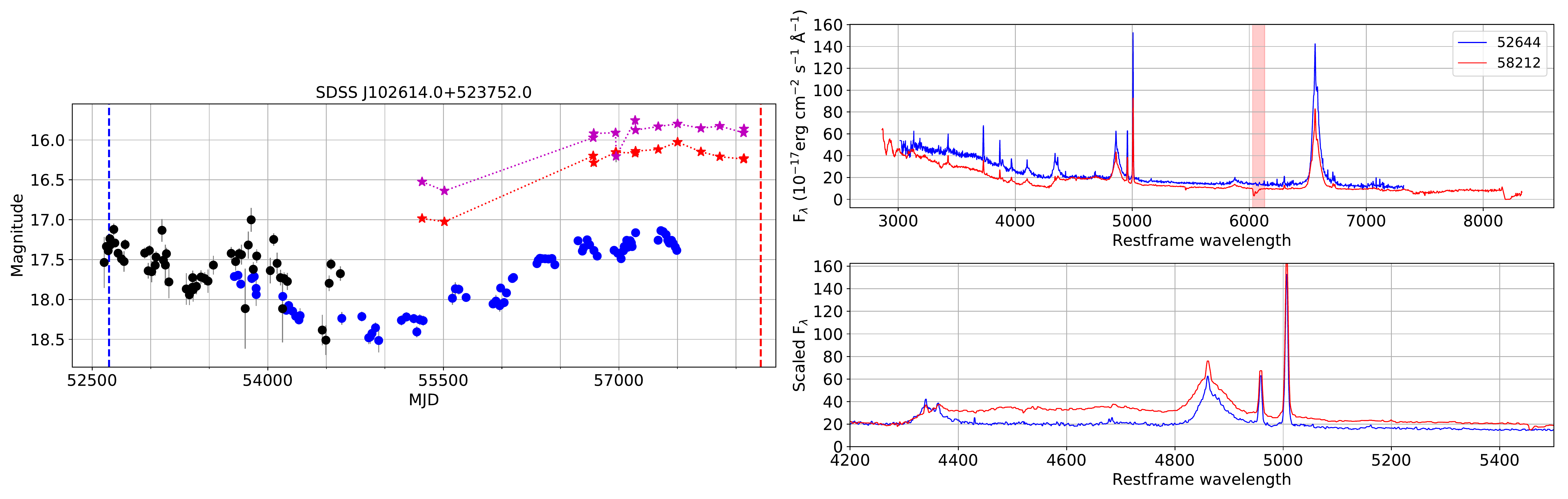}
\includegraphics[width = 7.0in]{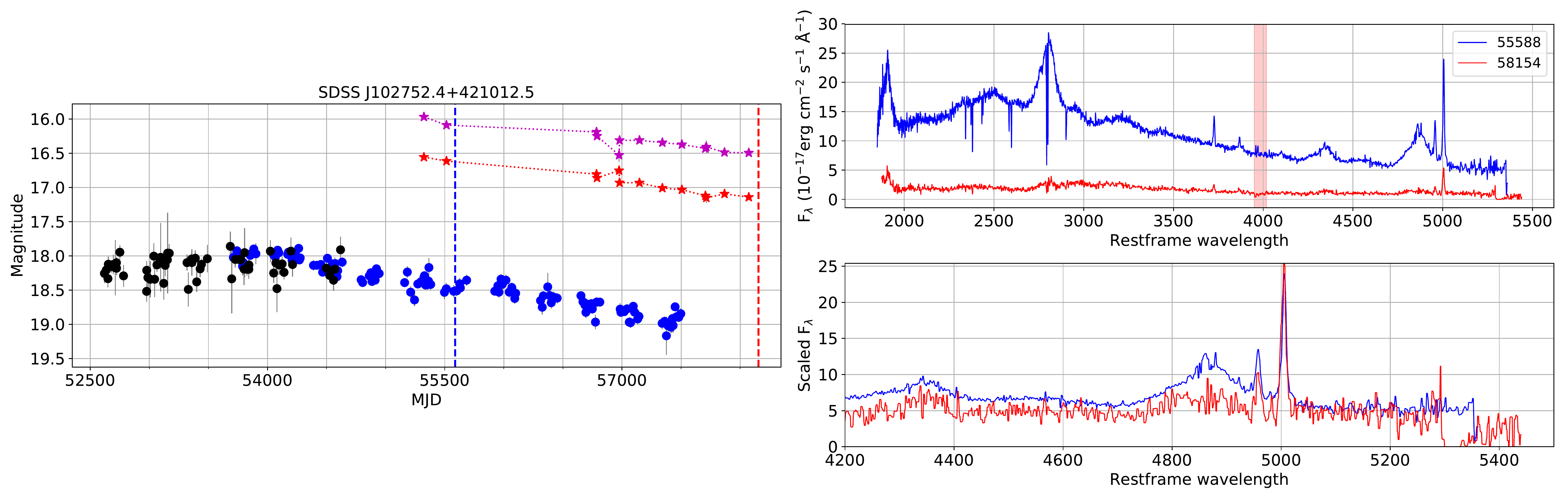}
\includegraphics[width = 7.0in]{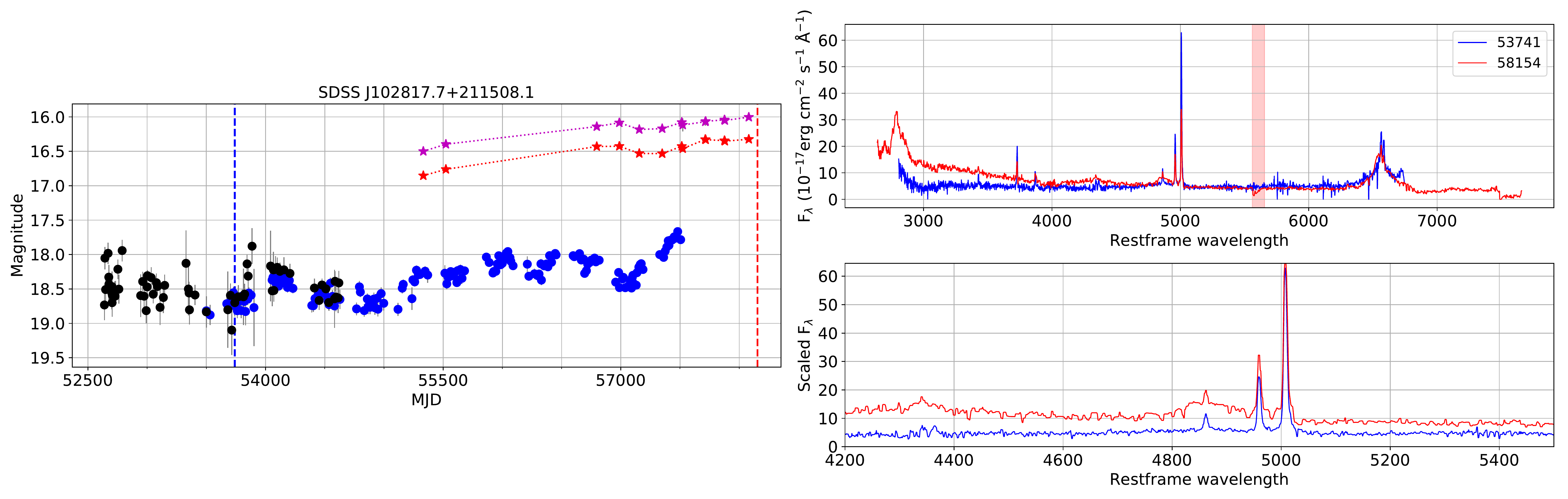}
\contcaption{}
\end{figure*}

\begin{figure*}
\centering
\includegraphics[width = 7.0in]{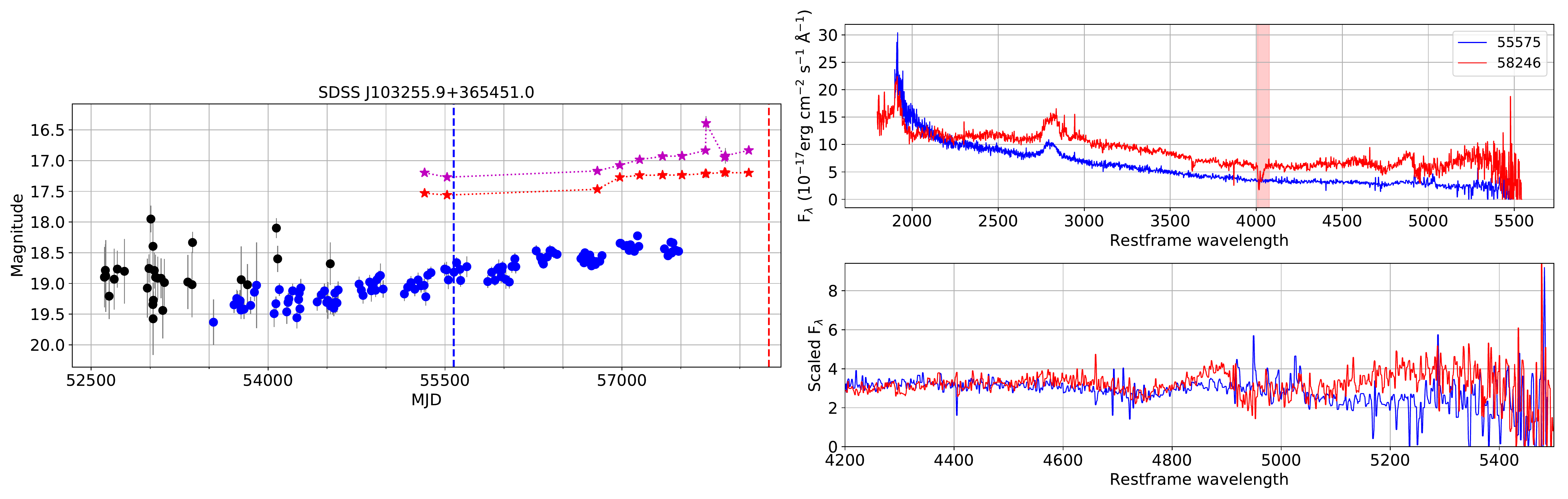}
\includegraphics[width = 7.0in]{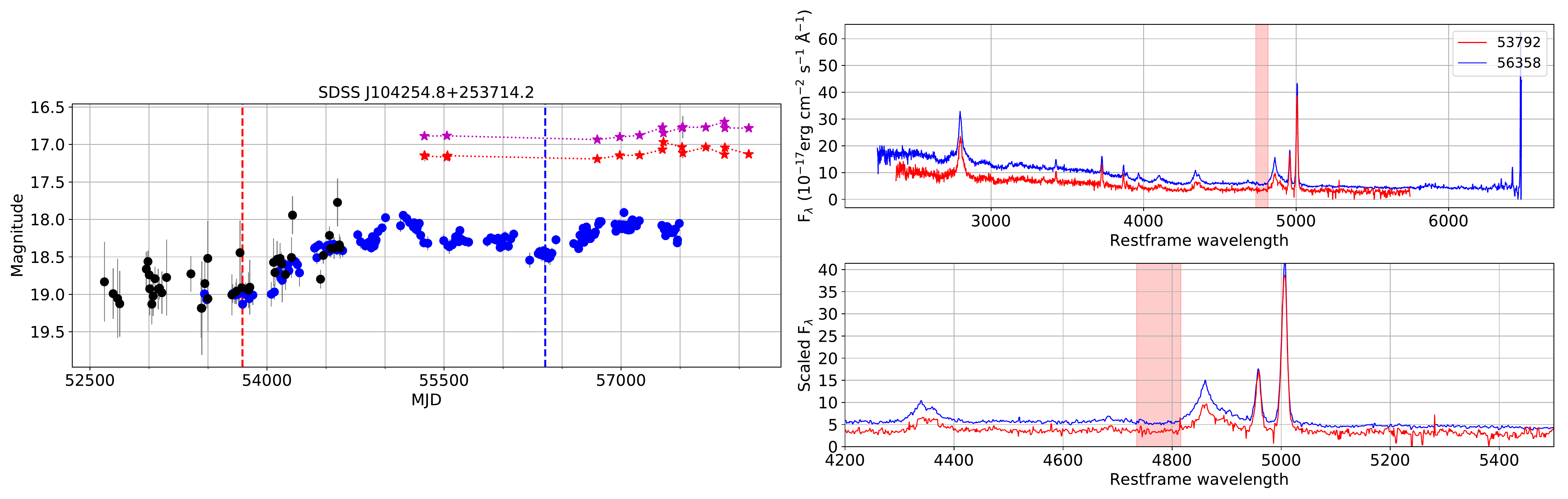}
\includegraphics[width = 7.0in]{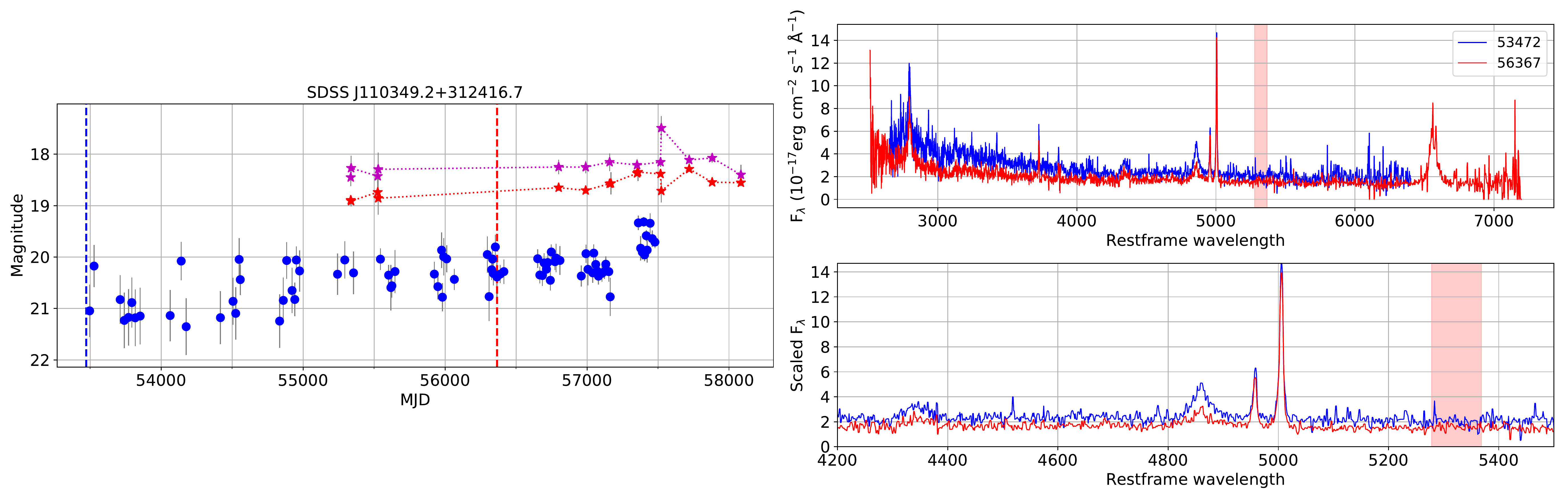}
\includegraphics[width = 7.0in]{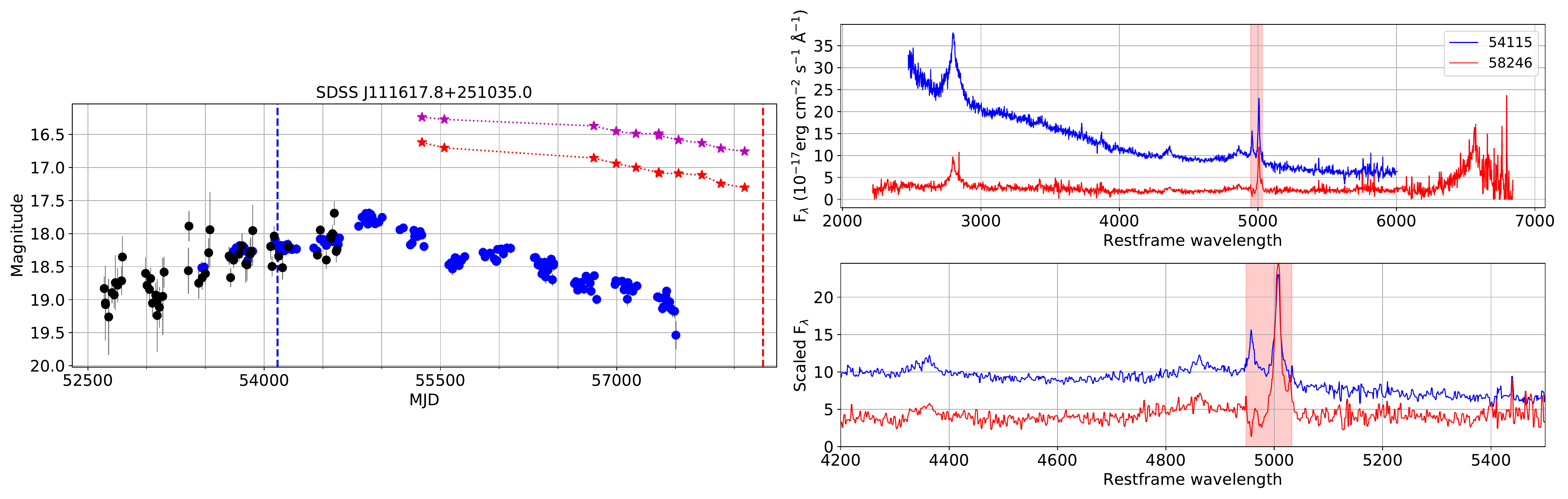}
\contcaption{}
\end{figure*}

\begin{figure*}
\centering
\includegraphics[width = 7.0in]{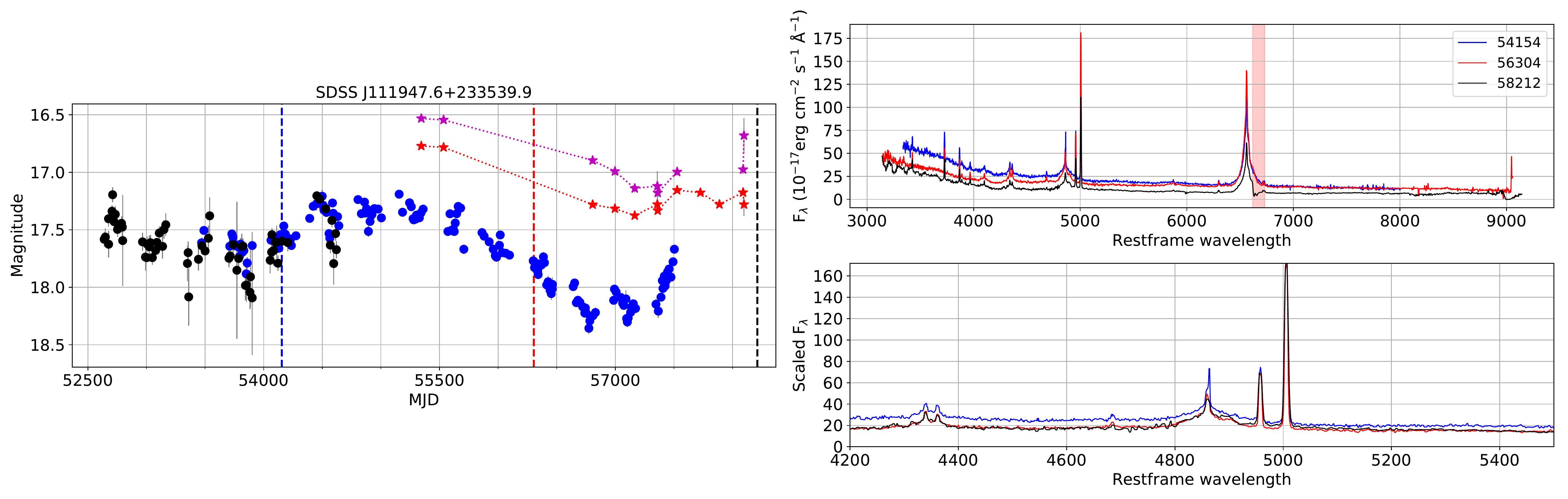}
\includegraphics[width = 7.0in]{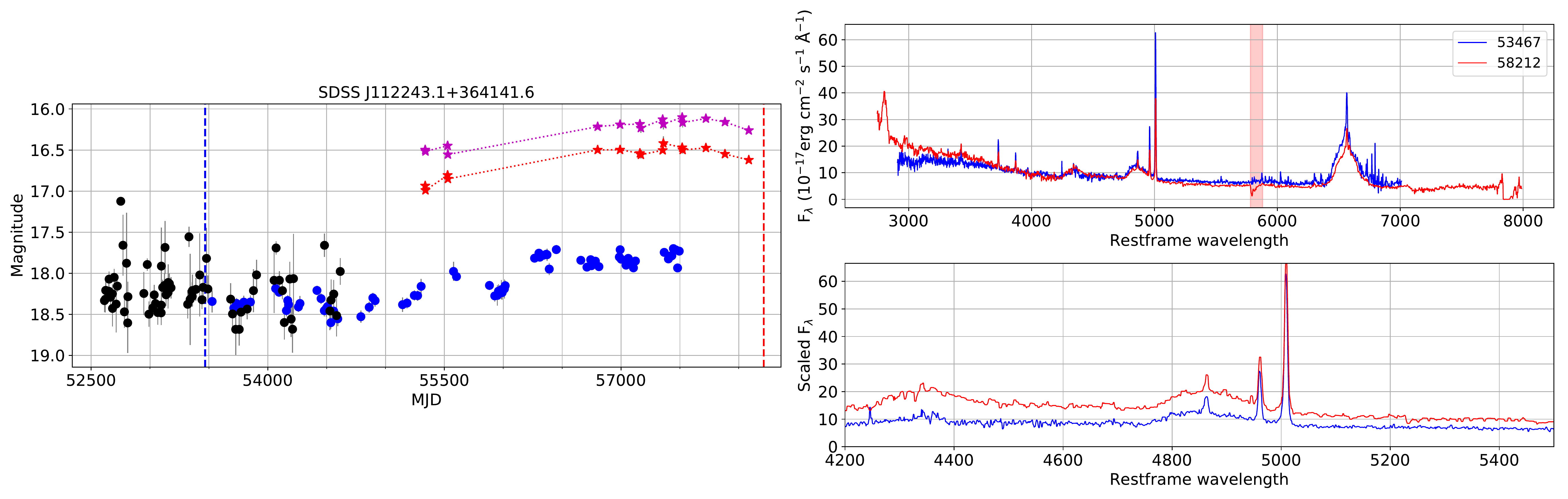}
\includegraphics[width = 7.0in]{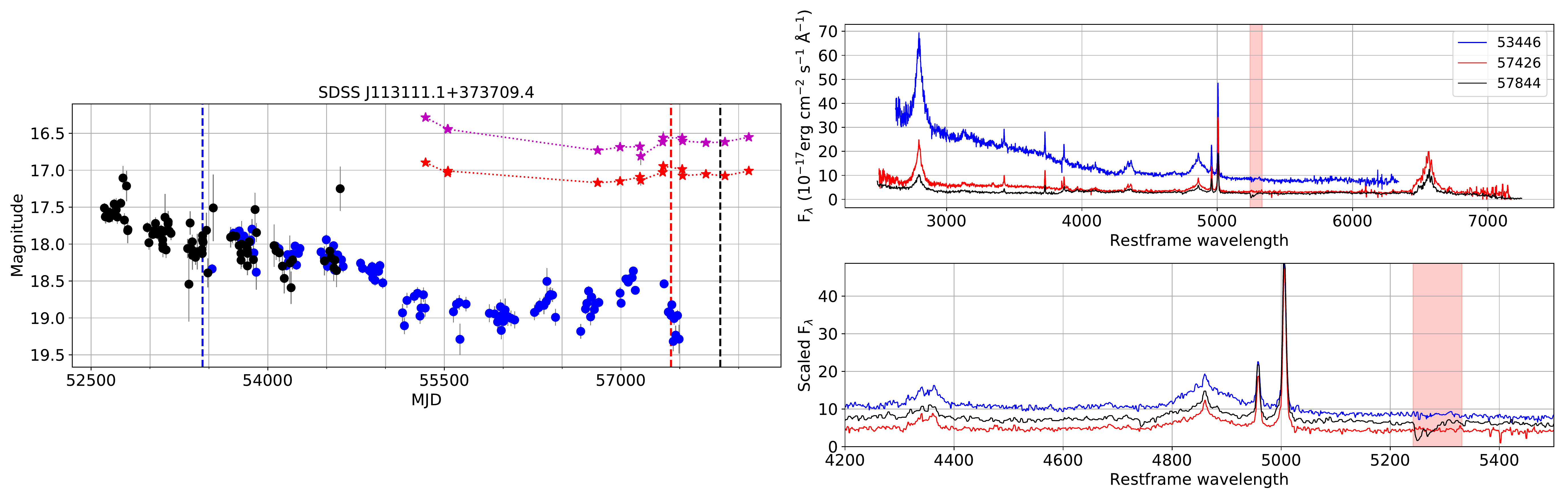}
\includegraphics[width = 7.0in]{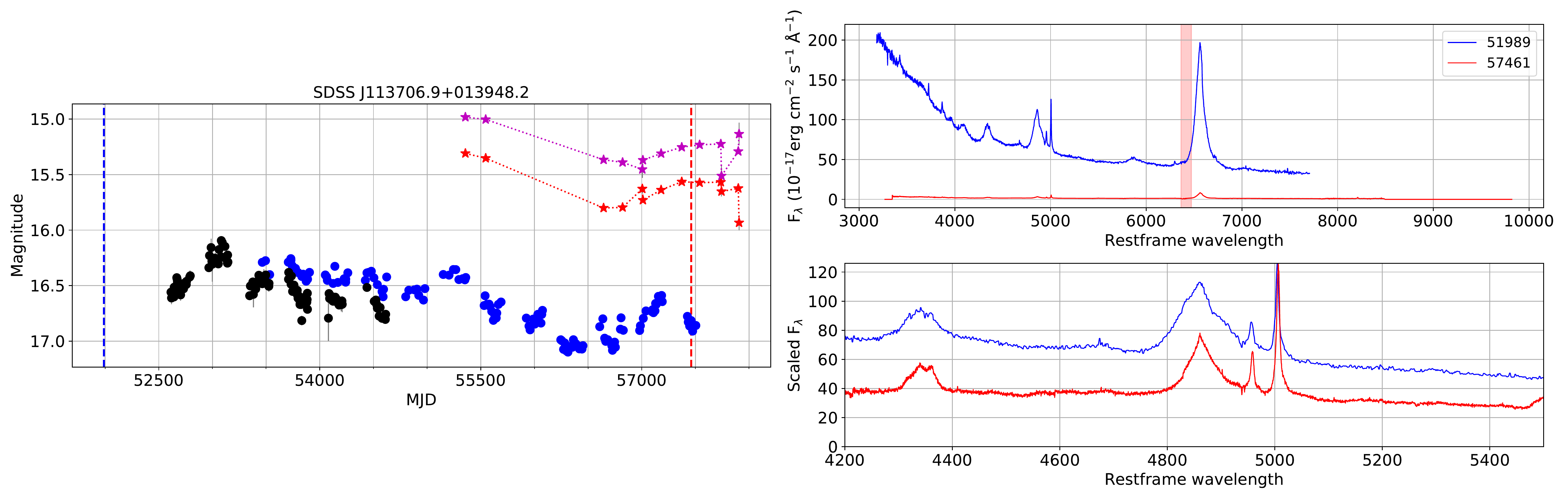}
\contcaption{}
\end{figure*}

\begin{figure*}
\centering
\includegraphics[width = 7.0in]{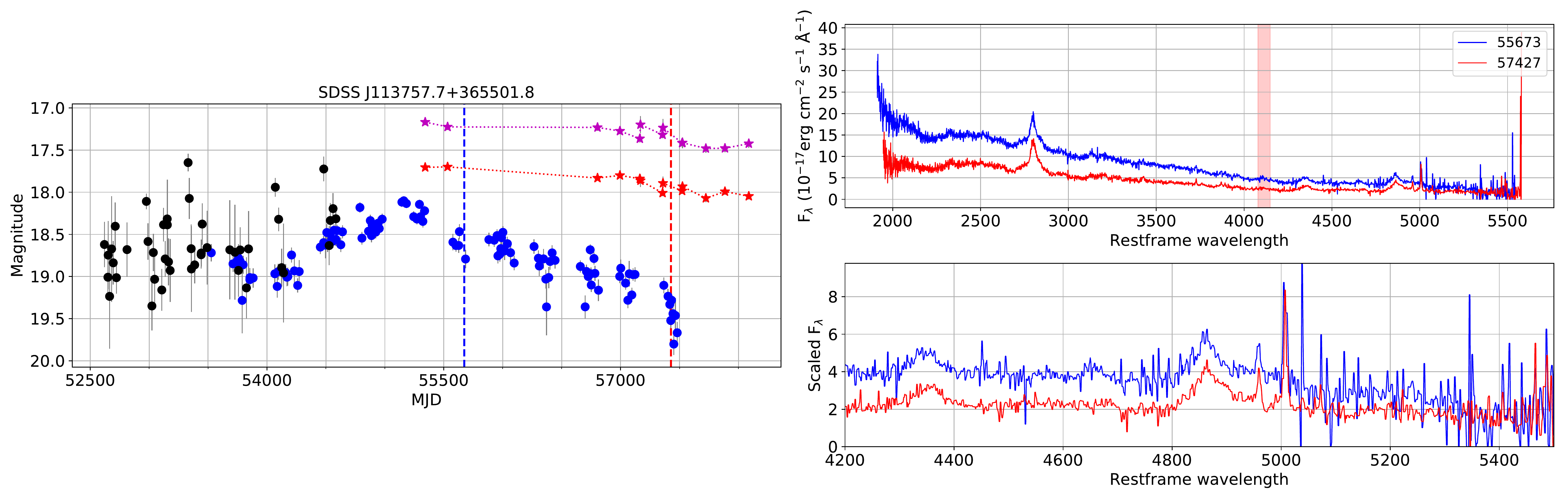}
\includegraphics[width = 7.0in]{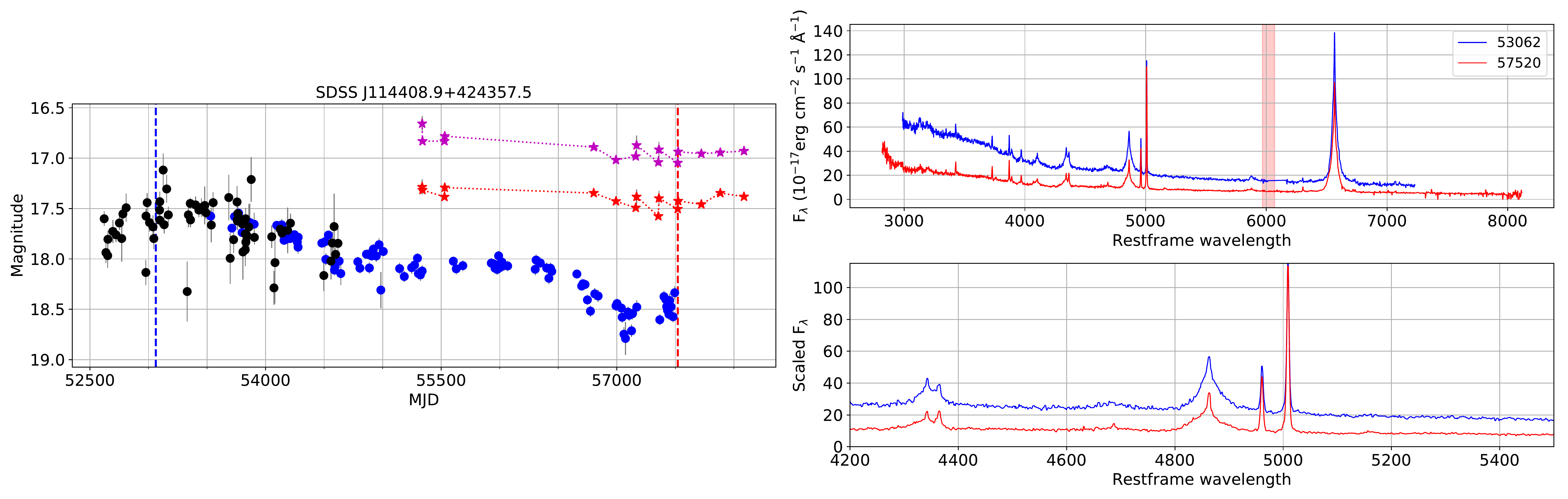}
\includegraphics[width = 7.0in]{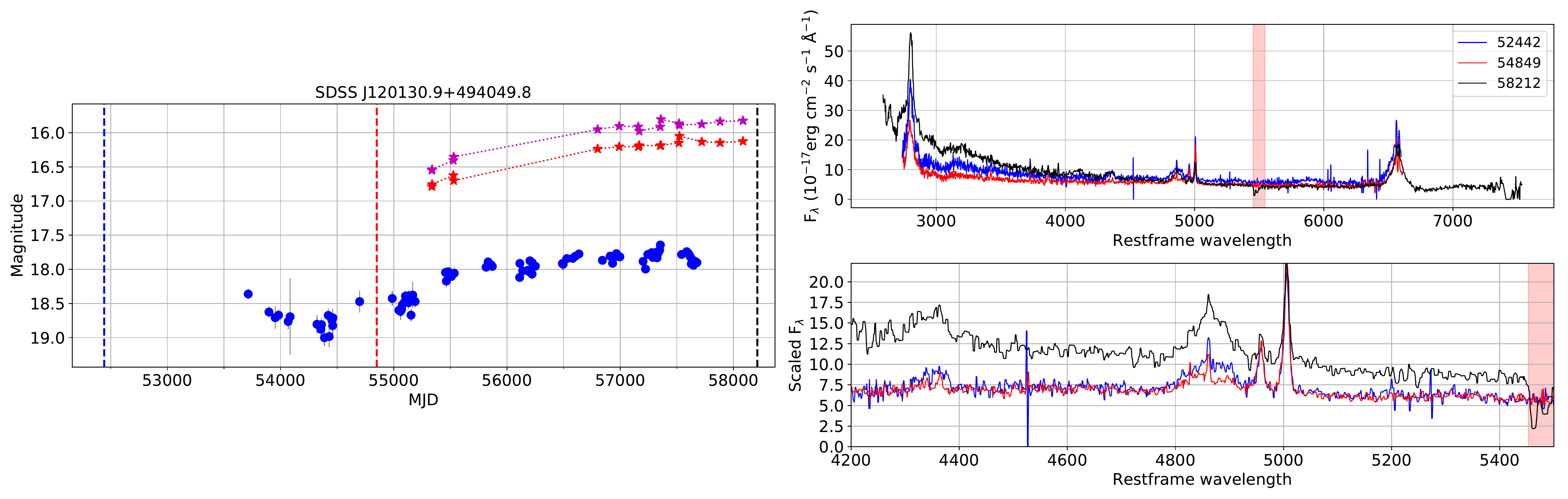}
\includegraphics[width = 7.0in]{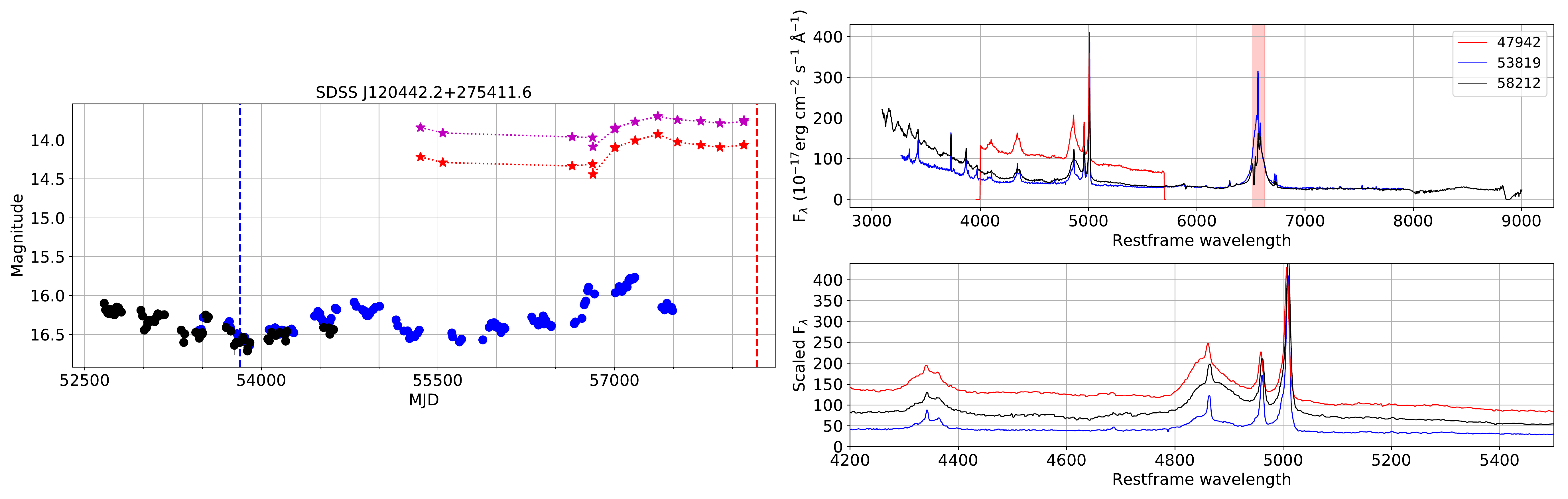}
\contcaption{}
\end{figure*}

\begin{figure*}
\centering
\includegraphics[width = 7.0in]{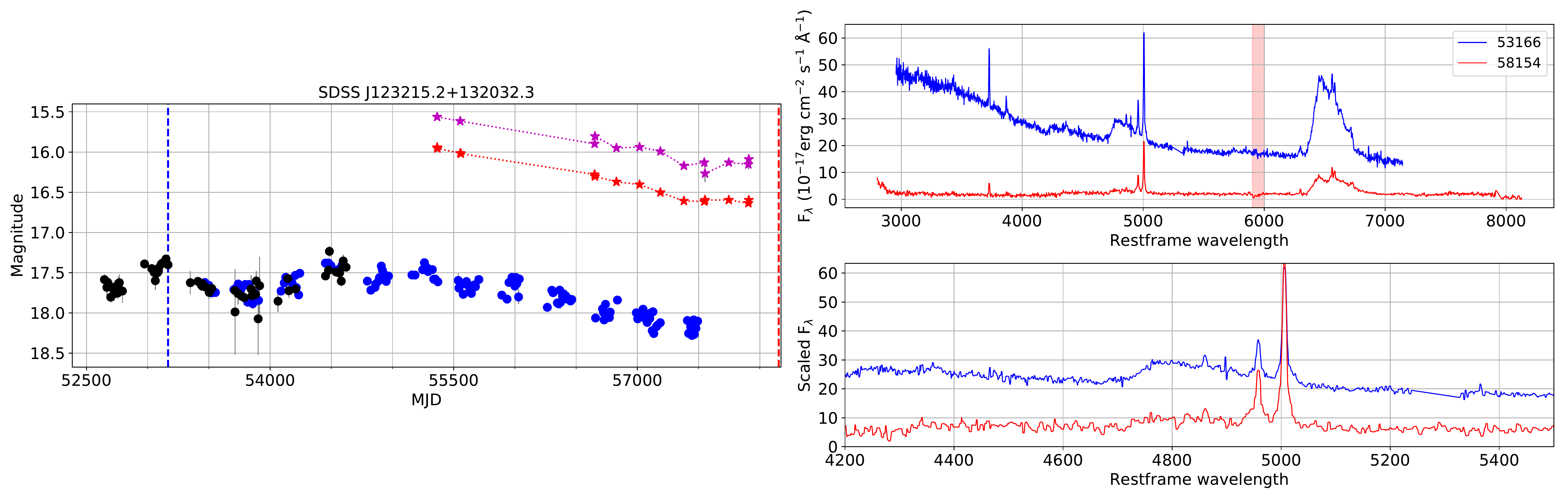}
\includegraphics[width = 7.0in]{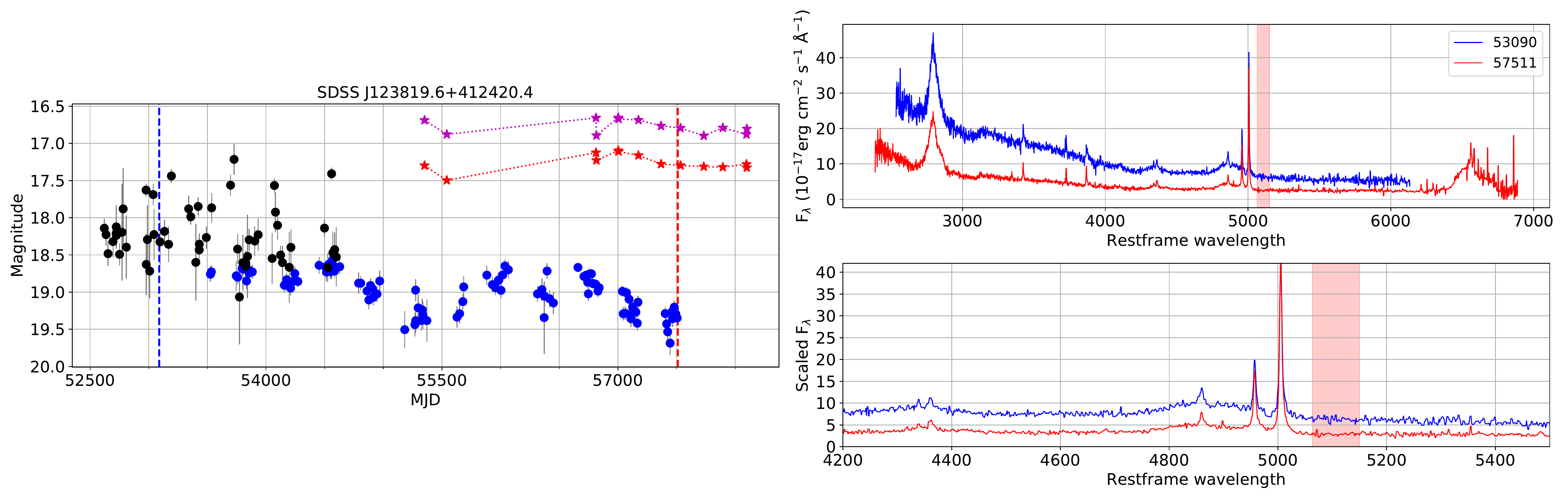}
\includegraphics[width = 7.0in]{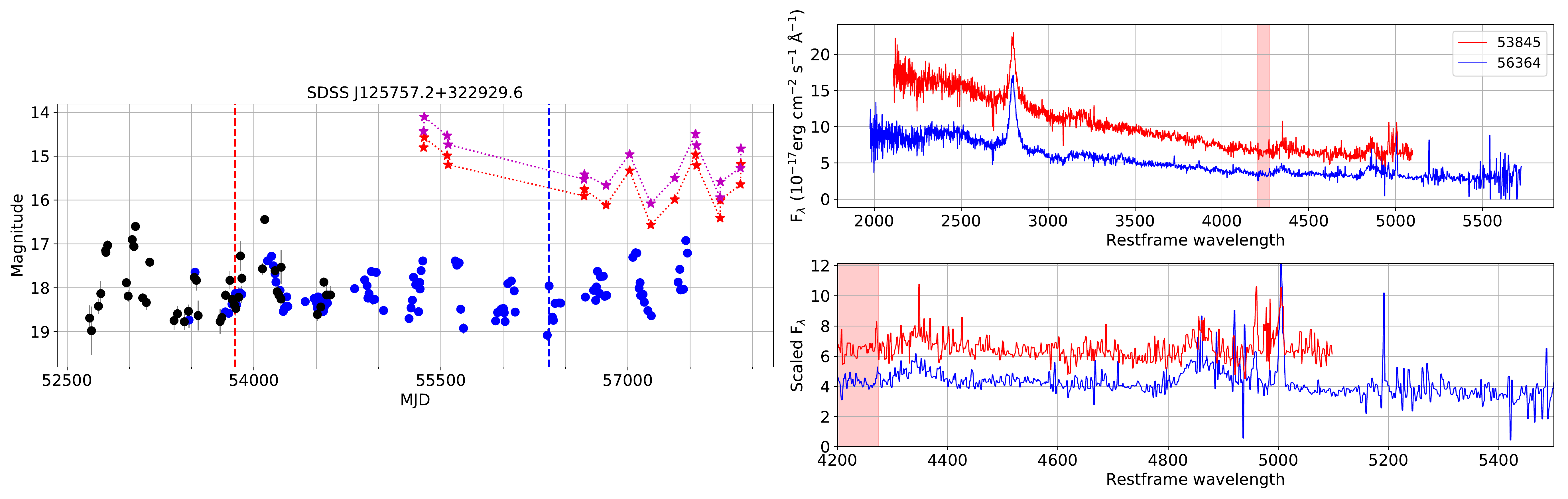}
\includegraphics[width = 7.0in]{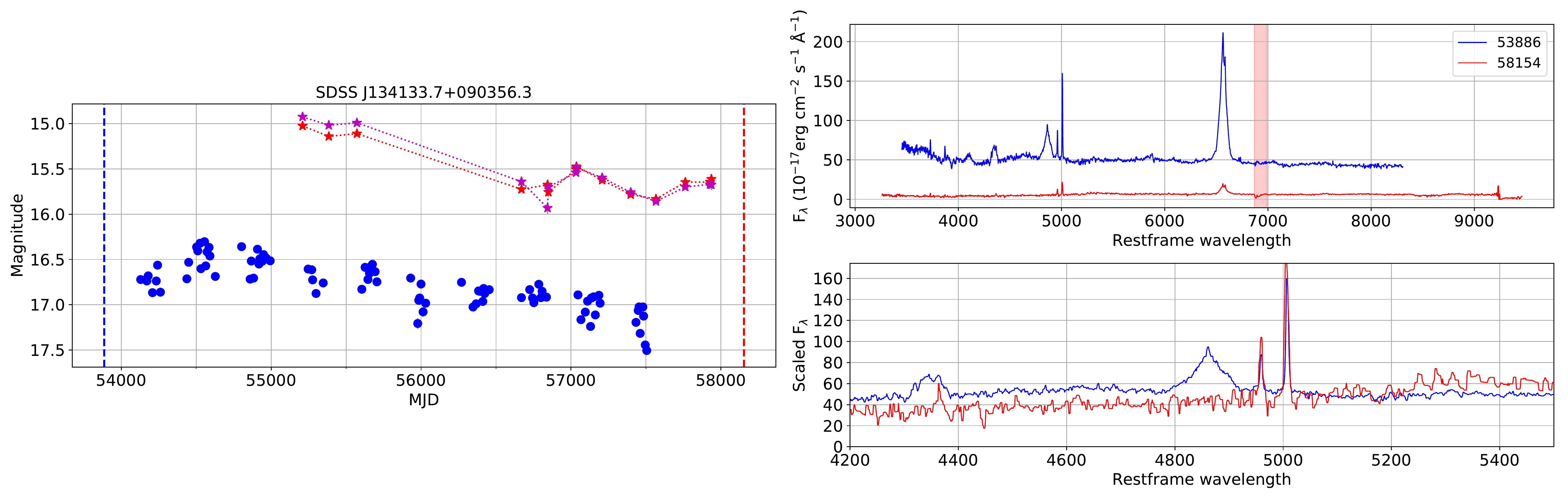}
\contcaption{}
\end{figure*}

\begin{figure*}
\centering
\includegraphics[width = 7.0in]{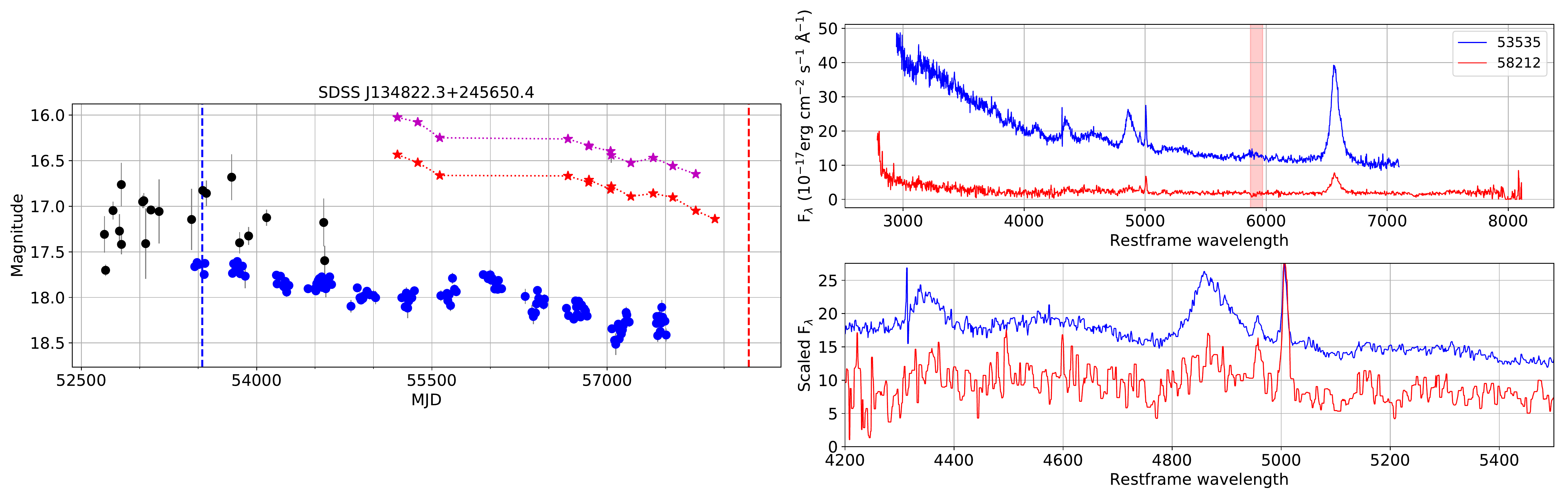}
\includegraphics[width = 7.0in]{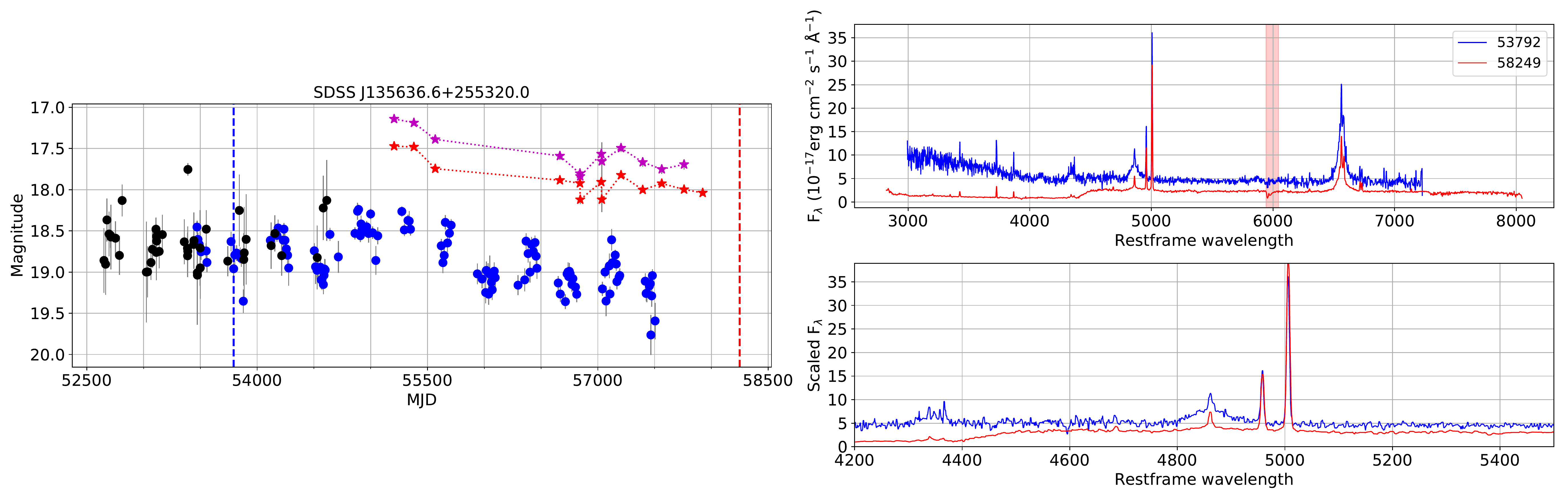}
\includegraphics[width = 7.0in]{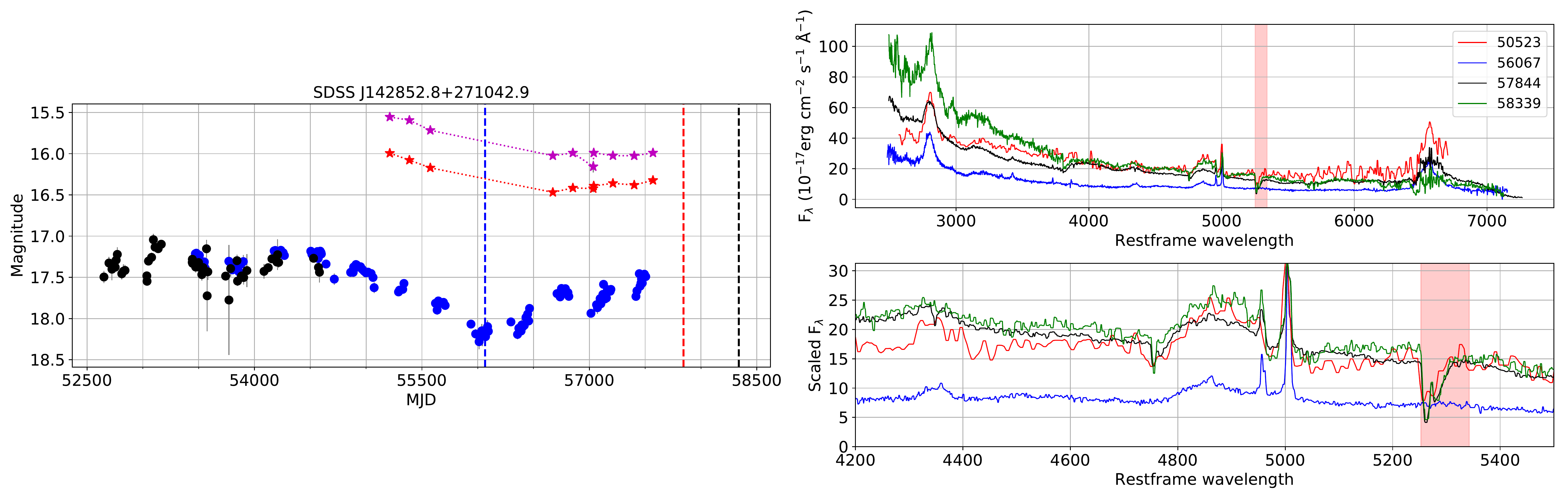}
\includegraphics[width = 7.0in]{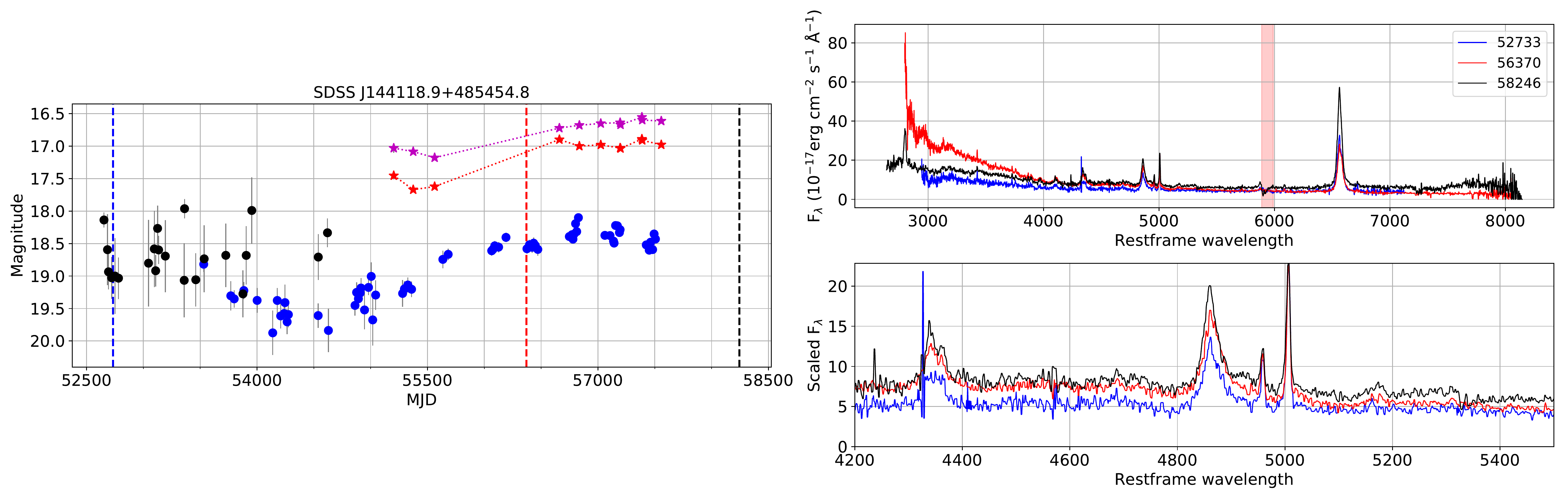}
\contcaption{}
\end{figure*}

\begin{figure*}
\centering
\includegraphics[width = 7.0in]{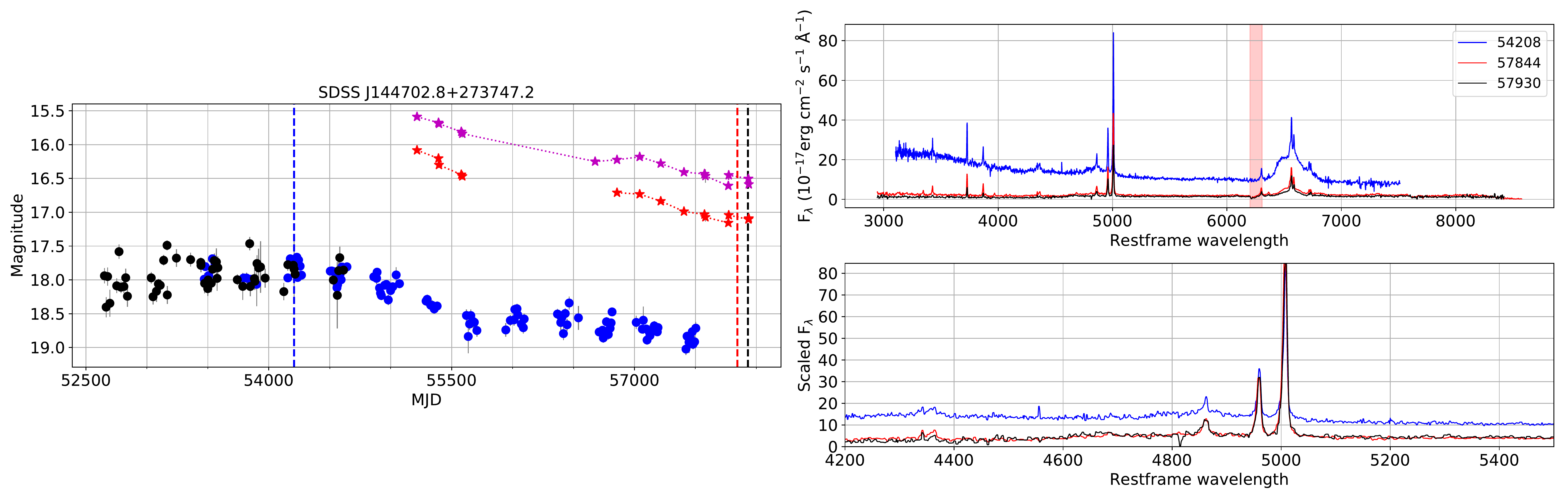}
\includegraphics[width = 7.0in]{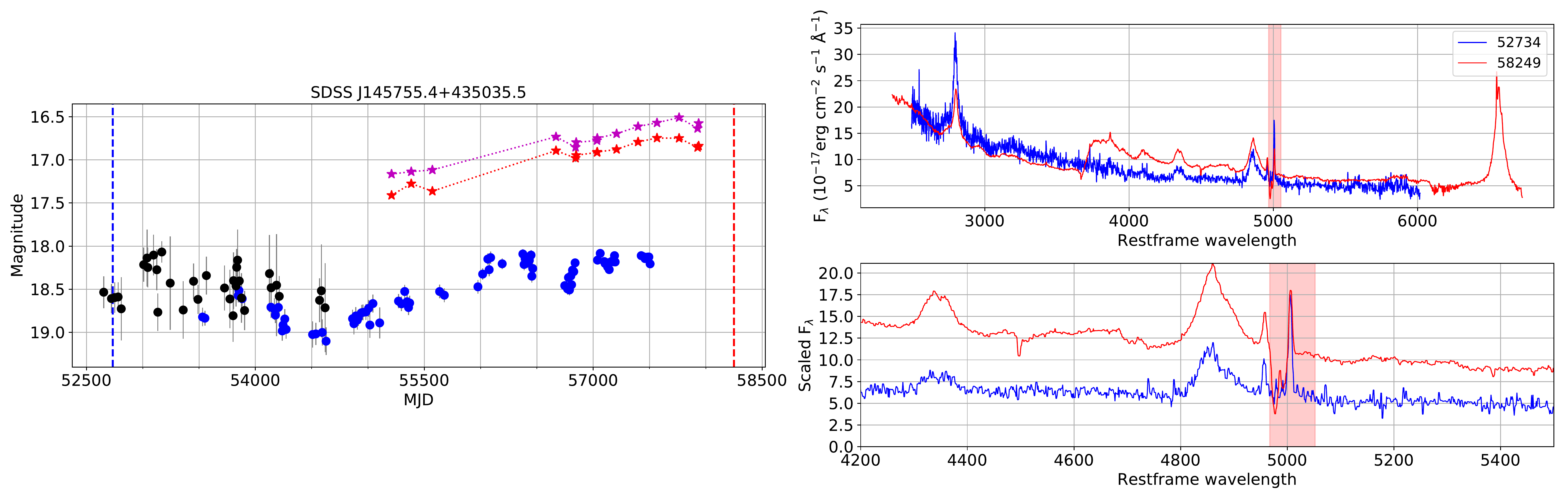}
\includegraphics[width = 7.0in]{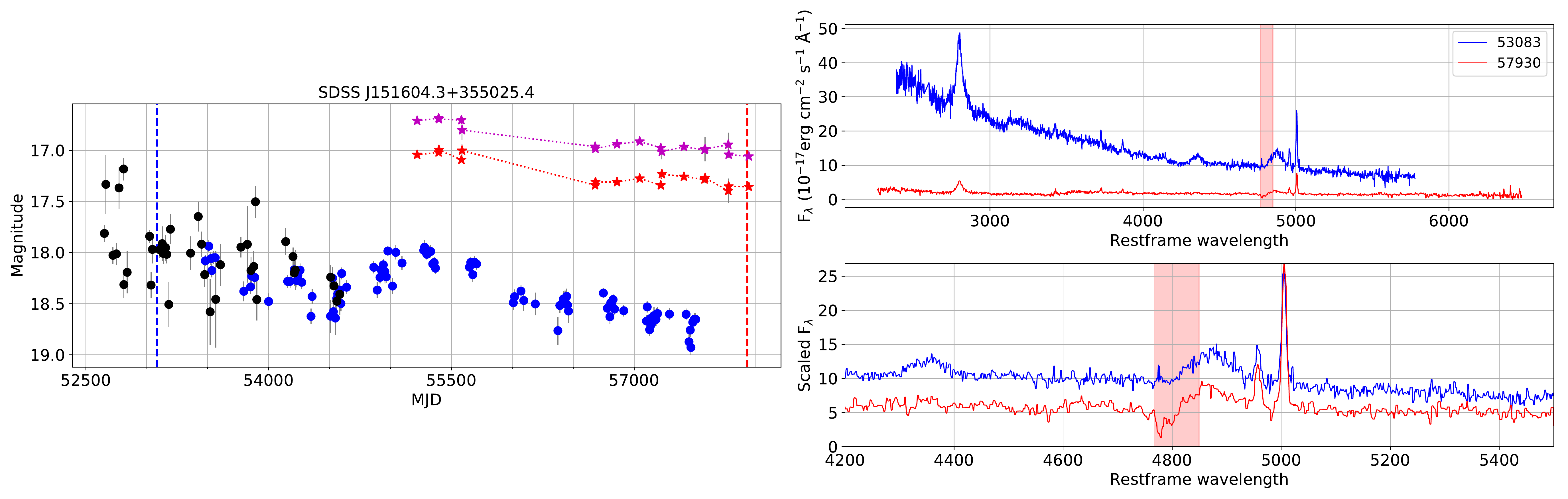}
\includegraphics[width = 7.0in]{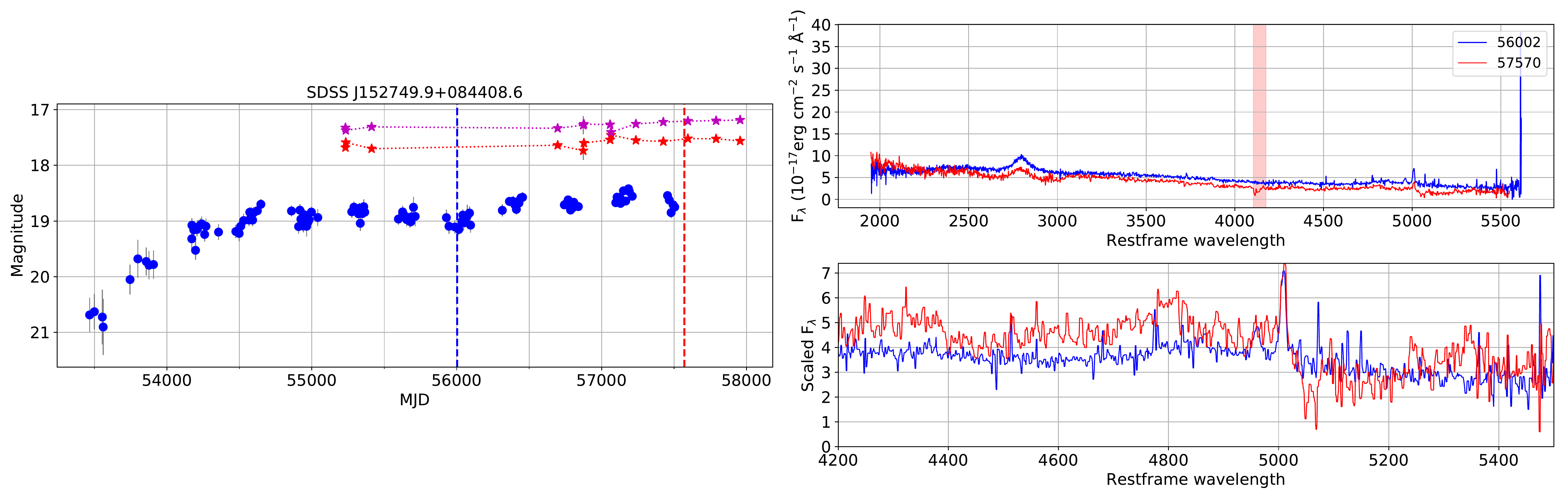}
\contcaption{}
\end{figure*}

\begin{figure*}
\centering
\includegraphics[width = 7.0in]{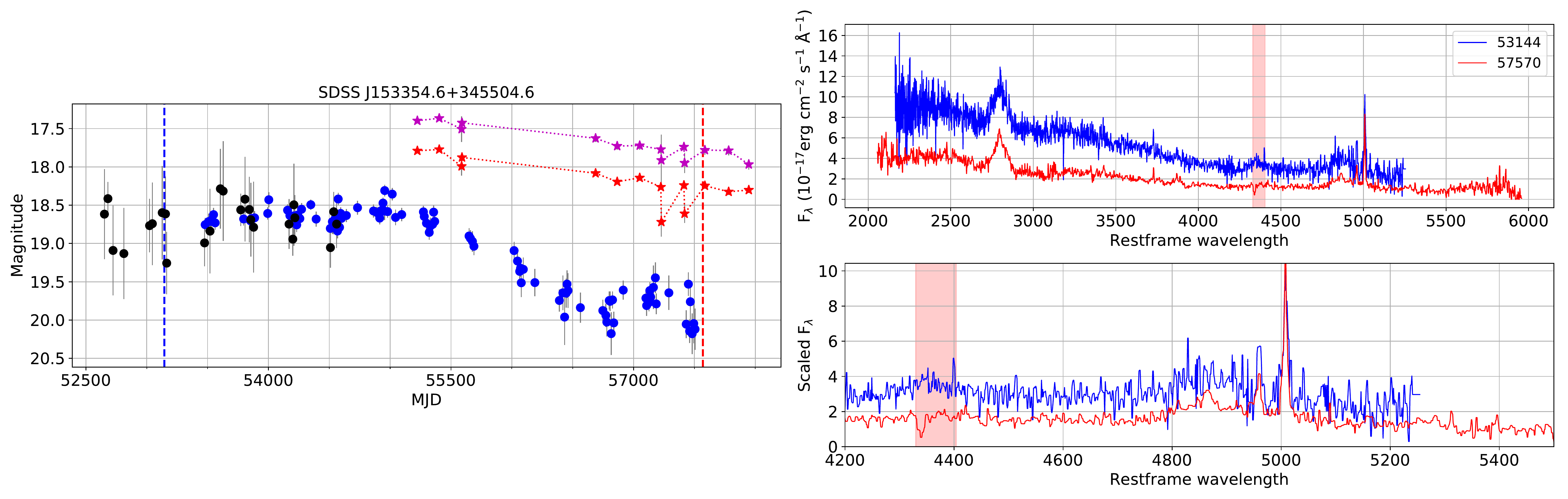}
\includegraphics[width = 7.0in]{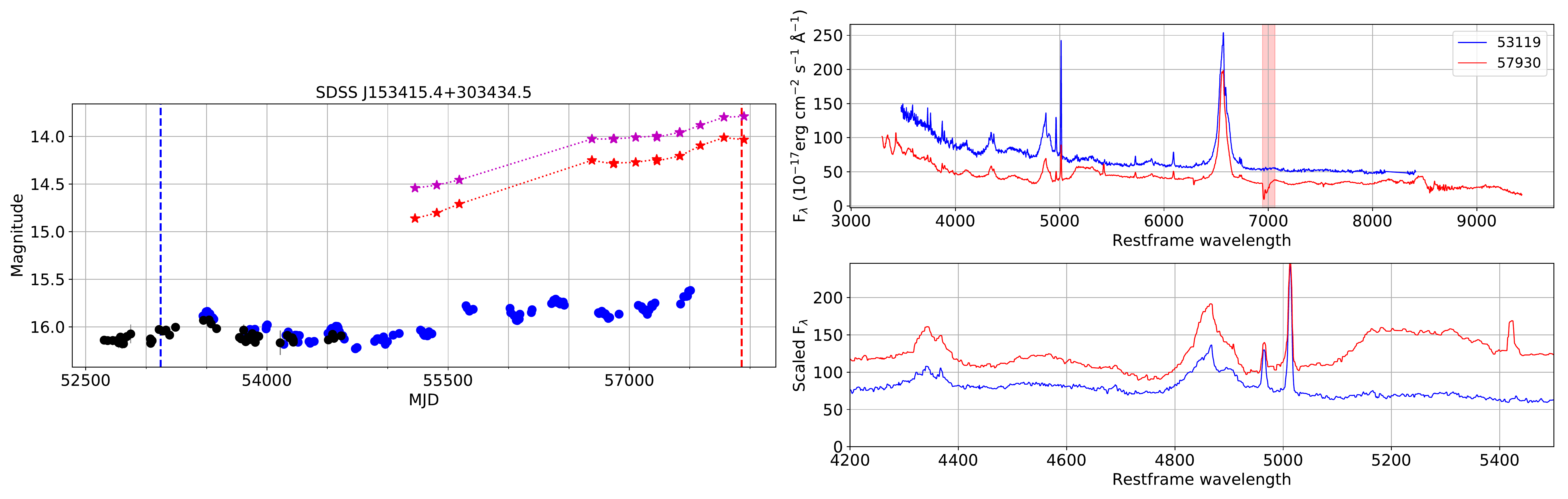}
\includegraphics[width = 7.0in]{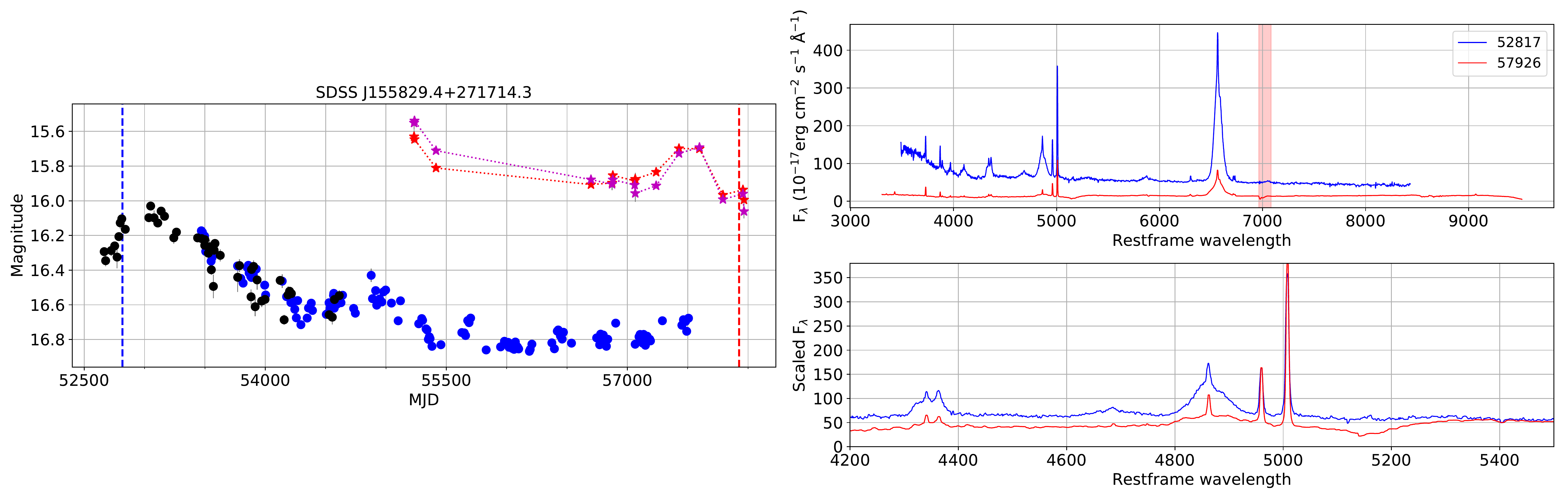}
\includegraphics[width = 7.0in]{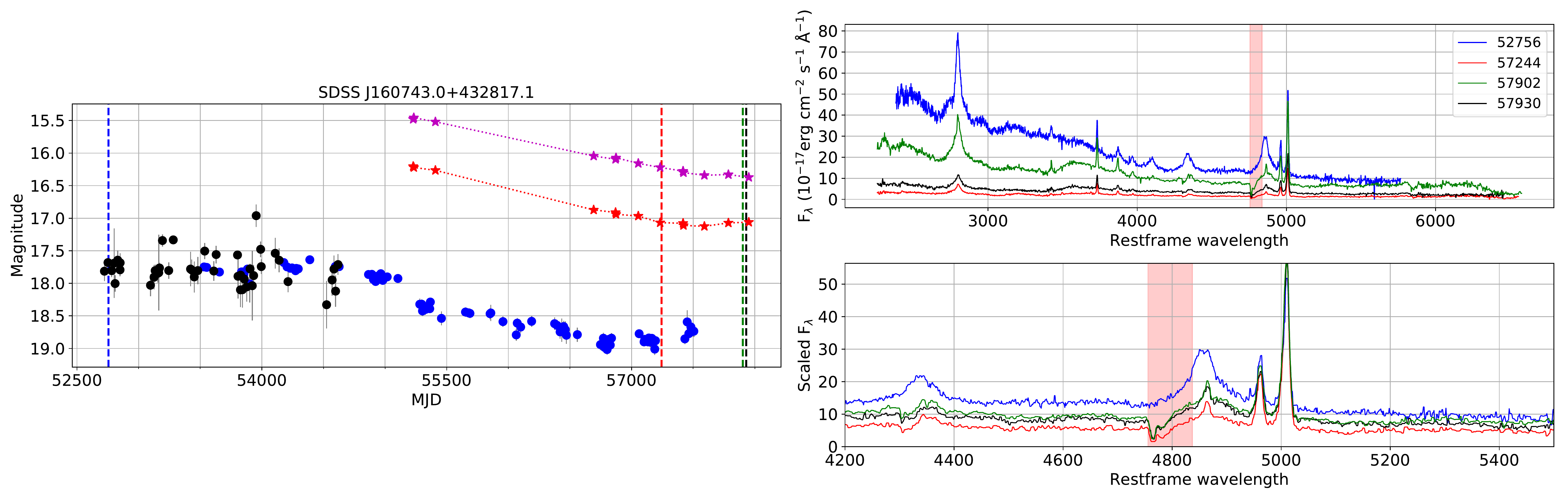}
\contcaption{}
\end{figure*}

\begin{figure*}
\centering
\includegraphics[width = 7.0in]{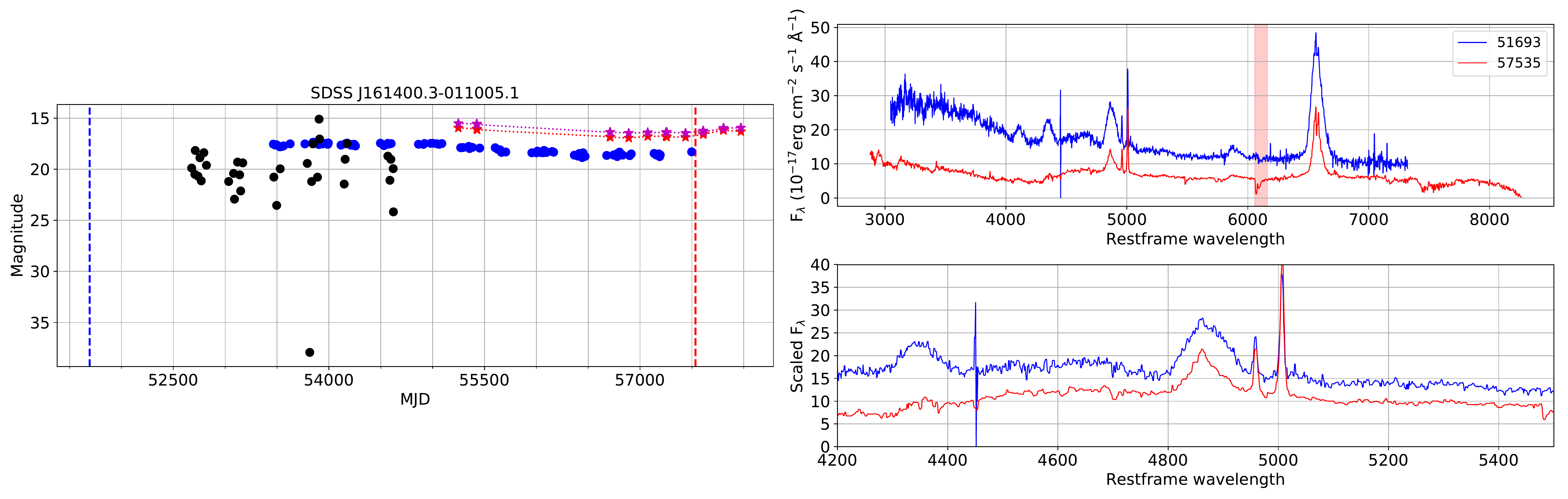}
\includegraphics[width = 7.0in]{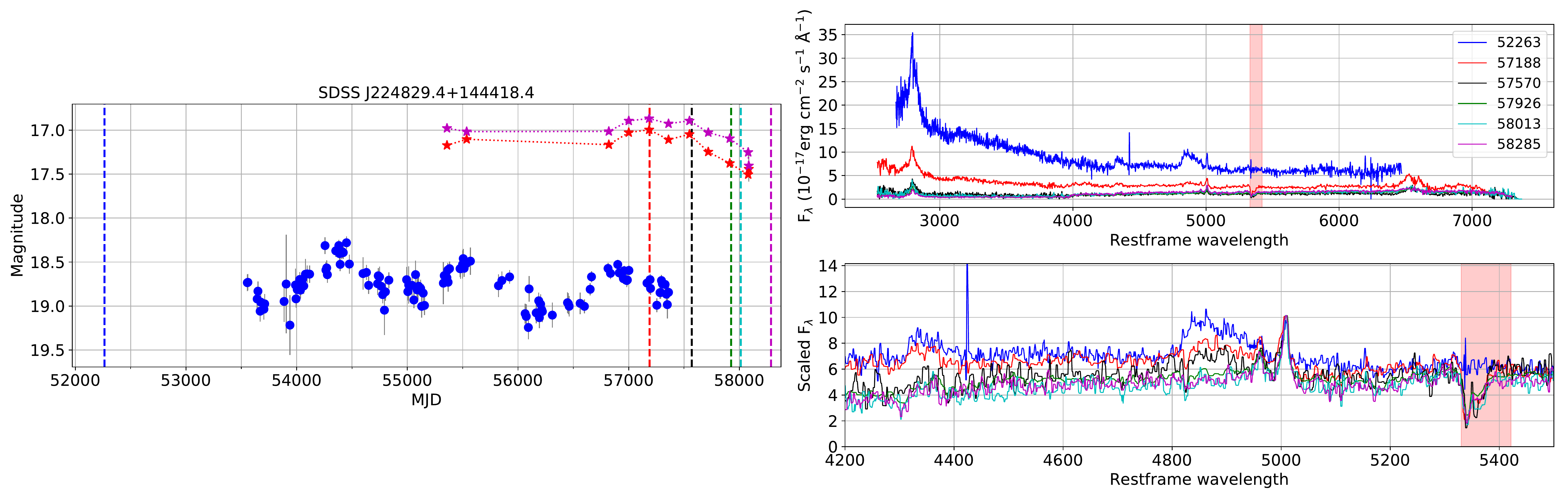}
\includegraphics[width = 7.0in]{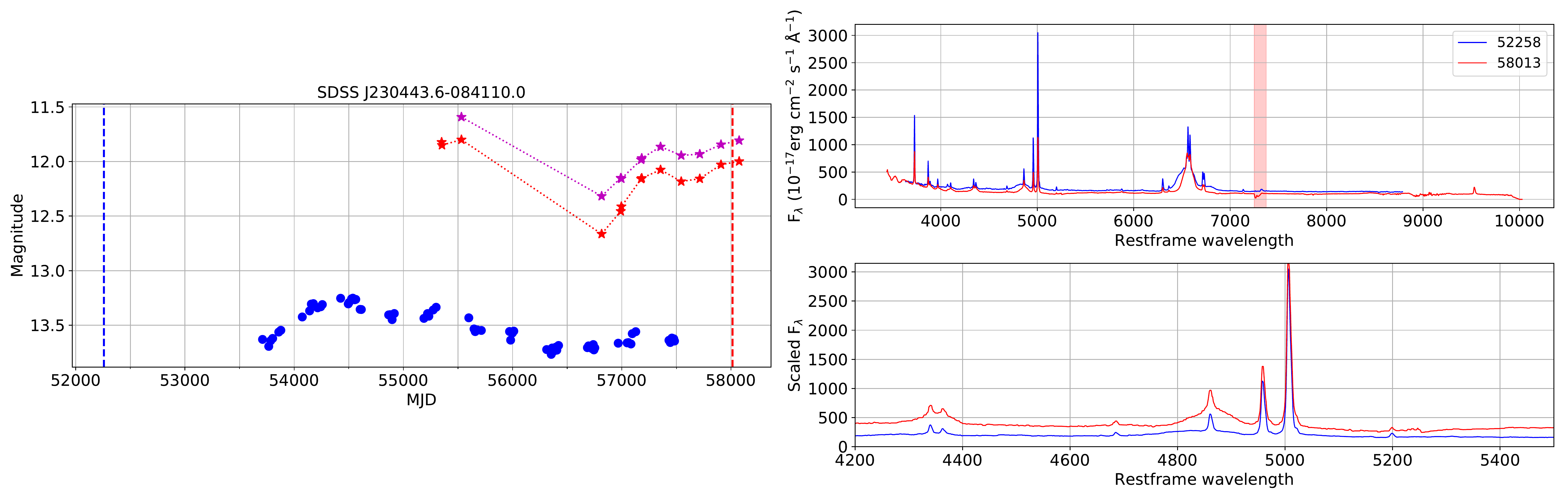}
\includegraphics[width = 7.0in]{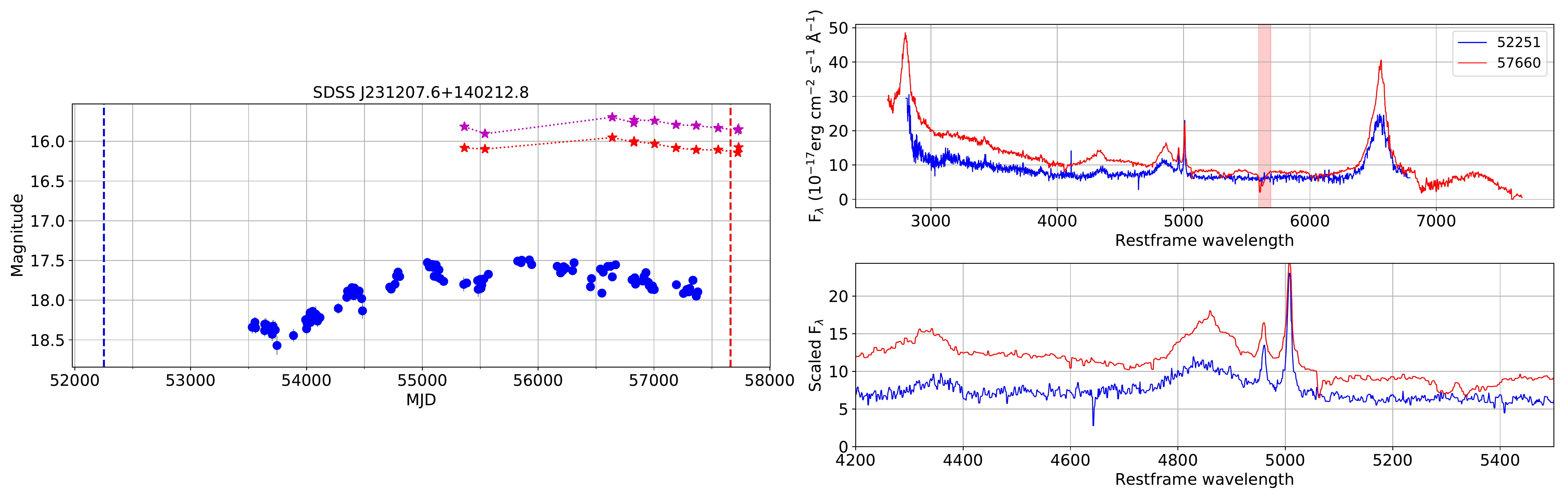}
\contcaption{}
\end{figure*}

\begin{figure*}
\centering
\includegraphics[width = 7.0in]{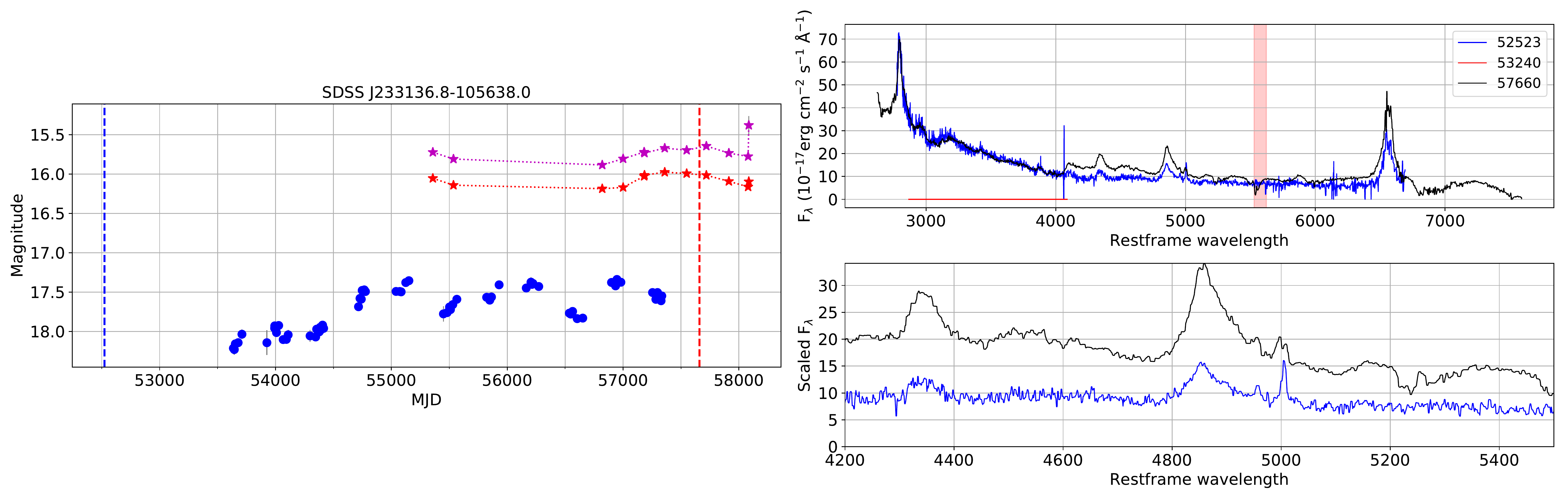}
\includegraphics[width = 7.0in]{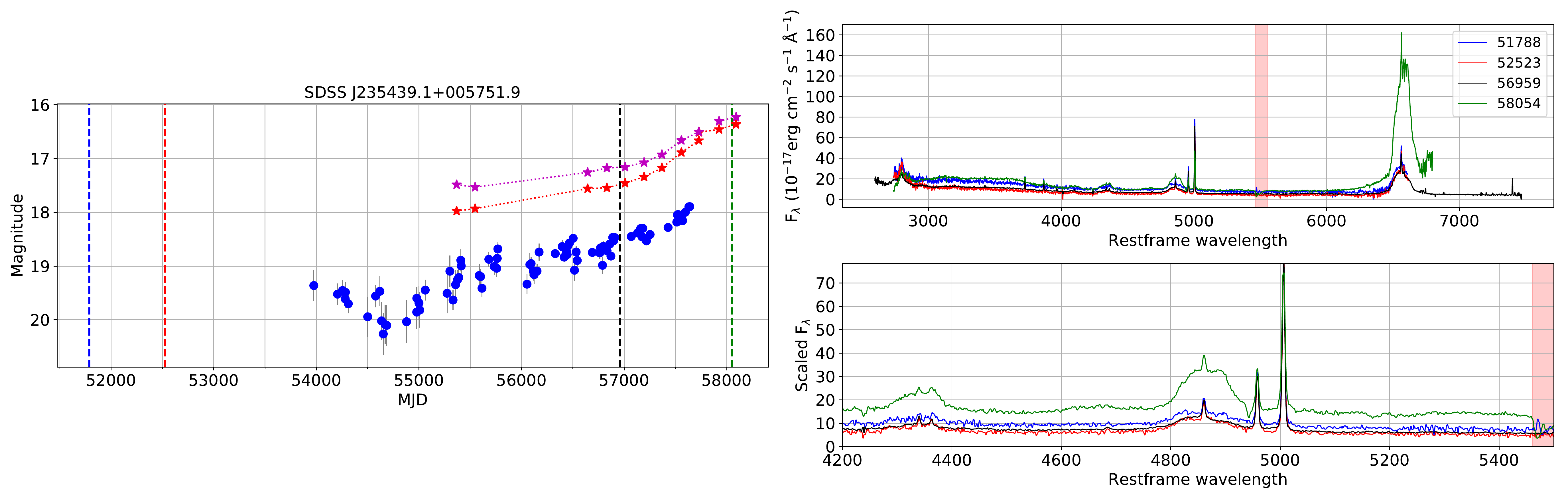}
\contcaption{}
\end{figure*}



\bsp	
\label{lastpage}
\end{document}